\documentclass{cernrep}
\usepackage{texnames}
\usepackage[T1]{fontenc}
\usepackage{varwidth}
\usepackage{xcolor}

\usepackage{dcolumn}
\newcolumntype{d}{D{.}{.}{2.5}}
\newcolumntype{s}{D{.}{.}{1.2}}

\begin{document}

\title{Beam Cleaning and Collimation Systems}

\author{S. Redaelli}

\institute{CERN, Geneva, Switzerland}

\maketitle 

\begin{abstract}
Collimation systems in particle accelerators are designed to dispose of un\-avoidable losses safely and efficiently during beam operation. Different
roles are re\-quired for different types of accelerator. The present
state of the art
in beam collimation is exemplified in high-intensity, high-energy super\-conducting
had\-ron colliders, like the CERN Large Hadron Collider (LHC), where stored beam
energies reach levels up to several orders of magnitude higher than the tiny ener\-gies
required to quench cold magnets. Collimation systems are   essen\-tial
systems for the daily operation of these modern machines. In this docu\-ment, the
design of a multistage collimation system is reviewed, tak\-ing the LHC as an example case
study. In this case, unprecedented cleaning perform\-ance has been
achieved, together with a system complexity compar\-able to no other accelerator.
Aspects related to collimator design and operational chal\-lenges of large
collimation systems are also addressed.
\end{abstract}



\section{Introduction}
\label{intro}
\begin{figure}
  \centering
  \includegraphics[width=110mm]{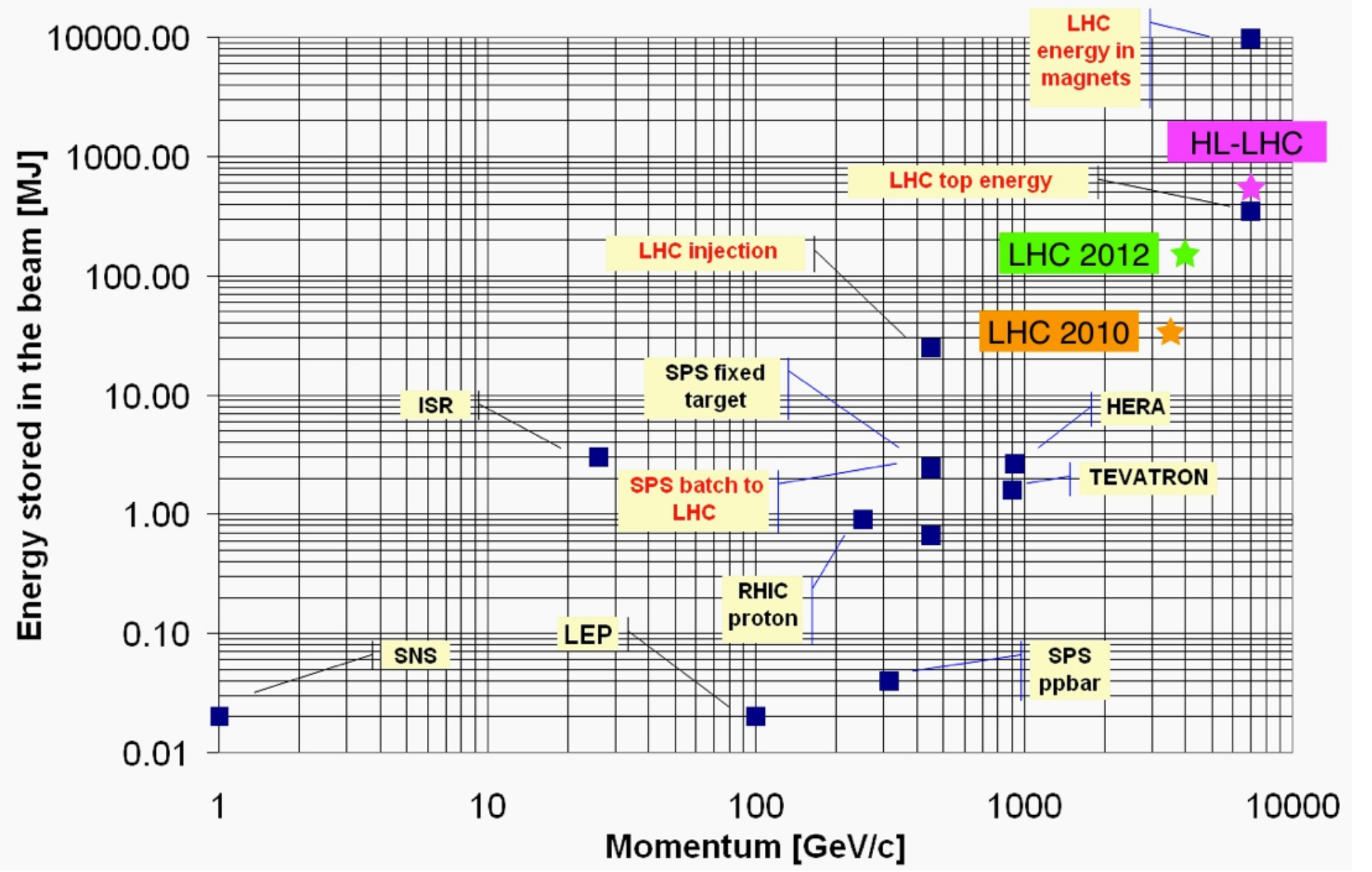}
  \vspace{-0.2cm}
  \caption{Livingston-like plot of beam stored energy achieved and planned for
    different present and future particle accelerators. Courtesy of J.~Wenninger.}
  \vspace{-0.2cm}
  \label{fig_eb}
\end{figure}

\begin{figure}
  \centering
  \includegraphics[width=100mm]{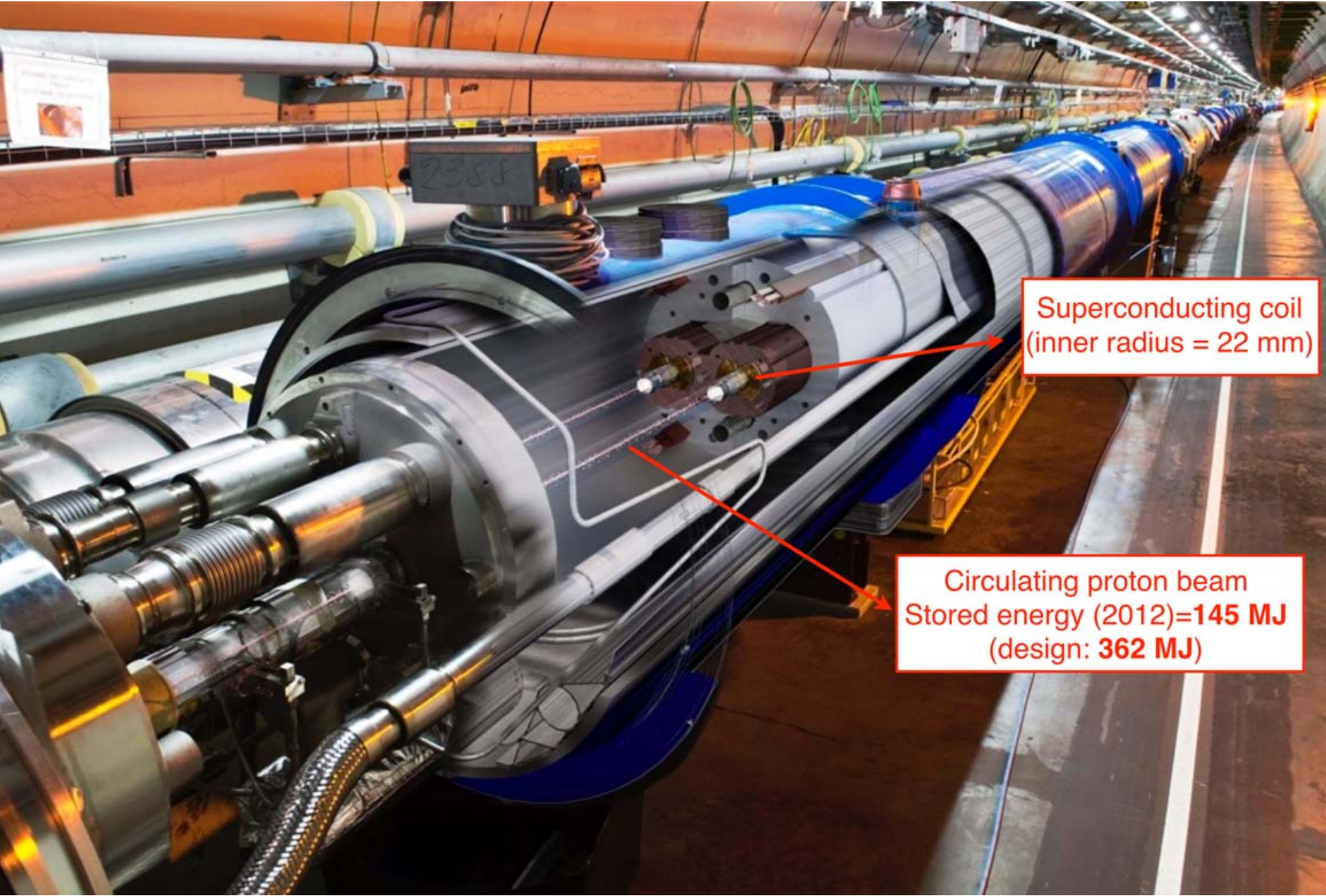}
  \vspace{-0.2cm}
  \caption{The LHC dipole in the tunnel, showing the cross-section
    of the magnet cold mass. The inner horizontal and vertical half dimensions
    of the dipole beam screen are 22~mm and 17~mm, respectively. }
  \label{fig_dip}
\end{figure}

The role of beam collimation systems in modern particle accelerators has become
increasingly important in the quest for higher beam energies and intensities.
For reference, the beam stored energy of recent and future particle accelerators
is shown in ~\Fref{fig_eb}, which includes the design (362\UMJ) and achieved
(150~\UMJ) values of the CERN Large Hadron Collider (LHC) \cite{lhc}, as well as
the $700\UMJ$ goal for its high-luminosity upgrade (HL-LHC) \cite{hl, hl2}.
High-power accelerators simply cannot operate
without adequate systems to control unavoidable losses in standard beam
operation.
The operation and physics goals of recent superconducting,
high-energy hadron colliders, such as the
Tevatron \cite{tev}, the Relativistic Heavy-Ion Collider \cite{rhic},
and the LHC, could not be fulfilled without adequate
beam collimation. With the LHC, the design complexity and the
performance of beam collimation has achieved unprecedented levels. This is
required to `clean' beam losses efficiently before they reach the small
apertures of superconducting magnets. As illustrated in
\Fref{fig_dip}, the inner aperture of LHC magnets sits only a few
centimetres apart from the circulating beams, which carry a total energy more
than a billion times larger than that necessary to perturb the operation of
superconductors.

In this document, the design of collimation systems for hadron accelerators
is presented, with a special focus on the requirements and design aspects of
high-energy and high-intensity machines.
The general scope of a collimation system is to dispose, safely and in a
controlled way, of beam losses that would otherwise occur at sensitive locations
or on accelerator equipment that is not designed to withstand beam losses.
In practice, this general definition finds its concrete implementations
depending on the specific design goals required for a given accelerator. For
example, collimation requirements are different for `warm' high-power
machines, where loss localization is crucial, than for `superconducting'
accelerators, where operation efficiency is ensured by keeping losses in cold
magnets below quench limits. The roles of collimation systems in accelerators are
discussed in Section~\ref{roles}. In Section~\ref{notation}, some basic
notation is introduced and the inputs to collimation design from machine
aperture and beam loss mechanisms are discussed. The design of a multistage
collimation system is outlined in Section~\ref{stages}.

The second part of this document is focused on the presentation of the
LHC collimation system as a case study. In Section~\ref{lhc-coll}, the system
layout is reviewed and operational challenges for beam collimation are
introduced, presenting the solutions deployed at the LHC. The LHC collimator
design is discussed in Section~\ref{design} and the collimation performance
achieved in LHC run~I \cite{runI}, at energies of up to $4\UTeV$ and
stored
beam energies of 150\UMJ, is reviewed in Section~\ref{lhc-perf}. The lecture is
concluded with a brief review of advanced collimation concepts that are being
considered for upgrading the present LHC collimation and for implementation
in future accelerators under study. This is presented in Section~\ref{advanced}.


\section{Roles of collimation systems in particle accelerators}
\label{roles}
Typical roles of collimation systems are summarized in the following.
\begin{itemize}
\item \textbf{Cleaning of betatron and off-momentum beam halos}: Unavoidable beam
  losses of halo par\-ticles must be intercepted and safely disposed of before
  they reach sensitive equipment. The required cleaning performance
  depends on the design of the accelerator. The most challenging require\-ments
  arise for superconducting accelerators, where loads from
  beam losses must remain  below the quench limits of superconducting magnets.
  For
  example, the LHC design beam stored energy of $362\UMJ$ has to be compared with
  typical quench limits of a few tens of mW/cm$^3$ \cite{q-paper}.
\item \textbf{Passive machine protection}: Collimators are the closest elements
 to the circulating beam and repre\-sent the first line of defence in various normal and abnormal loss cases, as discussed in sev\-eral companion
 lectures at this school. Owing to the damage potential of hadron beams, this
 functionality has become one of the most critical aspects of the operation
 of accelerators \cite{rs, jw}, as well as a crucial input to the design of
 collimators that must withstand design failures.
\item \textbf{Cleaning of collision products}: In colliders, this is achieved
  with dedicated movable collimators located in the outgoing beam paths of each
  high-luminosity experiment, to catch the products of collisions: direct
  collision debris and beam particles that emerge from the collision points
  with modified angles and energy.
\item \textbf{Optimization of the experiment background} (\ie minimization of
  halo-induced noise in de\-tector measurements): this is one of the classical
  roles of collimation systems in previous col\-liders.
  Beam tail scraping or local shielding at the detector locations
  can reduce spurious signals in detectors (see, for example, a recent
  report \cite{bckg}).
\item \textbf{Concentration of radiation doses}: for high-power machines, it
  is becoming increasingly import\-ant to be able to localize beam losses in
  confined and optimized `hot' areas rather than having a distribution of
  many activated areas along the machine. This is an essential design
  requirement for collimation systems,
  to allow easy access for maintenance in the largest fraction of the machine.
\item \textbf{Local protection of equipment for improved lifetime against radiation
  effects}: Dedicated mov\-able or fixed collimators are used to shield
  equipment locally. For example, passive absorbers are used in the LHC collimation
  inserts to reduce total doses, and to warm dipoles and quadrupoles
  that would otherwise have a short lifetime in the high-radiation environment
  foreseen during the nominal LHC operation. The exposure of radiation to equipment might not
  pose immediate limi\-tations to operation of a machine but its optimization is crucial
  to ensure long-term reliability.
\item \textbf{Beam halo scraping and halo diagnostics}: Though rarely a driving
  design criterion, the possi\-bility to scan the beam distribution actively can
  be a very useful functionality of a collimation system. Collimator scanning in
  association with sensitive beam loss monitoring systems proved a powerful method of probing the population of beam tails \cite{diff, mess},
  which are otherwise too small, compared with the beam core, to be measured by conventional
  emittance measurements. Thanks to their robustness, the LHC primary
  collimators can be efficiently used to scrape and shape the beams, as
  in Ref. \cite{2s}. Full beam scraping also provides precise, though destructive,
  measurements of beam sizes.
\end{itemize}

A collimation system might typically fulfil several roles. For example, the
concentration of radi\-ation losses or the reduction of experimental
background
are natural by-products of a very efficient beam collimation design. Conversely, before designing a collimation system, it is important to identify
the driving requirements for its design in a specific accelerator. For
the LHC, the driving design criterion is halo cleaning, which must be
excellent to operate the machine below the quench limit of the super\-conducting
magnets
at maximum beam energy. It is interesting to note that the
present LHC beam colli\-mation \cite{lhc, finalColl} is quite special in that
it fulfils all the roles listed, thanks to a careful design that has
been extended beyond the cleaning functionality.
The price to pay for this
performance is the unprecedented complexity, which poses important operational
challenges, as discussed in Section \ref{lhc-perf}.


\section{Inputs to collimation design from aperture and beam loss mechanisms}
 \label{notation}


\subsection{Basic definitions for collimation and beam halo}
\begin{figure}
  \centering
  \includegraphics[width=80mm]{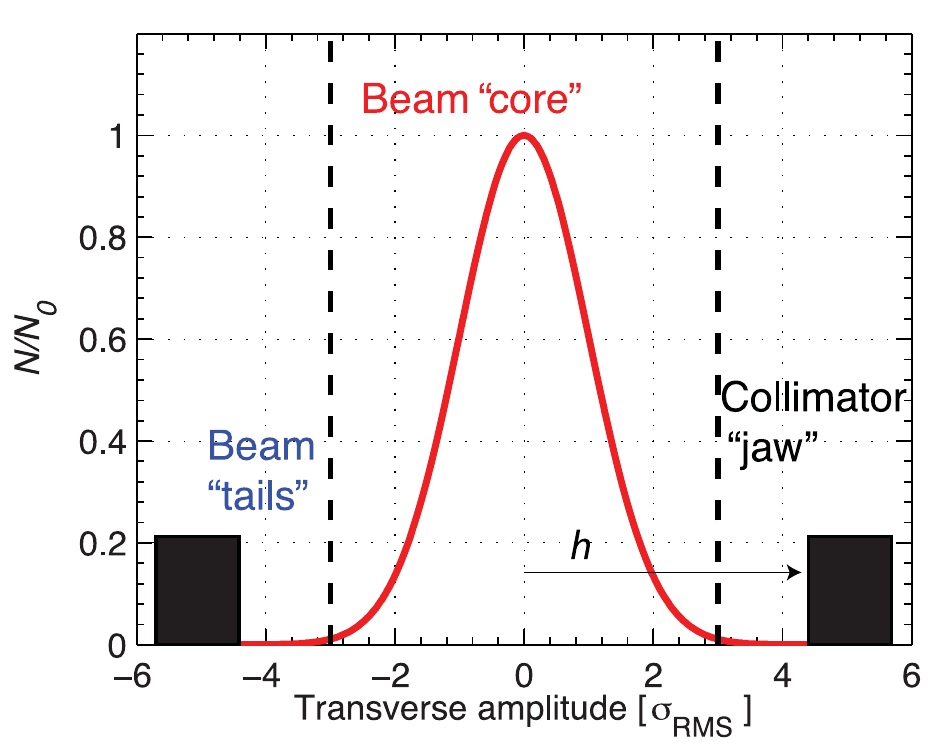}
  \vspace{-0.2cm}
  \caption{Gaussian distribution, which is typically
    adequate to model the particle distribution of the beam core (red line).
    Overpopulated tails may be intercepted by collimator jaws, which
    constrain particle motion at a given transverse betatron amplitude.}
  \label{fig_gau}
\end{figure}

Particles with transverse amplitudes or energy deviations significantly larger
than those of the reference particle are referred to as \emph{beam halo particles}.
One can distinguish between \emph{betatron} and \emph{off-momentum} halos, which are formed in
the case of larger-than-nominal transverse emittance or energy error, respect\-ively.
The transverse amplitude of a particle $i$ around a closed orbit,
$z\equiv(x,y)$, can be expressed as a function of the longitudinal curvilinear
coordinate $s$ for the Twiss parameters $\beta_z(s)$, $D_z(s)$ , and $\phi_z(s)$
as
\begin{equation}
z_i(s) = \sqrt{\beta_z(s)\epsilon_{z, i}}\sin[\phi_z(s)+\phi_{z,i,0}] +
\left(\frac{\delta p}{p}\right)_iD_z(s)~,
\label{eqMot}
\end{equation}
where $\epsilon_{z, i}$ is the single-particle emittance,
$\left(\delta p/p\right)_i$ is the energy error, and $\phi_{z,i,0}$ is
an arbitrary phase. The r.m.s.~size of the beam at location $s$ is then
given by
\begin{equation}
\sqrt{\beta_z(s)\epsilon_z+\left(\frac{\delta p}{p}\right)^2D_z^2(s)}~,
\end{equation}
where $\epsilon_z$ and $\delta p/p$ are the r.m.s.~transverse emittance
and energy spread of the beam. The notation to express the machine aperture and
the collimator settings will use, unless specified otherwise, the \emph{betatron
  beam size},
\begin{equation}
  \sigma_z(s)=\sqrt{\beta_z(s)\epsilon_z}~,
  \label{sz}
\end{equation}
which takes into account only the contribution to the beam size from the
betatron motion. Collimator settings might then be given in normalized units
as
\begin{equation}
  n_\sigma=\frac{h}{\sigma_z}~,
  \label{ns}
\end{equation}
where $h$ is the distance in millimetres between the collimator jaw and the circulating
beam (\eg the half gap of a two-sided collimator centred around the beam, as
shown in \Fref{fig_gau}).

The distinction between halo and core particles is, to a certain extent,
arbitrary. For Gaussian distributions, one may define as halo particles those
with amplitudes above three r.m.s.~deviations of the Gaussian (see dashed
lines in \Fref{fig_gau}), \ie with emittances larger than
$9\epsilon_z$.
For a beam with perfect two-dimensional Gaussian distributions in the $(z,z')$ plane,
about 1.1\% of the total beam particles have amplitudes above $3\sigma_z$
and 0.03\% above $4\sigma_z$, respectively. Particles this far out from the beam
core are rarely of any use for accelerators and are more likely to cause nuisances (beam
losses, irradiation of components, background in detectors, \etc).
For off-momentum halos, a similar definition could be adopted. For other
purposes in circular accelerators, one might consider as  halo the particles
outside the RF bucket that are lost when  beams are
accelerated or in the presence of synchrotron radiation (which is non-negligible at the
LHC).

Beam collimation is achieved by placing blocks of material, the \emph{collimator
  jaws}, close to the circu\-lating beams, to constrain the betatron amplitudes of
stray particles outside the core. This is shown schematically by the black
boxes in \Fref{fig_gau}. Collimation of off-momentum tails might be
achieved in a similar way as for betatron tails, by placing collimators at
locations of high dispersion, where the particle's energy shift results
in a transverse offset, as in the second term on the right-hand side of
\Eref{eqMot}.

How close to the beam a collimator should be depends
on various aspects that will be elaborated on in the rest of this
paper. In particular, it will be shown that the outer limit for a collimator
setting depends on the available machine aperture that needs to be protected
and on the cleaning performance that needs to be achieved. The inner limit
depends essentially on how collimators perturb beam stability through an
increase in the machine impedance. Tighter settings also typically lead to
higher beam losses and tighter positioning tolerances against orbit movements
and optics
errors. Thus, collimators should not be set closer to the beams than
is strictly necessary.

\subsection{Collimation cleaning inefficiency}
The cleaning performance of a collimation system is measured by the
\emph{collimation efficiency}, a figure of merit that expresses the fraction
of halo particles `caught' by the system over the total lost from the beam.
A perfect beam collimation provides 100\% cleaning, \ie there is no beam loss
at sensitive equipment. Alternatively, the
\emph{cleaning inefficiency}, $\eta_\text{c}$, can be introduced as the relative
fraction of beam that `leaks' to other accelerator components, $A_{\text{lost}}$,
compared with what is intercepted and safely disposed of by the
collimators, $A_{\text{coll}}$:
\begin{equation}
  \eta_\text{c}=\frac{A_{\text{lost}}}{A_{\text{coll}}}~.
\end{equation}
The relevant measure of `beam loss', indicated as $A$ in this equation,
has to be identified for the specific design criteria that the collimation
system addresses.

The LHC beam collimation requirements are driven by the challenge to keep beam
losses below the quench limits of the superconducting magnets. In this case, the
inefficiency $\eta_\text{c}$ is defined as the number of protons lost as a fraction of
the total number of particles absorbed by the collimation system. The {\it
  local cleaning inefficiency}, $\tilde\eta_\text{c}\equiv\tilde\eta_\text{c}(s)$, is defined
as a function of the longitudinal coordinate $s$ as the fractional loss per unit
length,
\begin{equation}
  \tilde\eta_\text{c}=\frac{\eta_\text{c}}{L_{\rm \text{dil}}} =
  \frac{N(s\rightarrow s+\Delta s)}{N_{\text{abs}}}\frac{1}{\Delta s}~,
  \label{eta}
\end{equation}
where $N(s\rightarrow s+\Delta s)$ is the number of particles lost over the distance
$\Delta s$, \ie in the longitudinal range $(s, s+\Delta s)$, and $N_{\text{abs}}$
is the number of particles absorbed by the collimation system.

This definition has the advantage that it can be directly compared with the quench
limits of super\-conducting magnets if a proper \emph{dilution length} is
chosen. Indeed, for the LHC it was estimated \cite{RalphCham12} that
the quench limits in units of proton lost per unit length, $R_\text{q}$, are
\begin{eqnarray}
R_\text{q}^{\text{inj}}&=&7.0\times10^{8}\:{\text{protons}/(\UmZ\cdot\UsZ)}\:\:{(450\UGeV)}~, \\
R_\text{q}^{\text{top}}&=&7.6\times10^{6}\:{\text{protons}/(\UmZ\cdot\UsZ)}\:\:{(7\UTeV)}~,
\label{q}
\end{eqnarray}
for beams at injection (${\text{inj}}$) and top (${\text{top}}$) energies, respectively. These
approximate figures were used early in the LHC design phase and
in first collimation performance estimates \cite{cham2005}.
Although nowadays detailed simulation tools and more adequate models are
available to compare peaks of energy deposition in the magnet coils
directly against quench limits of superconducting cables (see, for example, Ref.~\cite{q-paper}), the formalism introduced here is very useful to introduce challenges for collimation design, as discussed next.

\subsection{Beam lifetime and loss modelling}
There are various mechanisms that lead to losses in particle accelerators, as
also discussed in companion lectures at this school \cite{rs, jw}. One can
distinguish between \emph{regular} and \emph{abnormal beam losses}, referring to unavoidable losses that
occur during standard operation, as opposed to losses caused by failures of
accelerator systems or by wrong beam manipulations. For both categories,
losses might occur over a broad range of
time-scales, from a fraction of a single turn to tens of seconds.

In circular colliders, a main source of loss comes from the collisions
of the opposing beams that cause burn-off of beam particles. Other sources of
loss are interactions with
residual gas, intrabeam scattering, beam instabilities of various types
(single-bunch, collective, beam--beam, \etc), the noise of feedback systems used
to stabilize beams, transverse and longitudinal
resonances, include RF noise. Other losses inherent to operation phases of the
accelerator, such as capture losses at the beginning of the ramp, injection and
dump losses, losses during dynamics changes of the operational cycle (orbit
drifts, optics changes, energy ramp, \etc), are referred to as `operational
losses' \cite{vk}.


Ignoring, for the moment, very fast loss scenarios and their
impact on collimator design \cite{rs, jw}, let us consider the requirements for beam
collimation in the presence
of \emph{diffusive losses}. In this case, the transverse increase
of particle action per turn is much smaller than one sigma of the r.m.s.~distribution. Rather than treating each loss
mechanism in detail, losses are modelled by considering the beam lifetime.

\begin{figure}
  \centering
  \includegraphics[width=75mm]{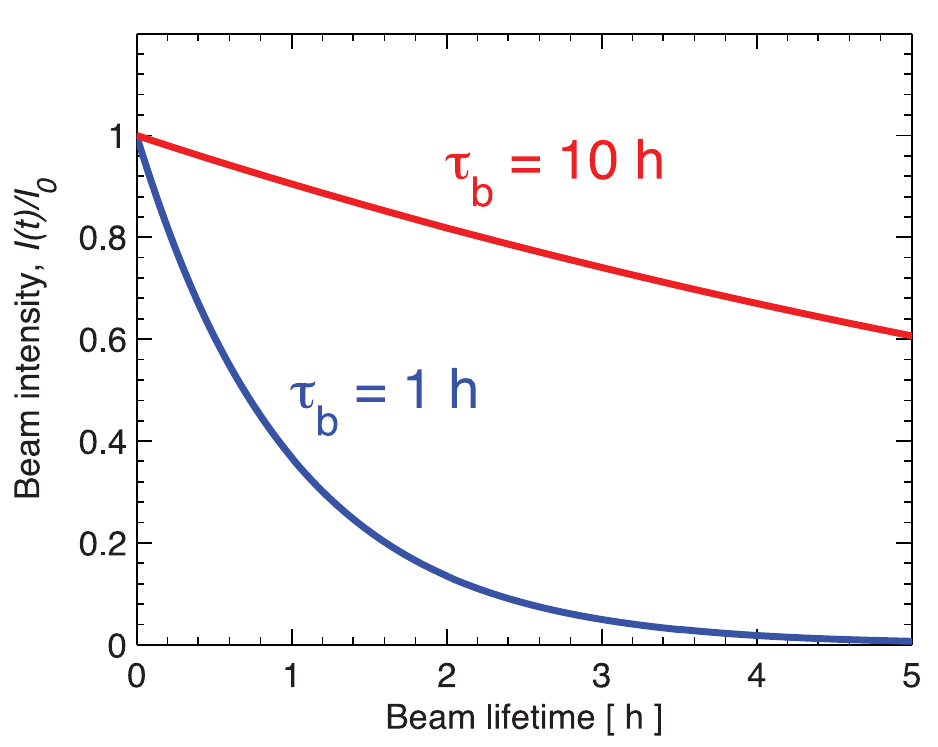}
  \vspace{-0.2cm}
  \caption{Relative reduction of beam current versus time, $I(t)/I_0$, for
  beam lifetime values of $1\Uh$ and $10\Uh$}
  \label{fig_lt}
\end{figure}

\begin{figure}
  \centering
  \includegraphics[width=120mm]{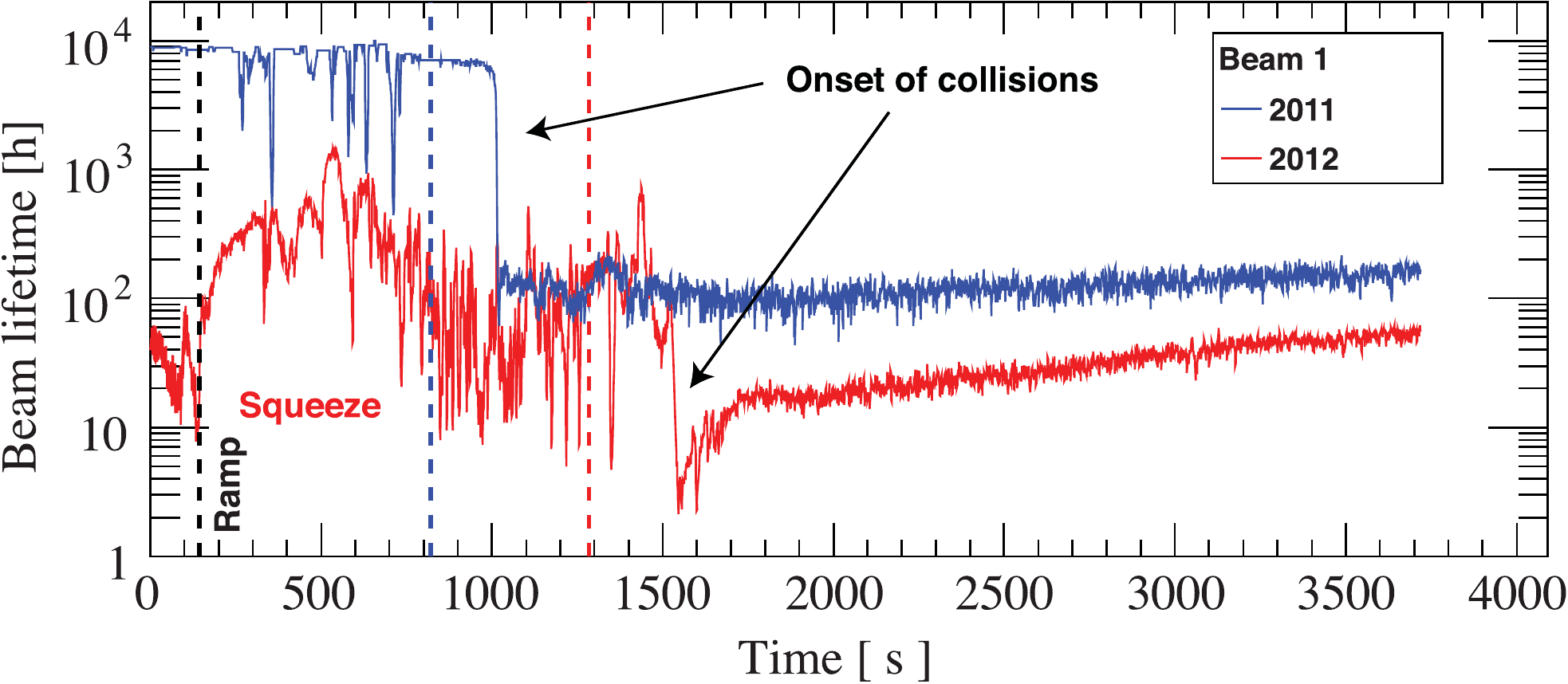}
  \vspace{-0.2cm}
  \caption{Measured beam lifetime LHC during two typical LHC
    physics fills in 2011 at $3.5\UTeV$ (blue) and in 2012 to $4\UTeV$ (red),
    as a function of time in the cycle. The ramp and squeeze durations were
    different, so the onset of collisions (reduction in $\tau_\text{b}$ indicated by black
    arrows) started at different times.
    The primary halo cut in the betatron cleaning insertion changed from
    $5.7\sigma_z$ to $4.3\sigma_z$ for $\epsilon_z=3.5\Uum$. Courtesy of B.~Salvachua.}
  \label{fig_lt_meas}
\end{figure}

The time-dependent circulating beam intensity, $I(t)$, can, for most practical
purposes, be modelled by an exponential decay function whose time constant,
$\tau_\text{b}\equiv\tau_\text{b}(t)$, defines the \emph{beam lifetime} as
\begin{equation}
I(t) = I_0\mathrm{e}^{-t/\tau_\text{b}},
\end{equation}
for an initial beam current $I_0$. After a time $\tau_\text{b}$, the total beam
current is reduced to about 37\%. Example profiles of relative beam intensity versus time, $I(t)/I_0$,  are shown in \Fref{fig_lt} for lifetime
values of $1\Uh$ and $10\Uh$. In a linear approximation, beam loss rates, $\text{d}I/\text{d}t$,
are inversely proportional to $\tau_\text{b}$ and can be calculated as
\begin{equation}
  -\frac{1}{I}\frac{\text{d}I}{\text{d}t} = \frac{1}{\tau_\text{b}}~.
\end{equation}
It is important to emphasize that $\tau_\text{b}(t)$ is indeed a function of time
and is not constant through the oper\-ational cycle. The sources of beam losses
introduced previously---operational losses and other acceler\-ator physics
mechanisms---occur at
different times and might become apparent as drops of beam lifetime at given
times in the cycle. An example of measured $\tau_\text{b}$ during LHC fills for
physics is shown in \Fref{fig_lt_meas}. In 2012, proton beams were
accelerated to $4\UTeV$, whereas in 2011 the maximum energy was 3.5\UTeV. The machine
configuration and TCP settings were different in these runs.

\begin{figure}
  \centering
  \includegraphics[width=129mm]{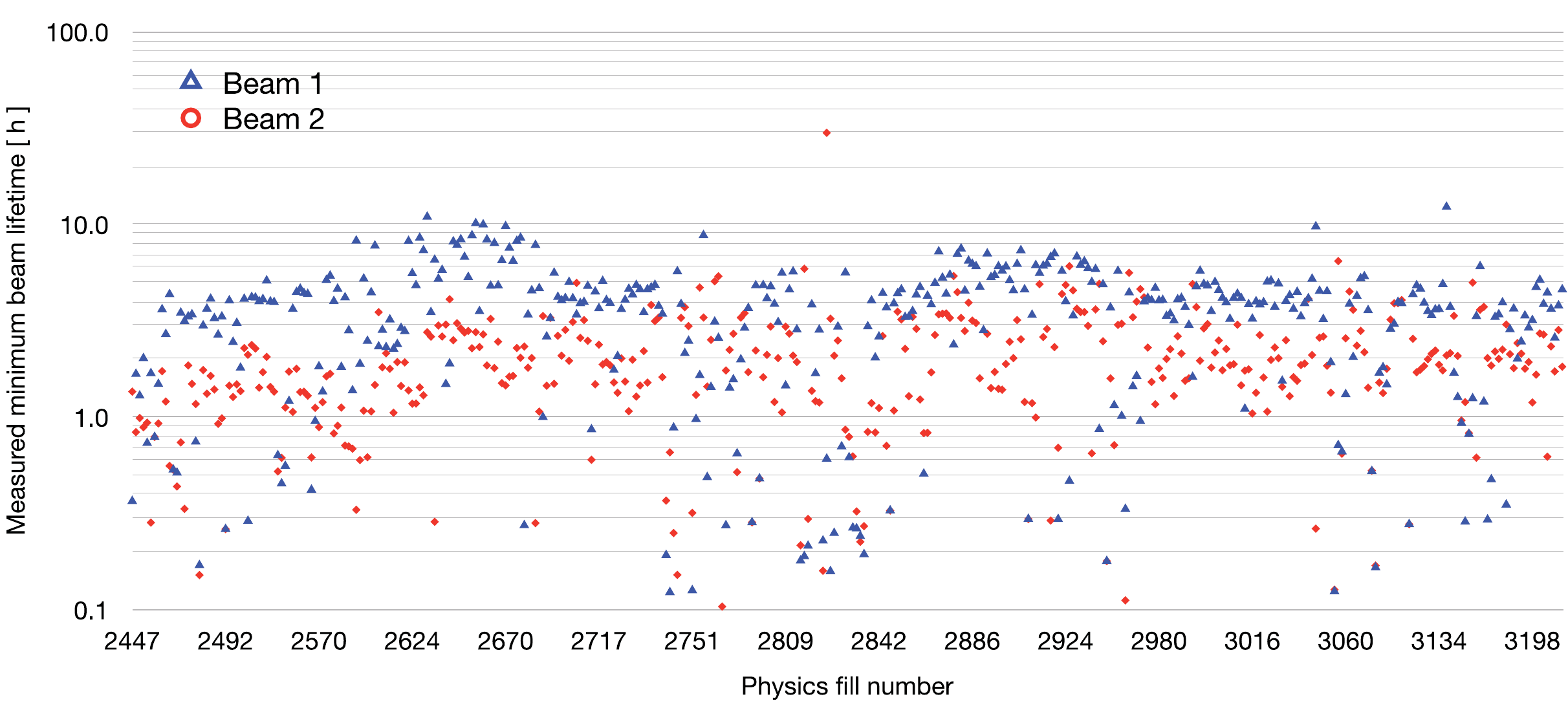}
  \vspace{-0.2cm}
  \caption{Minimum lifetime measured during the squeeze of physics fills in
    2012 as a function of fill number (to be considered as arbitrary
    unit). Courtesy of B.~Salvachua.}
  \label{fig_lt_sq}
\end{figure}

A collimation system must be designed to cope with the maximum expected rates of
beam loss. This is determined by the \emph{minimum allowed beam lifetime},
$\tau_\text{b}^{\min}$, throughout the operational cycle, most not\-ably
during phases at maximum energy (flat-top, squeeze, collision
preparation, and physics data recording) when the beam stored energy is largest.
The design value used to specify the LHC collimation system is
$\tau_\text{b}^{\min}=0.2\Uh$ for up to a maximum time of $10\Us$ \cite{RalphCham12}.
The minimum lifetime measured during the squeeze process in 2012 is shown in
~\Fref{fig_lt_sq}. Values of $\tau_\text{b}$ below $1\Uh$ were recorded on a regular
basis, with several cases even below $0.2\Uh$. The same behaviour at higher beam
intensity and energy, as expected in 2015, will cause frequent beam dumps,
with a severe impact on LHC operation.


\section{Design of a multistage collimation system}
\label{stages}
\subsection{Design requirements and work flow}

\begin{figure}
  \centering
  \includegraphics[width=90mm]{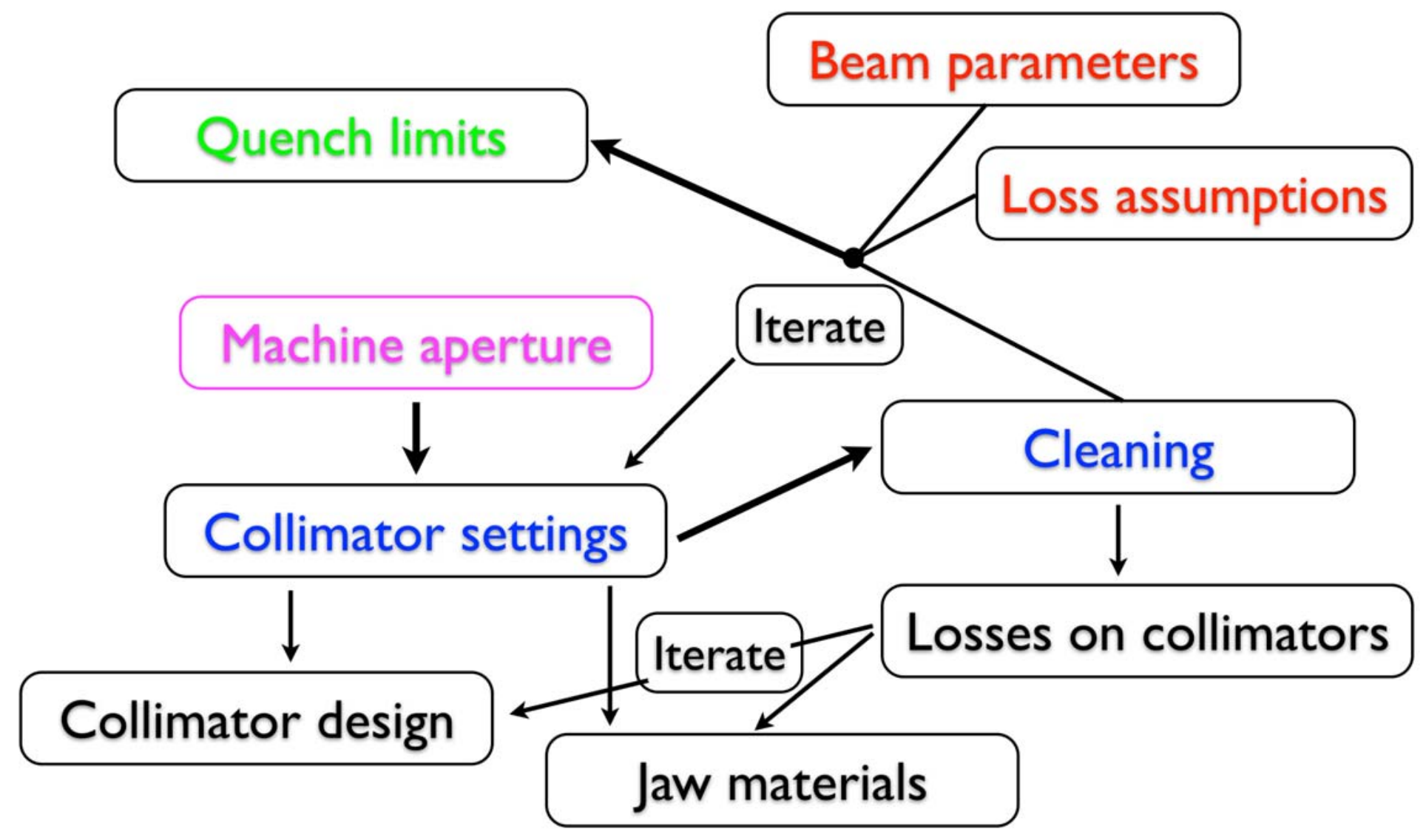}
  \vspace{-0.2cm}
  \caption{The various ingredients and competencies
    required to design a complete collimation system}
  \label{fig_workflow}
\end{figure}

The LHC case of beam collimation in the presence of high-energy and high-intensity
proton beams is considered here. For the collimation system to fulfil the
required cleaning goals, it must be ensured that:
\begin{itemize}
\item[(1)] the aperture bottlenecks of the accelerators are geometrically
  shielded such that, for all loss scen\-arios, primary beam losses hit first
  collimators;
\item[(2)] the total energy carried by the beam, \ie out-scattered beam particles
  and the secondary prod\-ucts of beam particles' interactions with the collimator
  matter, is absorbed within the collimation region, with tolerable leakage
  to sensitive equipment, notably to cold magnets, for any relevant loss
  scenarios;
\item[(3)] the collimators themselves and other equipment installed in the
  collimation regions must with\-stand, without damage, beam losses for different
  design scenarios;
\item[ (4)] the contribution to machine impedance from the collimator materials
  approached to the beam must be tolerable and ensure that the high-intensity
  beams remain stable.
\end{itemize}
This last aspect is particularly critical when it comes to designing collimators
and absorbers. For more details, see a companion lecture \cite{ab}.
The complete design of a complex system like that of the LHC re\-quires several
steps and iterations between different domains that go well beyond the field of accelerator physics. \Figure[b]~\ref{fig_workflow}  shows relevant steps towards a complete
design of a collimation system. As indicated, several iterations
are required.

The understanding of the machine aperture and the required cleaning determines a
specification of the collimator settings that can be used to calculate the collimation cleaning. With the input of loss assumption and machine parameters,
the first cleaning estimates are compared with the quench limits. This initial
design phase is worked out in this section. Closing this loop provides a first
conceptual design, which is then used for detailed collimator design. Cleaning
simulations also provide distributions of losses on the collimators that are
used, in an iterative process, to specify adequate collimator design and
jaw material choices. This aspect will not be discussed in this document
(see Ref. \cite{ab} for more detail).

\subsection{Beam cleaning specifications}
The total design beam intensity of the LHC beams is
$I_{\text{tot}}=2808\times1.15\times10^{11}$ protons, \ie $3.2\times10^{11}$
protons,
where $n_\text{b}=2808$ is the number of bunches and $I_\text{b}=1.15\times10^{11}$~protons
is the bunch popu\-lation. At the minimum allowed lifetime of $0.2\Uh$, this
corresponds to a proton loss rate of $4.4\times10^{11}~\text{protons}/\UsZ$.
At 7\UTeV, the beam stored energy is $362\UMJ$ and loss rates  approach 500\UkW. By
expressing the quench limits in the approximate formulation of \Eref{q},
one can derive a specification for the local cleaning inefficiency in a cold magnet
as
\begin{equation}
  \tilde\eta_\text{c}\le\frac{1}{10000} {\rm [1/m]}~.
  \label{cleanSpecs}
\end{equation}
Although simulation tools are now available to combine particle tracking
and energy deposition simu\-lations, so as to evaluate precisely the energy lost in the
superconducting magnet coils, it is very useful to follow this approximate
approach. This formalism provides a powerful tool for designing a
collimation system. The number of protons lost per unit length can be simulated
with a fast and accurate set-up \cite{grd}, which provides an essential
design optimization tool. Final validation of collimation solutions
then follows, using more sophisticated tools that also involve energy deposition
simulations \cite{fc}.

\subsection{Machine aperture and collimator settings}
\begin{figure}
  \centering
  \includegraphics[width=110mm]{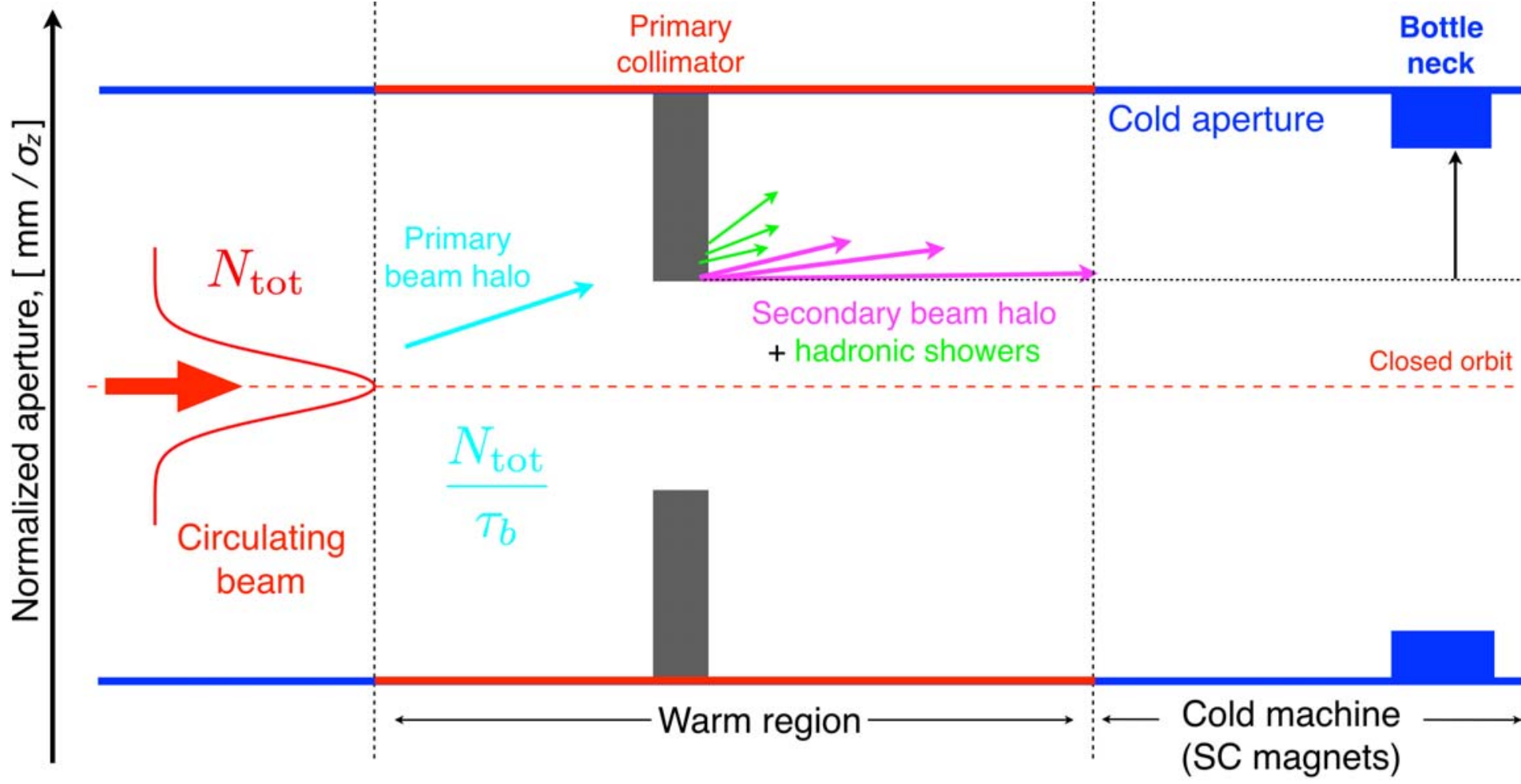}
  \vspace{-0.2cm}
  \caption{Single-stage collimation system: SC, superconducting}
  \label{fig_single}
\end{figure}

To design a betatron cleaning system, one must first compute the
available aperture of the accelerator. Let us assume that the
circulating beam sees an isolated \emph{aperture bottleneck}, $A_{{\min},z}$,
in the transverse plane $z$. This is defined as the smallest normalized
transverse aperture at any location around the ring. For convenience, the
aperture is normalized by the local betatron beam size $\sigma_z$. The
limiting location is calculated as
\begin{equation}
  \hat{A}_{{\min},z}={\min}\left[\frac{A_z(s)}{\sigma_z(s)}\right]~,
  \label{a}
\end{equation}
where the minimum is calculated for all locations $s$ around the ring and
$A_z(s)$ is the distance in milli\-metres between the circulating beam and
the mechanical aperture. This quantity can be measured directly in an accelerator
\cite{ap2, ap3}. While, operationally, it is convenient to express
$A_{{\min},z}$ for a given beam emittance, what matters is actually the
\emph{aperture acceptance} $A_z(s) / \sqrt{\beta_z}$. A
cold aperture bottleneck is indicated in \Fref{fig_single} as a blue box, and projects into the nominal aperture.

The minimum aperture calculated during the LHC design phase \cite{cham2005}
for both planes and beams at injection and top energy are listed in
Table~\ref{tab_bottleneck_inj}. Calculations relied on a conservative approach
\cite{n1} that ensured adequate margins during beam commissioning. While
measurements during LHC run~I \cite{apEvian2011} indicated that the
LHC aperture is indeed larger than assumed, such conservative figures are
considered in this lecture for the collimation design.

\begin{table}
  \caption{Minimum horizontal and vertical apertures at injection
  (450\UGeV) and top energy (7\UTeV, $\beta^*=0.55\Um$) for warm and cold
  elements, as estimated in the LHC design phase \cite{cham2005}. }

  \begin{center}
    \begin{tabular}{lddcdd}
       \hline\hline
      & \multicolumn{2}{c}{\textbf{450\,GeV}}&&\multicolumn{2}{c}{\textbf{7\,TeV}}\\
     \cline{2-3}\cline{5-6}
      &  \multicolumn{1}{l}{\textbf{Warm}}  &  \multicolumn{1}{l}{\textbf{Cold}} & &  \multicolumn{1}{l}{\textbf{Warm}}  &  \multicolumn{1}{l}{\textbf{Cold}}  \\
      \hline
     \textbf{Beam 1}\\
      Horizontal   & 6.8 & 7.9 && 28 & 8.9\\
      Vertical     & 7.7 & 7.8 && 8.3 & 8.4\\

      \textbf{Beam 2}\\
      Horizontal   & 6.7 & 7.7 && 28 & 8.1\\
      Vertical     & 7.7 & 7.6 && 8.7 & 8.8\\
      \hline\hline
    \end{tabular}
  \end{center}

  \label{tab_bottleneck_inj}
\end{table}

\subsection{Single-stage collimation}

Designing a collimation system involves finding an optics solution and an
arrangement of collimators that ensure that losses in cold magnets remain below
the quench limits for all design loss rates. In an ideal machine
without beam losses, there would be no need for beam collimation, if
the minimum machine aperture were at a safe distance from the beam core.
In practice, various beam loss mechanisms cause outwards drifts of halo
particles, which eventually hit the aperture  if there is no mechanism
to intercept them. The
deposited energy at this location would then depend on
the primary beam loss rates, $N_{\text{tot}}/\tau_\text{b}$.

One could build a simple single-stage collimation system by placing a primary
collimator (TCP; `target collimator, primary') that intercepts beam losses.
Preferably, collimators are
placed in a warm region, as far as possible from superconducting magnets. The
collimator jaws must be set at a transverse aperture below that of the
machine bottleneck,
$\hat{A}_{\text{TCP}}\le\hat{A}_{{\min},z}$. This
simple system would work if the TCP were a black absorber that
could stop all the primary particles at their first passage through the jaw.
Also note that, because of the mixing of positive and negative amplitudes
of halo particles from the betatron motion, a single-jaw collimator suffices
to protect the aperture against slow diffusive losses. (For standard
  losses, impact parameters in the submicrometre range are expected
  \cite{seidel}. At this scale, particles do not see the full jaw length
  at their first passage because of jaw flatness and surface roughness errors.
  This increases the inefficiency of a single-stage cleaning system, as more
  turns are required before particles accumulate enough interactions with the
  TCP.)

%

The single-stage system of \Fref{fig_single} does not
provide sufficiently efficient halo cleaning. The halo protons that are
out-scattered before being absorbed by the jaw material leave the collimator
at larger normal\-ized amplitudes and modified energies. These particles
populate the so-called \emph{secondary beam halo}, which risks being lost
in the machine before interacting again with the collimator in subsequent turns.
In addition, the products of hadronic and electromagnetic showers are not
contained in the colli\-mator volume and might reach sensitive elements
without additional downstream collimators or ab\-sorbers.

\begin{figure}
  \centering
  \includegraphics[width=130mm]{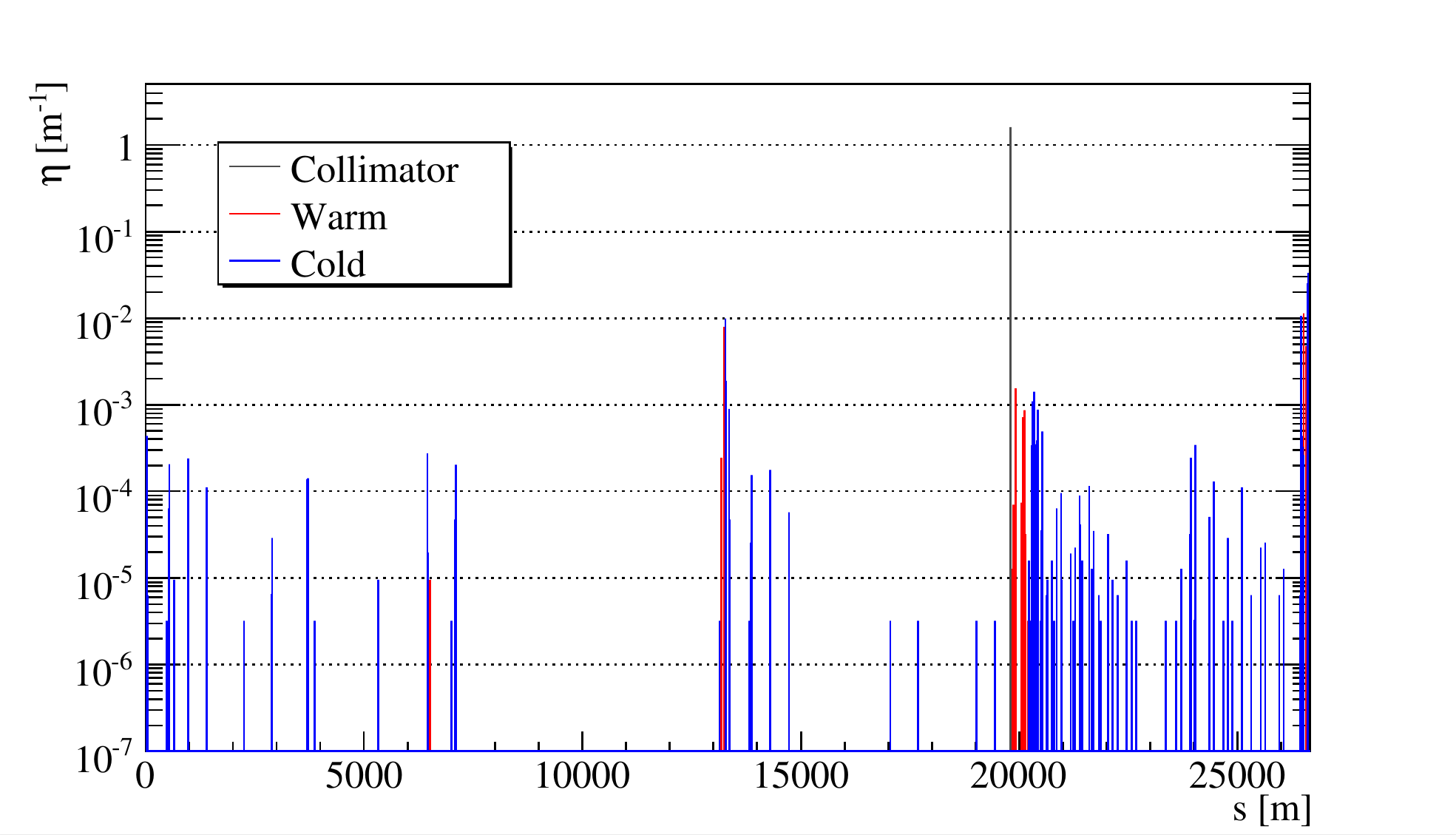}
  \vspace{-0.2cm}
  \caption{Simulated cleaning inefficiency at the LHC for a single-stage
    collimation system achieved with one hori\-zontal primary collimator (TCP)
    located at the beginning of the LHC warm betatron cleaning insert. The
    position of the existing primary collimators, \ie $s = 19.8\Ukm$, is used. Courtesy of D. Mirarchi.}
  \label{fig_ss-cleaning}
\end{figure}

\begin{figure}
  \centering
  \includegraphics[width=110mm]{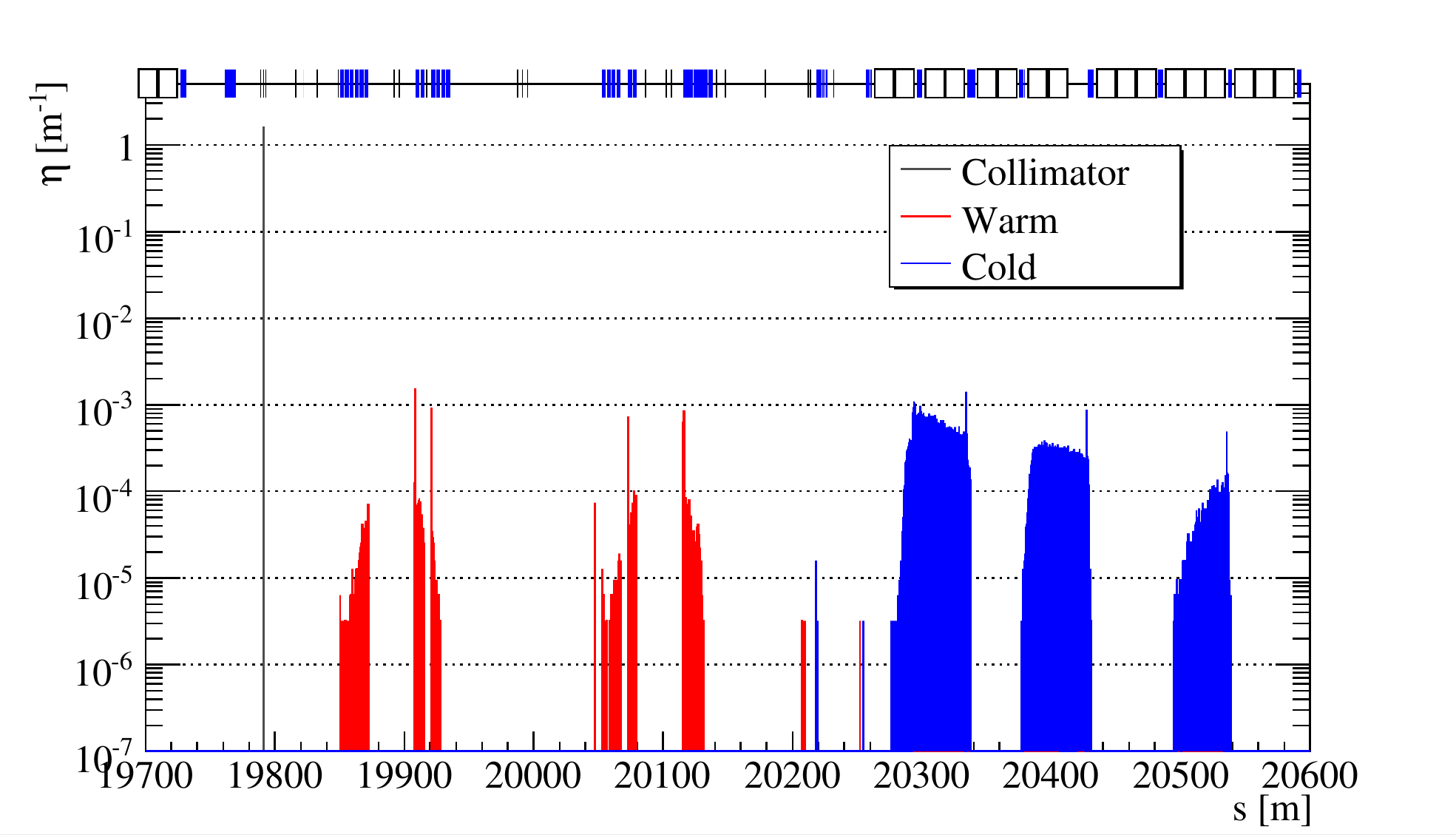}
  \includegraphics[width=110mm]{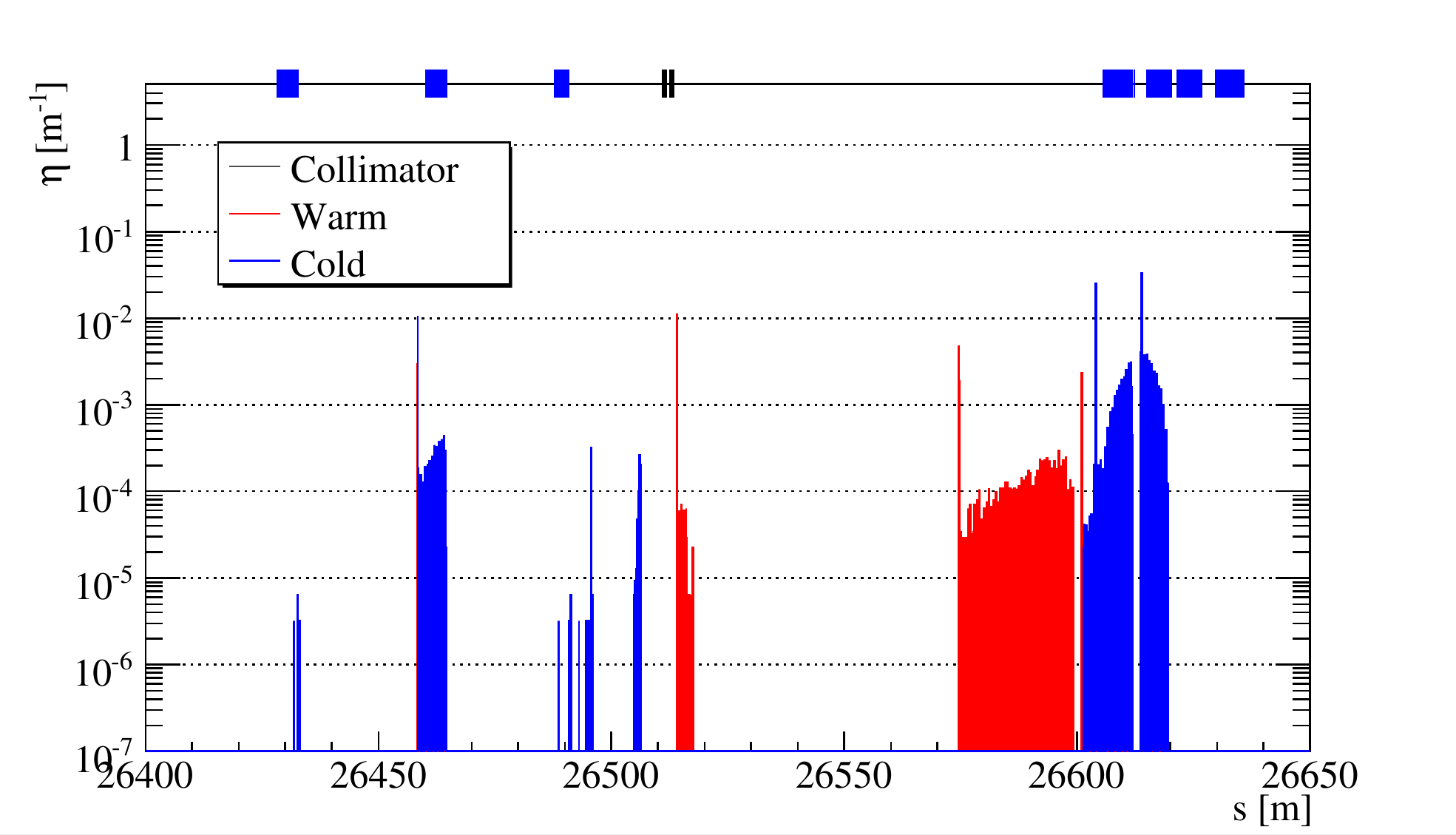}
  \vspace{-0.2cm}
  \caption{Enlargement of \Fref{fig_ss-cleaning}
    in the regions immediately downstream of the cleaning insertion (top) and
    upstream of the ATLAS experiment (bottom). Courtesy of D. Mirarchi.}
  \label{fig_ss-cleaning_z}
\end{figure}

The cleaning performance of the single-stage system described here was
simulated under the assump\-tion that a horizontal TCP is installed in the
current LHC betatron cleaning insert. The tools in Ref. \cite{grd} allow one to
calculate the number of halo protons lost in the collimators and machine aper\-ture. Simulations properly model the proton tracking through the
magnetic elements and the scattering in the collimator materials. In Figs. \ref{fig_ss-cleaning} and \ref{fig_ss-cleaning_z}, the  predicted local cleaning inefficiency of
\Eref{eta} is given as a function of the longitudinal coordinate $s$.

In these cleaning inefficiency plots, black peaks indicate losses
at collimators (only one TCP in this case), blue peaks indicate losses at cold magnets and
red peaks indicate losses at warm elements. It can be seen, by looking at   \Fref{fig_ss-cleaning_z}, which shows zoomed
plots around various interaction points, that cold losses reach cleaning inefficiency levels of up to $0.01\,/\UmZ$. This estimate, which is made for a perfect
machine without errors, and which does not take into
account the energy deposited by hadronic showers, indicates losses at least
two orders of magnitude higher than the value specified in Eq.~(\ref{cleanSpecs}).
One can therefore conclude that a single-stage collimation system is
inadequate for high-intensity super\-conducting machines, such as the LHC.

\subsection{Multistage collimation}
\begin{figure}
  \centering
  \includegraphics[width=110mm]{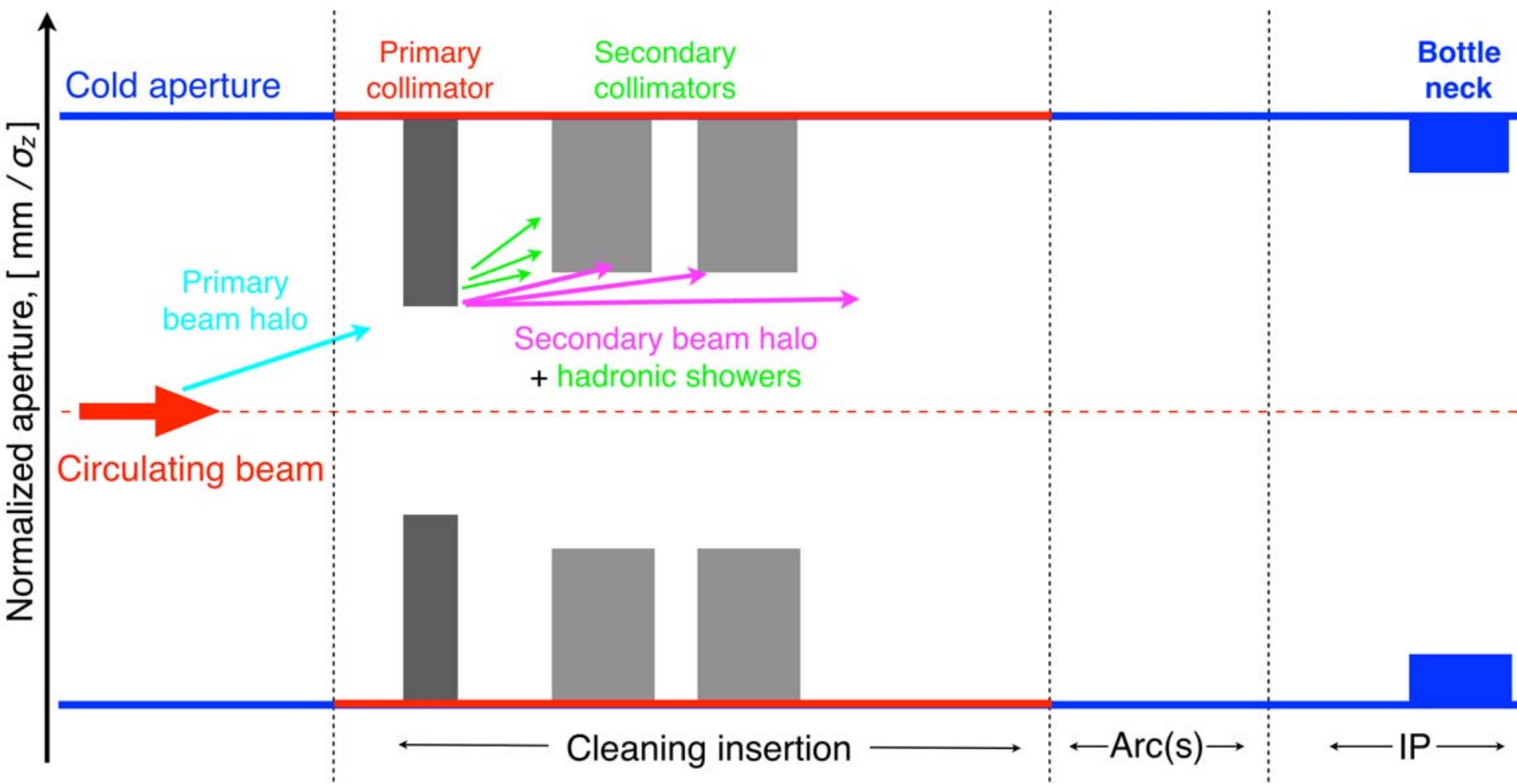}
  \vspace{-0.2cm}
  \caption{Two-stage beam collimation system, obtained by adding a set
    of secondary (TCS) collimators to the single-stage cleaning system of
    \Fref{fig_single}. IP, interaction point.}
  \label{fig_2stages}
\end{figure}

\begin{figure}
  \centering
  \includegraphics[width=70mm]{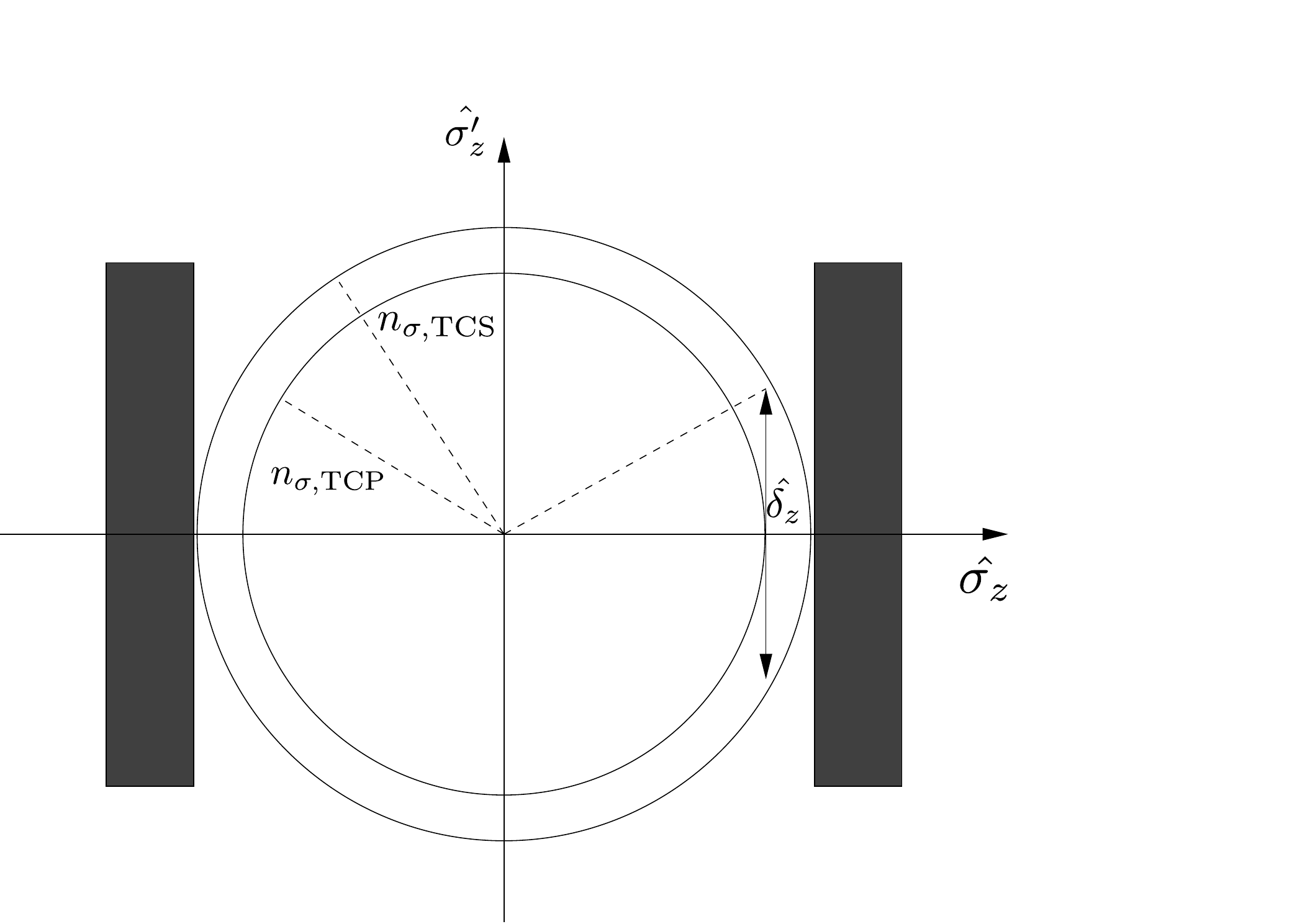}
  \vspace{-0.2cm}
  \caption{Normalized phase space with the circumferences radii
    $n_{\sigma, {\rm TCP}}$ and $n_{\sigma, {\rm TCS}}$. A normalized kick
    $\hat{\delta'}$, as in \Eref{d}, is necessary for halo particles impinging on the TCP to reach the TCS aperture.}
  \label{fig_delta}
\end{figure}

\begin{figure}
  \centering
  \includegraphics[width=80mm]{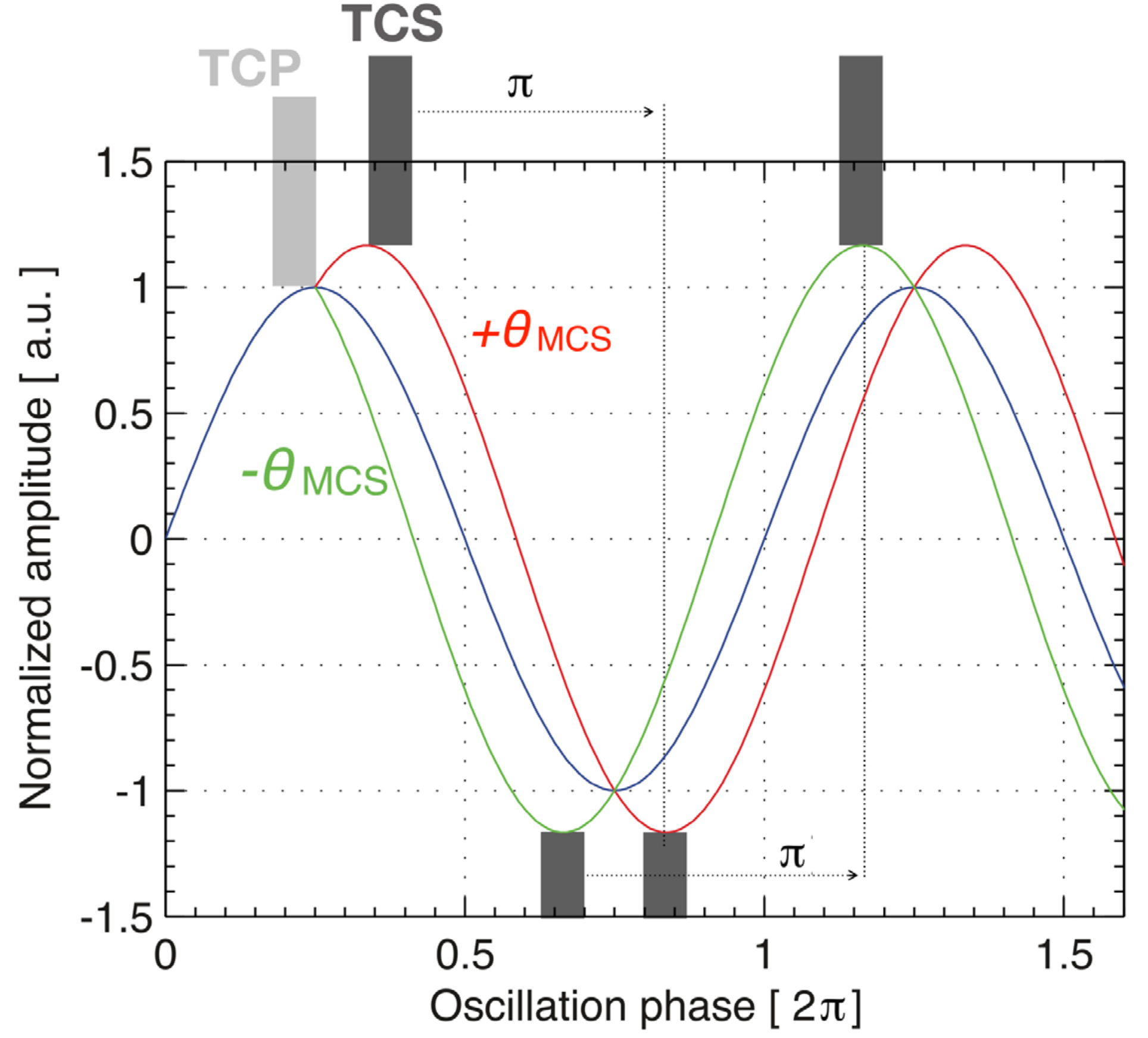}
  \vspace{-0.2cm}
  \caption{Qualitative definition of optimum locations for secondary
    collimators in a two-stage system, in which TCSs must intercept beam particles
    out-scattered at the primary collimators. In this one-dimensional model, two
    phase locations exist, where the amplitudes caused by multiple Coulomb
    scattering are a maximum for the two signs of the scattering angle,
    $\pm\theta_{\rm MCS}$.}
  \label{fig_phases}
\end{figure}

The performance of a single-stage cleaning system can be improved
with additional collimators down\-stream of the TCP to catch the secondary halo
particles, as shown in \Fref{fig_2stages}. These are called secondary collimators (target collimators, secondary; TCSs)  and are typically longer than TCPs, to maximize
the absorption of particles out-scattered at the TCPs. On the one hand,
the TCS aperture must be larger than that of the TCP, to ensure that the
\emph{collimation hierarchy} is respected without the risk of a TCS becoming closer to the beam than the TCP, which would result in a single-stage system similar to the one discussed earlier. On the other hand, the TCS aperture should be
small enough to maximize its efficiency in catching the particles out-scattered
by the TCPs. From \Fref{fig_delta}, one can calculate the kick of particles impinging on the TCP necessary to reach the amplitude of the TCS as
\begin{equation}
  \hat{\delta'}=\frac{\delta'}{\sigma'}
  =\sqrt{n_{\sigma, {\text{TCS}}}^2-n_{\sigma, {\text{TCP}}}^2}~,
  \label{d}
\end{equation}
where $\sigma'=\sqrt{\epsilon/\beta}$ is the r.m.s.~divergence. Such a kick is
typically accumulated after multiple passages through the TCP.
For a given TCS--TCP retraction, the longitudinal positions of the
TCS collimators must be optimized to intercept secondary halo particles. This is illustrated in \Fref{fig_phases} for a one-dimensional case. This condition is respected at betatron phase advances, where the multiple Coulomb  scattering angle translates into maximum offsets in the collimation plane.

The problem of optimum phase locations for a two-stage collimation system
is worked out in detail in Ref. \cite{jbj}. Finding a solution is more complicated
than appears in \Fref{fig_phases} because scattering occurs in all
directions. A one-dimensional model is thus not adequate. However, it
can be demonstrated that an arrangement of primary and secondary collimators
in three planes (horizontal, vertical and skew) can be found to ensure
satisfactory multiturn cleaning \cite{jbj}.

\begin{figure}
  \centering
  \includegraphics[width=110mm]{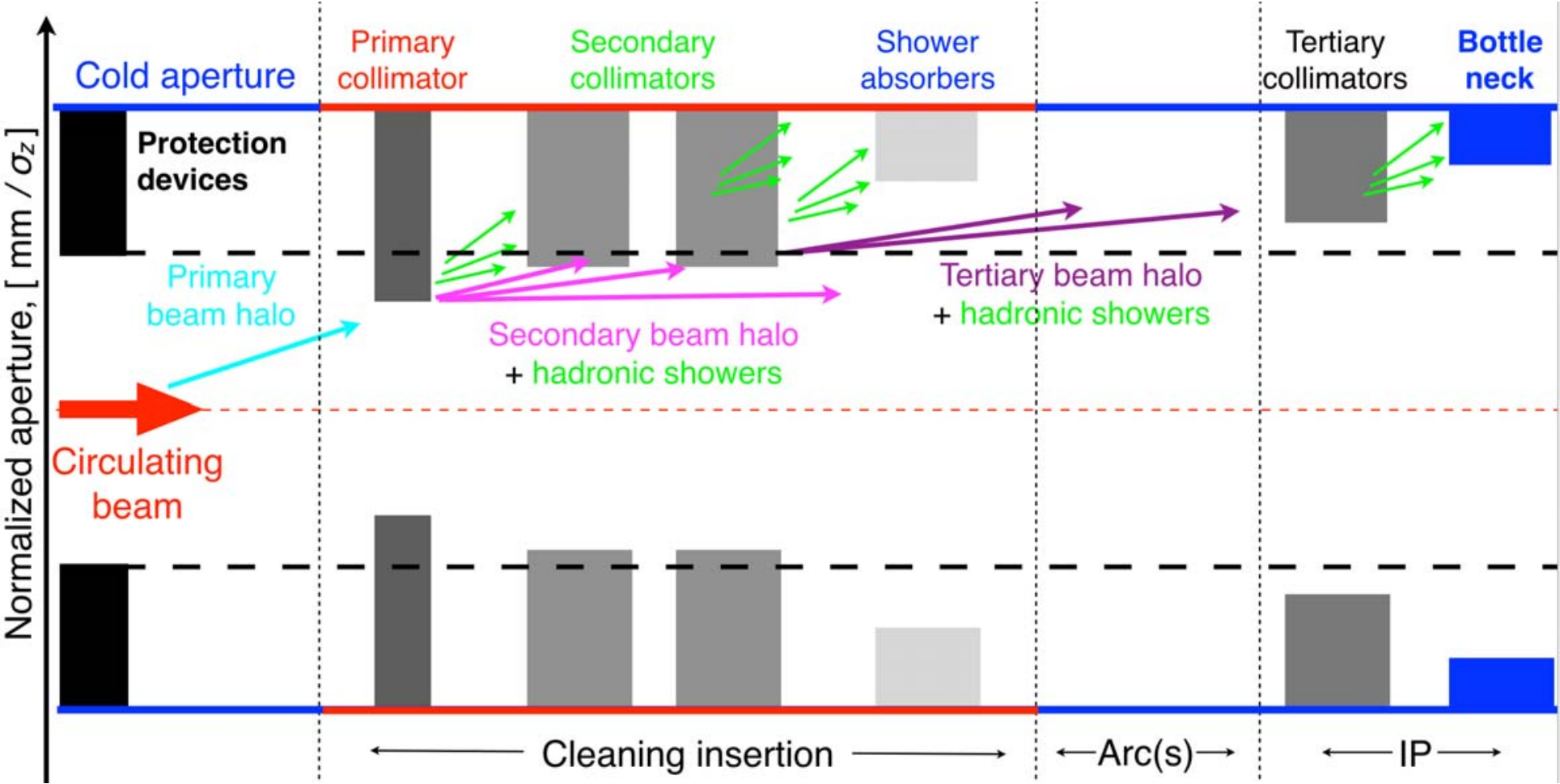}
  \vspace{-0.2cm}
  \caption{Key elements of the LHC multistage collimation
    system: IP, interaction point}
  \label{fig_multi}
\end{figure}

Detailed performance analysis of a two-stage cleaning process for the LHC was
conducted in the design phase \cite{cham2005}. While this scheme can ensure efficient shielding of the LHC aperture from transverse halo losses, it is not
sufficient to absorb products of hadronic showers before they reach cold magnets
downstream of the cleaning insert.
Moreover, a two-stage system localized in a single insertion is not
adequate for the local protection of critical bottlenecks that might be
exposed to losses, notably the triplet magnets around the experiments that
become critical during the squeeze.
The collimation system of the LHC has therefore evolved into a \emph{multistage
  collimation system} that includes, in addition to TCPs and TCSs,
tertiary collimators (target collimators, tertiary; TCTs) in front of critical bottlenecks, shower
absorbers in the warm cleaning inserts, and protection devices in the dump
region, to shield the machine in case of dump kicker failures. The LHC multistage collimation system is shown in \Fref{fig_multi}.

\begin{figure}
  \centering
  \includegraphics[width=120mm]{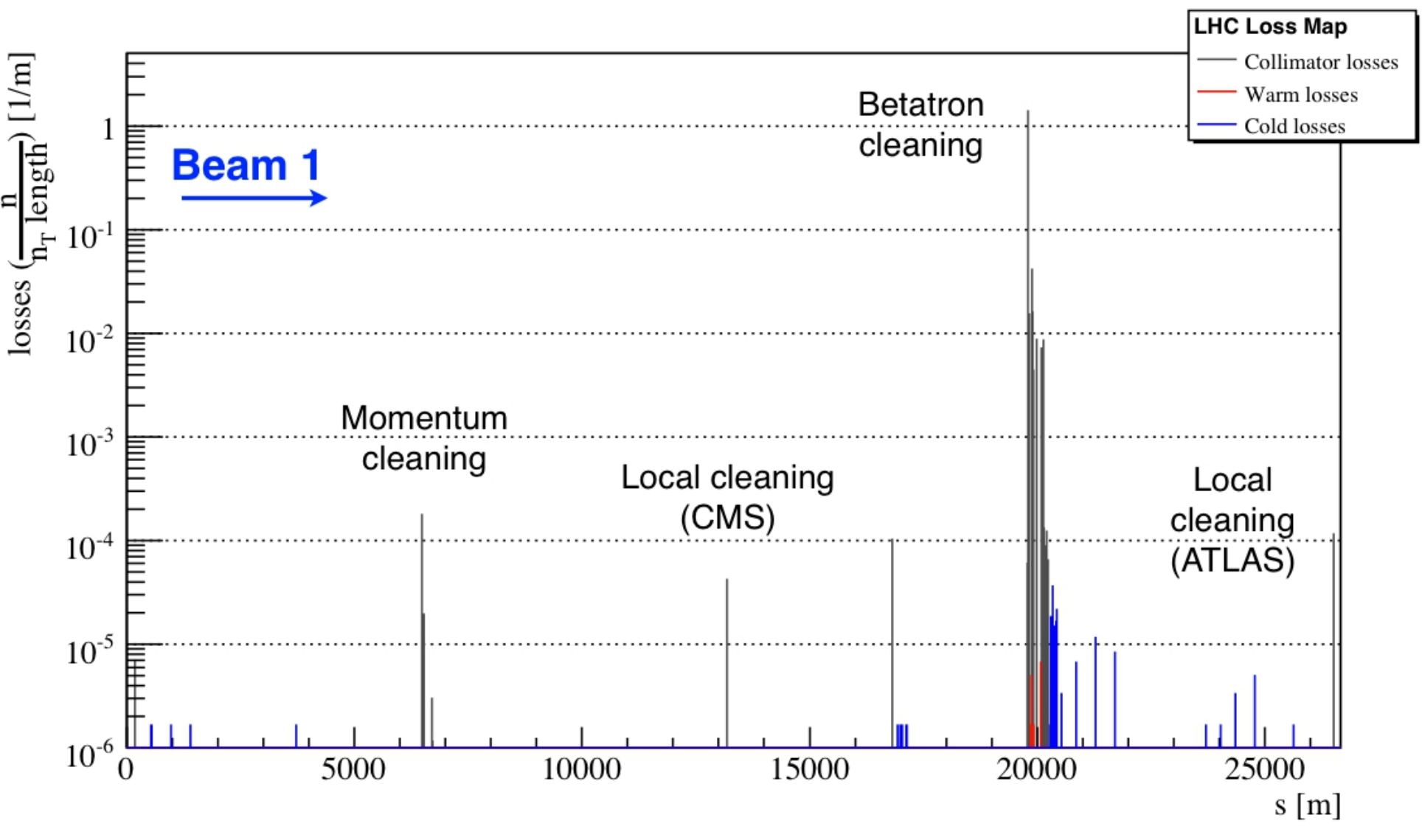}
  \vspace{-0.2cm}
  \caption{Local cleaning inefficiency as a function of $s$ for the final
    collimation system of the LHC run~I. Loss distributions are simulated for
    the LHC beam~1 at $7\UTeV$ for a perfect machine, with the collision
    optics squeeze to $\beta^*=0.55\Um$ in IR1 and IR5. Courtesy of
      D.~Mirarchi.}
  \label{fig_cl}
\end{figure}

\begin{figure}
  \centering
  \includegraphics[width=120mm]{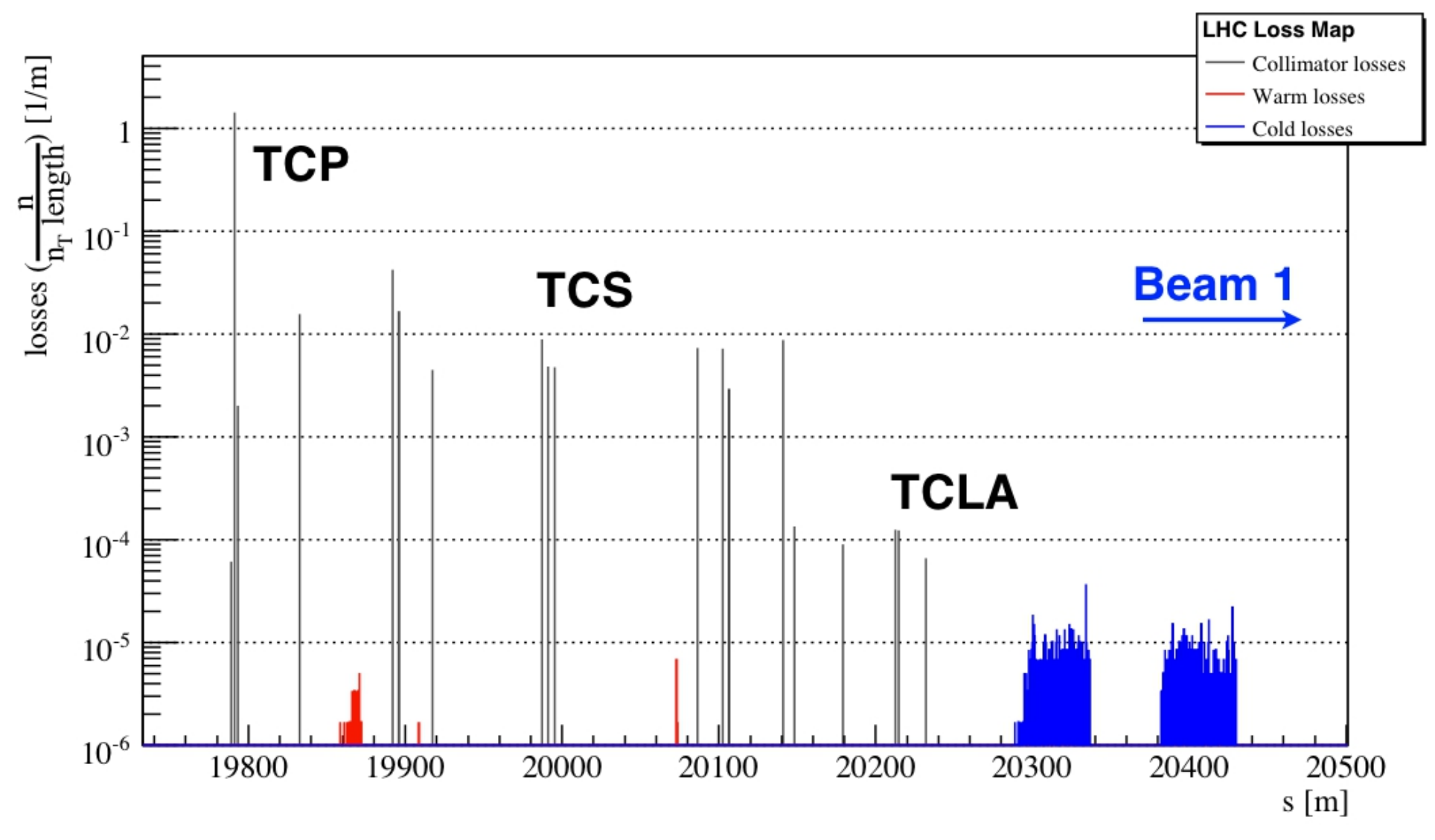}
  \vspace{-0.2cm}
  \caption{Enlargement of the IR7 region of the cleaning inefficiency plot of
    ~\Fref{fig_cl}. Labels indicate the approximate locations of the three
    families of collimator in IR7. TCLA, target collimator long absorber; TCP, target collimator (primary); TCS, target collimator (secondary).}
  \label{fig_cl_z}
\end{figure}

The cleaning performance of the final LHC collimation system
\cite{finalColl} is shown in ~\Fref{fig_cl}. While the system is described in detail in the next section, the simulations are shown here for
a direct comparison with the single-stage system. The insertion regions
(IRs) where the largest losses occur are the betatron (IR7) and momentum (IR3) cleaning, ATLAS (IR1) and CMS (IR5). This simulation is for beam 1 (B1), nominally
$7\UTeV$,  in collision conditions. An enlargement of the
loss map around the betatron cleaning insert is shown in
\Fref{fig_cl_z}. For a perfect machine, cold losses are now below $\sim10^{-5}$. The highest peaks are localized in the dispersion suppressor regions
downstream of IR7.

\section{The LHC collimation system}
\label{lhc-coll}
The LHC collimation system was designed to handle proton beams of a
stored energy of $362\UMJ$ and  is now being upgraded to cope with the design
HL-LHC goal of about $700\UMJ$ per beam.
A complex and distributed system is needed to achieve the excellent halo
cleaning required to operate the LHC below quench limits.
In this section, the
collimation layout is presented and the collimator design is reviewed. Operational challenges for the collimation at the LHC are then introduced, presenting
the solutions produced to set the system up for optimum performance in all
operational phases.

\subsection{LHC ring collimation layout}
\begin{figure}
  \centering
  \includegraphics[width=120mm]{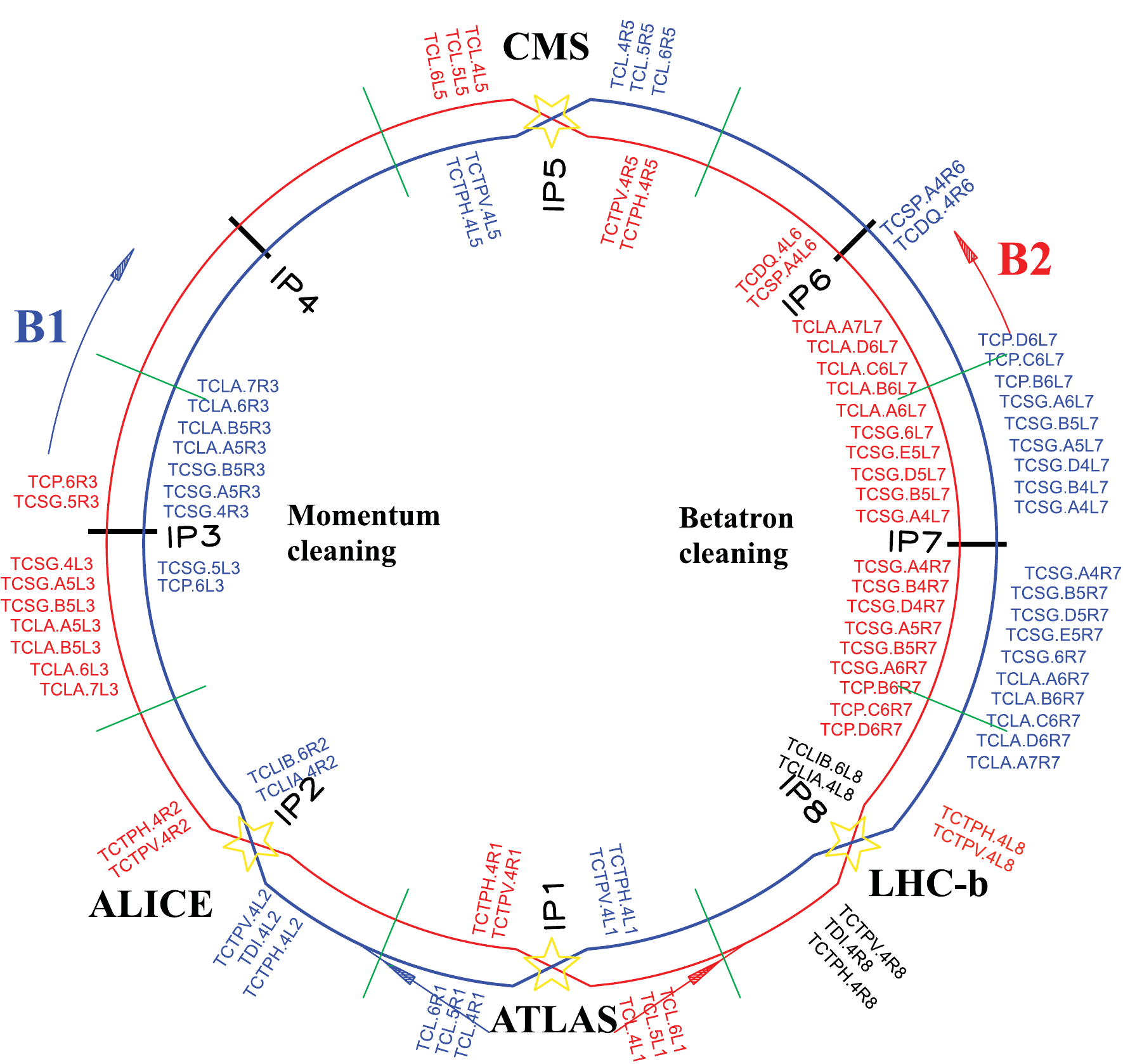}
  \vspace{-0.2cm}
  \caption{Layout of the LHC, showing the collimator locations around the
    ring}
  \label{figLayout}
  \vspace{-.3cm}
\end{figure}

\Figure[b]~\ref{figLayout} shows the LHC layout and the positions of the collimators around the
ring. A list of collimator types, with a
description of their functionality (primary, secondary, \etc) and key
collimator properties is given in Table~\ref{tabList}. Including the dump
protection block (target collimator dump quadrupole, TCDQ) and the injection protection collimator (target dump, injector TDI), the system deployed for the 2015 LHC operation comprises 110 movable collimators
installed in the LHC ring and its transfer lines.

\begin{table}
  \caption{List of movable LHC collimators for run~II. CFC, carbon fibre composite; H, horizontal; S, skew; V, vertical.}

  \begin{center}
    \begin{tabular}{lllrl}
      \hline\hline
      \textbf{Functional type }    & \textbf{Name}    & \textbf{Plane} & \textbf{Number} & \textbf{Material} \\
      \hline
      Primary IR3         & TCP     & H     & 2  & CFC \\
      Secondary IR3       & TCSG    & H     & 8  & CFC \\
      Absorbers IR3       & TCLA    & H,V   & 8  & W alloy\\
      Primary IR7         & TCP     & H,V,S & 6  & CFC \\
      Secondary IR7       & TCSG    & H,V,S & 22 & CFC \\
      Absorbers IR7       & TCLA    & H,V   & 10 & W alloy\\
      Tertiary IR1/2/5/8  & TCTP    & H,V   & 16 & W \\
      Physics debris absorber& TCL     & H     & 12  & Cu/W alloy \\
      Dump protection     & TCSP    & H     & 2  & CFC \\
                          & TCDQ    & H     & 2  & C \\
      Injection protection (lines)    & TCDI    & H,V   & 13 & CFC \\
      Injection protection (ring)   & TDI     & V     & 2  & C \\
                          & TCLI    & V     & 4  & CFC \\
                          & TCDD    & V     & 1  & CFC \\
      \hline\hline
  \end{tabular}

  \end{center}

 \label{tabList}
\end{table}

Halo collimation is achieved by the multistage cleaning system introduced
in \Sref{stages}.
This comprises three stages in IR3 (momentum cleaning) and IR7 (betatron
cleaning), where the primary colli\-mators (TCPs), closest to the beam, are
followed by secondary collimators (TCSs) and active ab\-sorbers (TCLAs). For
optimal performance, the particles in the beam halo should first hit a TCP,
and the TCSs should only intercept secondary halo particles that have been
already scattered by, and escaped from, upstream collimators. The TCPs and
TCSs, which are the closest collimators to the beam and hence intercept large
beam losses, are made of a carbon fibre composite (CFC) to ensure high
robustness. These
collimators are also more likely to be hit by the beam if there is a failure.
The TCLAs catch tertiary halo particles scattered out
of the TCSs, as well as showers from upstream collimators. The TCLAs are made
of a tungsten alloy, in order to stop as much as possible of the incoming
energy. However, they are not as robust as the CFC collimators and
should therefore never intercept primary beam losses. The setting hierarchy is
chosen to ensure that this condition is respected in all operation state.

In addition to the dedicated inserts in IR7 and IR3, there are
collimators in most other IRs. A pair of tertiary collimators (target collimators, tertiary, pick-up; TCTPs), made of
a tungsten
alloy, are installed in both beams about $150\Um$ upstream of the collision points
for all experiments, one TCTP in the horizontal plane (TCTPH) and one in the
vertical (TCTPV). They provide local protection of the quadrupole triplets in
the final focusing system, which are the limiting cold apertures during physics
operation. They are also important for decreasing the experimental background.
Downstream of the high-luminosity experiments, ATLAS and CMS, there are three
TCLs (target collimator, long) per beam, to intercept the collision debris. Furthermore, at
the beam extraction in IR6, dump protection collimators are installed as a
protection against miskicked beams in the case of extraction failures.
Similarly, there are injection protection collimators in IR2 and IR8.

During the long LHC shutdown in 2013 and 2014, 18 new collimators based on a
beam position moni\-tor design \cite{carra}, in which beam position monitor pick-ups are embedded in the
jaws to measure the beam position at the collimator location, have
been installed. They replaced the TCSGs (target colli\-mator, secondary, graphite) in IR6 and the
tertiary collimators in all experiments, as these locations are considered
more critical for orbit control, in order to enhance LHC performance
\cite{roderikBeta}. These collimators are called TCTP and TCSP, where `P' stands
for pick-up.

\subsection{Optics and layout of cleaning inserts}
\begin{figure}
  \centering
  \includegraphics[width=119mm]{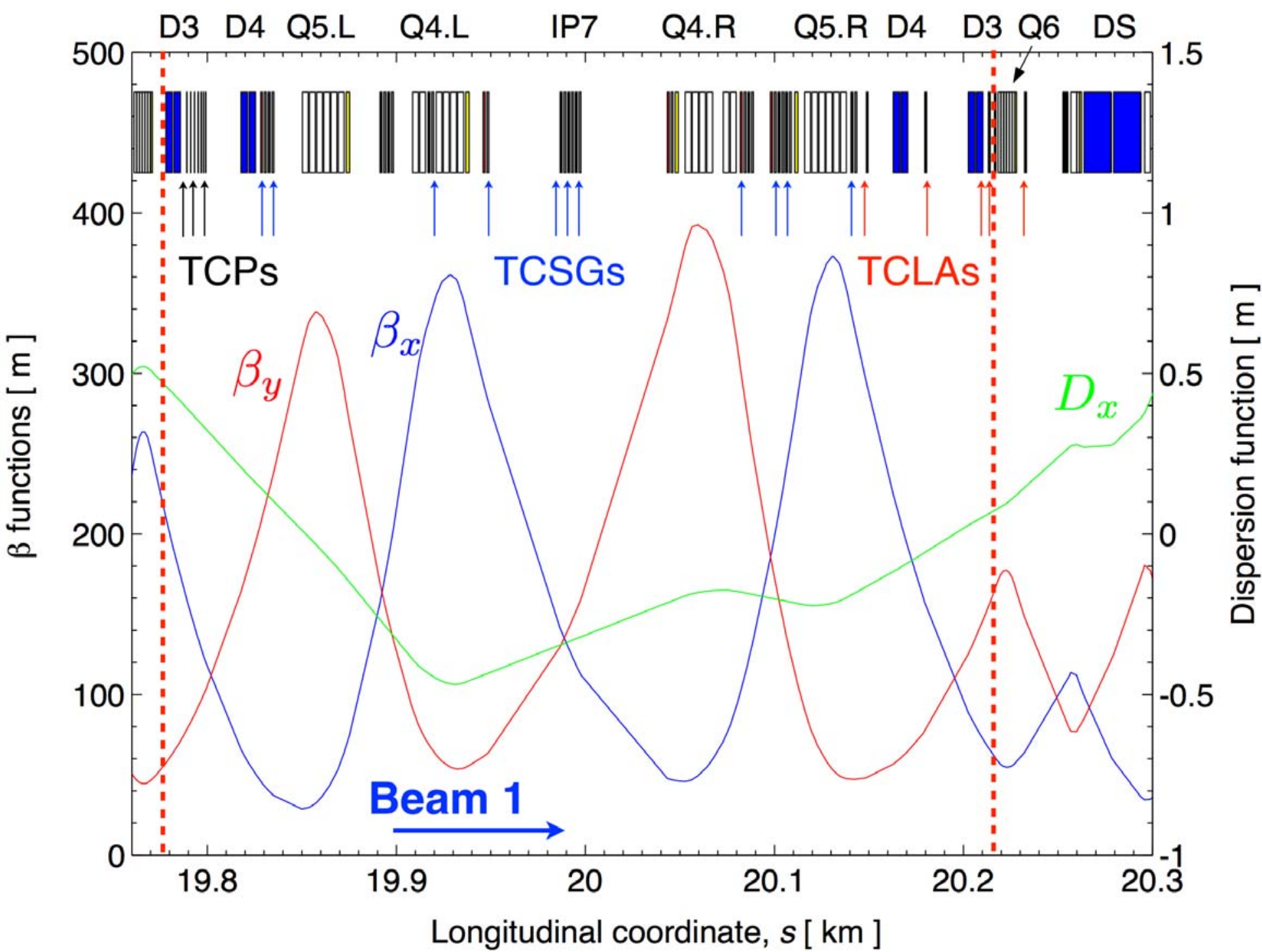}
  \vspace{-0.2cm}
  \caption{Betatron ($\beta_x$, $\beta_y$) and dispersion ($D_x$) functions
    as a function of $s$
    in the LHC betatron cleaning insertion IR7. The main layout elements
    are also shown: quadrupoles (white boxes), dipoles (blue), and collimators
    (black). Vertical arrows indicate the installed collimators: 3 TCPs, 11 TCSGs;
    5 TCLAs. Vertical red dashed lines indicate the limits of the warm regions (Q6 magnets at either side of IP7 are the first cold magnets). D, dipole magnet; IP, interaction point; L, left; R, right; Q, quadrupole magnet; TCLA, target collimator long absorber; TCP, target collimator (primary); TCSG, target collimator (secondary, graphite).}
  \label{figLayoutIR7}
  \vspace{-.3cm}
\end{figure}

\begin{figure}
  \centering
  \includegraphics[width=119mm]{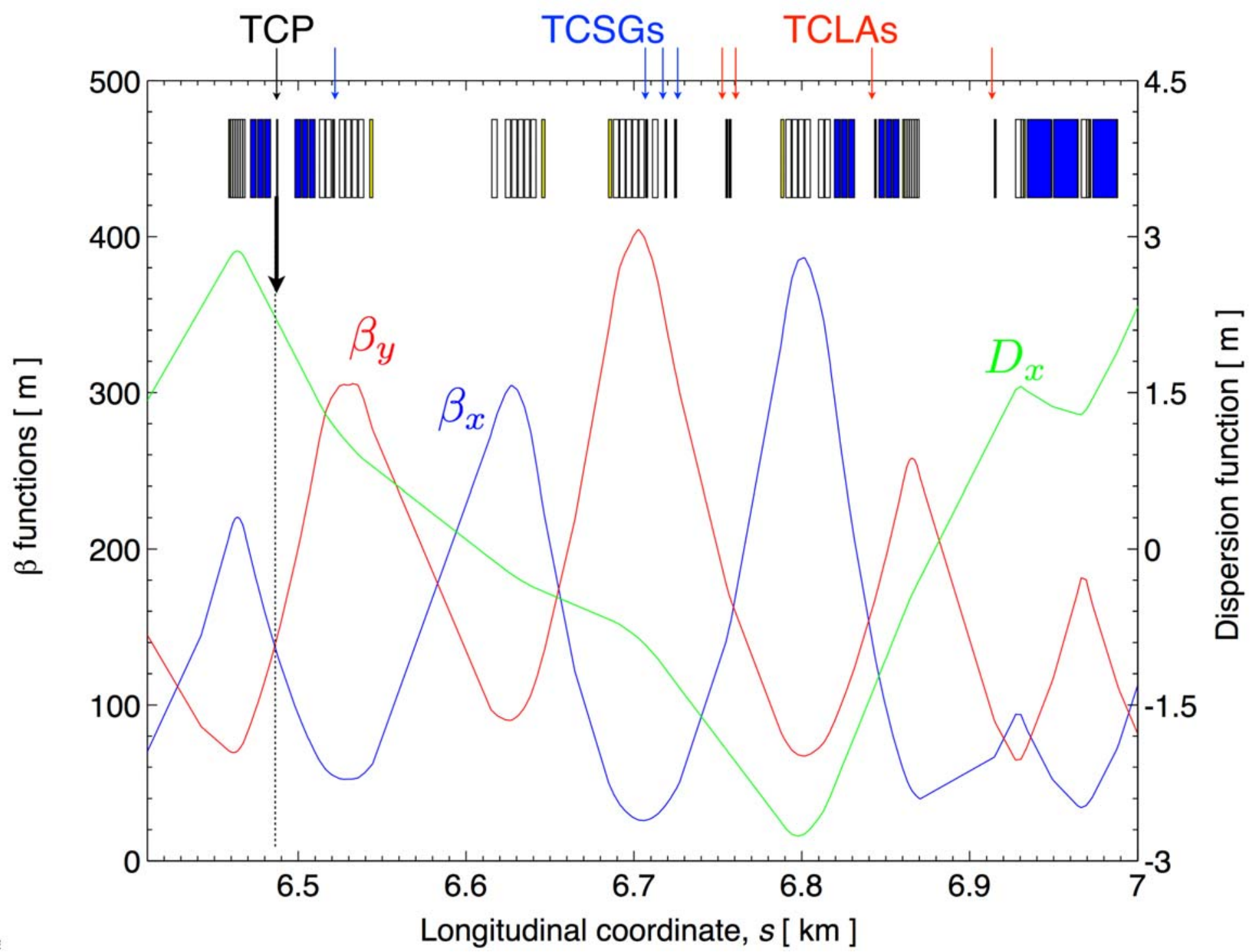}
  \vspace{-0.2cm}
  \caption{Betatron ($\beta_x$, $\beta_y$) and dispersion ($D_x$) functions
    as a function of $s$
    for B1 in the LHC momentum cleaning insertion IR3. Vertical arrows indicate
    the installed collimators: 1 TCP, 4 TCSGs;
    4 TCLAs. The main layout elements
    are also shown: quadrupoles (white boxes), dipoles (blue), and collimators
    (black). TCLA, target collimator long absorber; TCP, target collimator (primary); TCSG, target collimator (secondary, graphite).}
  \label{figLayoutIR3}
  \vspace{-.3cm}
\end{figure}

The optics and layout of the betatron and momentum cleaning inserts are
shown in Figs.~\ref{figLayoutIR7} and \ref{figLayoutIR3}, respectively. In
both inserts, four \emph{dog-leg} dipoles, called D4 and D3, are placed symmetrically on either side of the `IP7', and are used to enlarge the
beam--beam separation from $194\Umm$ to $224\Umm$, making more transverse space
for collimators. The two D4 magnets also delimit the
${\approx}500\Um$
long warm insert, which  comprises the warm quadrupoles Q4 and Q5. The Q6
quadrupoles on either side of the D4 dipoles are the first superconducting
magnets before the beam enters the cold arc.

In IR7, three primary collimators intercept horizontal, vertical, and skew
halos. They are located in the region between the D3 and D4 dipoles, \ie
on the upstream side of the warm insertion for each beam. This maximizes the
length of the warm section downstream of the primary loss location. A similar
implementation is adopted in IR3, where, however, only one horizontal TCP is
needed, placed at a location with large
normalized dispersion, $D_x/\sqrt{\beta_x}$, to intercept particles
with energy deviations. Momentum cleaning in one plane is sufficient, as
at the LHC, vertical dispersion is negligible. The IR3 primary collimator
needs to be at larger transverse betatron amplitudes than those of the IR7,
to decouple the functionalities of the two inserts. Typical transverse
betatron amplitudes expressed in units of $\sigma_x$ as in \Eref{sz} are
2.5--3 times larger than in IR7, to ensure that IR3 does not act as a betatron
system for particles with small energy errors.

The collimators of IR3 and IR7 are indicated in Figs.~\ref{figLayoutIR7} and
\ref{figLayoutIR3} by black boxes. Eleven TCS collimators are used in IR7,
whereas four are used in IR3, since collimation occurs in one plane only. Five
active absorbers (TCLAs) are used in IR7 and four in IR3. These devices, of
types TCP, TCSG, and TCLA (see Table~\ref{tabList}) are all two-sided
collimators. Even if a one-sided collimator might be sufficient for a multiturn
cleaning process, two-sided collimators are crucial for precise alignment of the
circulating beams.

\begin{figure}
  \centering
  \includegraphics[width=\linewidth]{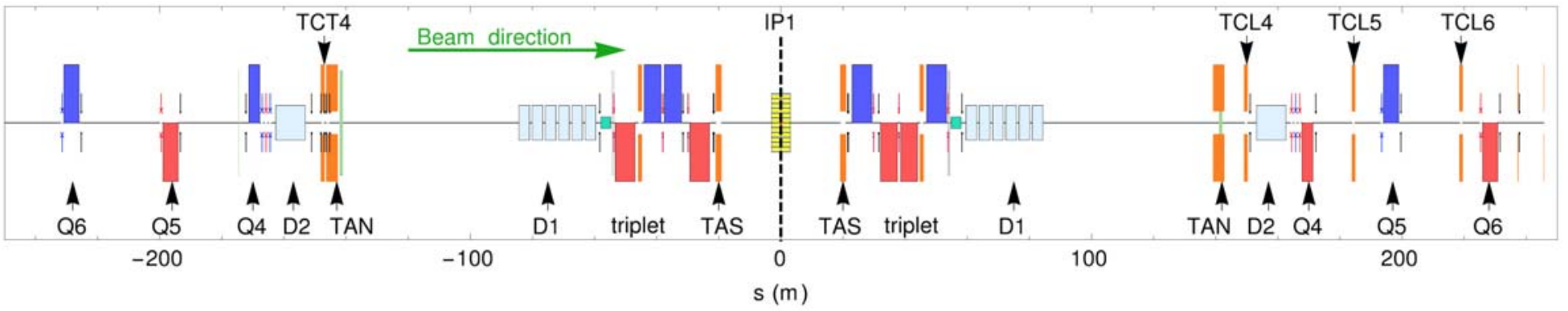}
  \vspace{-0.8cm}
  \caption{Layout elements around IR1 (ATLAS) in the 2015 configuration of the
    LHC collimation system. D, dipole; IP, insertion point; Q, quadrupole; TAN, target absorber (neutral); TAS, target absorber (secondary); TCL, target collimator (long); TCT, target collimator (tertiary). Courtesy of R.~Bruce.}
  \label{figIR}
  \vspace{-.3cm}
\end{figure}

The layout of IR1 (ATLAS) is shown in \Fref{figIR}. A pair
of horizontal and vertical TCTPs protect the triplet from incoming beam losses.
Three TCL-type physics debris absorbers protect the magnets
downstream of the IR from collision products. The other
high-luminosity experiment, CMS in IR5, has an equivalent layout. For IR2 (ALICE) and
IR8 (LHCb), there is no need for a TCL collimator because the lower luminosity
values do not put the matching sections at risk of quenching.

In addition to the movable collimators, 10 passive absorbers are also mounted in front of the most exposed warm magnets of each collimation insert: the D3
magnets downstream of the TCPs and the first modules of the Q5 and Q4
quadrupoles. These fixed-aperture collimators, called TCAPs, dramatically reduce the radiation doses to magnet coils, increasing their lifetimes
by a factor of 10  or more (chapter 18 of \cite{lhc}).

\subsection{Operational challenges and beam-based set-up}
\subsubsection{LHC operational cycle and recap. of machine configurations}
The main phases of the LHC operational cycle, which is
periodically run to prepare for periods of physics data acquisition
(`stable beam' mode), are injection, energy ramp, betatron squeeze, and
preparation of collisions (`adjust' mode). The squeeze, in which the optics
around the interaction points are changed to reduce the colliding beam sizes, has so far been performed at constant flat-top energy. In this phase, the
betatron function is enlarged at the inner triplets as required to reduce
the $\beta^*$ values, \ie the beta functions at the collision points.

The
LHC design value of $\beta^*$ for the high-luminosity points IP1 (ATLAS) and IP5
(CMS) is $55\Ucm$ for a beam energy of $7\UTeV$, limited by the available triplet
aperture. During LHC run~I, a $\beta^*$ value of $60\Ucm$ was achieved at $4\UTeV$.
The first year, 2015, of LHC run~II started with a $\beta^*$ value of $80\Ucm$
at $6.5\UTeV$, to ease recommissioning after the 2 year shutdown
\cite{beatCham2014}
but it is planned to move to a $\beta^*$ value close to $40\Ucm$ in 2016.
These excellent results were achieved thanks to a better aperture than had been anticipated during the LHC design phases, which was also better than the one
used to specify various LHC systems. For the scope of this lecture, it
is still useful to review the system design by starting from the design values.

\subsubsection{Collimation settings strategy in the LHC operational cycle}

The LHC aperture was reviewed in \Sref{notation}; see
Table~\ref{tab_bottleneck_inj}. With an injection stored energy of $22\UMJ$, \ie
not only above the quench limit but also significantly above the damage limit of
metals \cite{ab}, beam
colli\-mation is required in every phase of the LHC operational cycle, from
injection to collision. Particularly challenging are the dynamic
phases (energy ramp, betatron squeeze, and change of orbit configurations),
when collimator movements must be synchronized precisely with other
accelerator systems, such as power converters and RF units. This operation
mode imposes tight constrains on the collimator control design.

At the injection, distributed aperture bottlenecks are expected in the arcs, as
the magnet aperture was designed to fit the beams at injection \cite{lhc}.
At $7\UTeV$, the
arc aperture is no longer critical because the betatron amplitudes are damped
at larger beam energies. The aperture is now limited by the
triplets, where $\beta$ functions of up to ${\approx}4500\Um$ are required to
achieve small beam sizes at the interaction points. By design (see
Table~\ref{tab_bottleneck_inj}), the normalized apertures, $\hat{A}_{{\text{min}},z}$,
are actually similar for the two extreme cases. Thus, even if the
accelerator physics motivations are different, similar collimator
settings are deployed at injection and in physics conditions. This involves
moving collimators to follow the shrinking beam envelope.

\begin{figure}
  \centering
  \includegraphics[width=125mm]{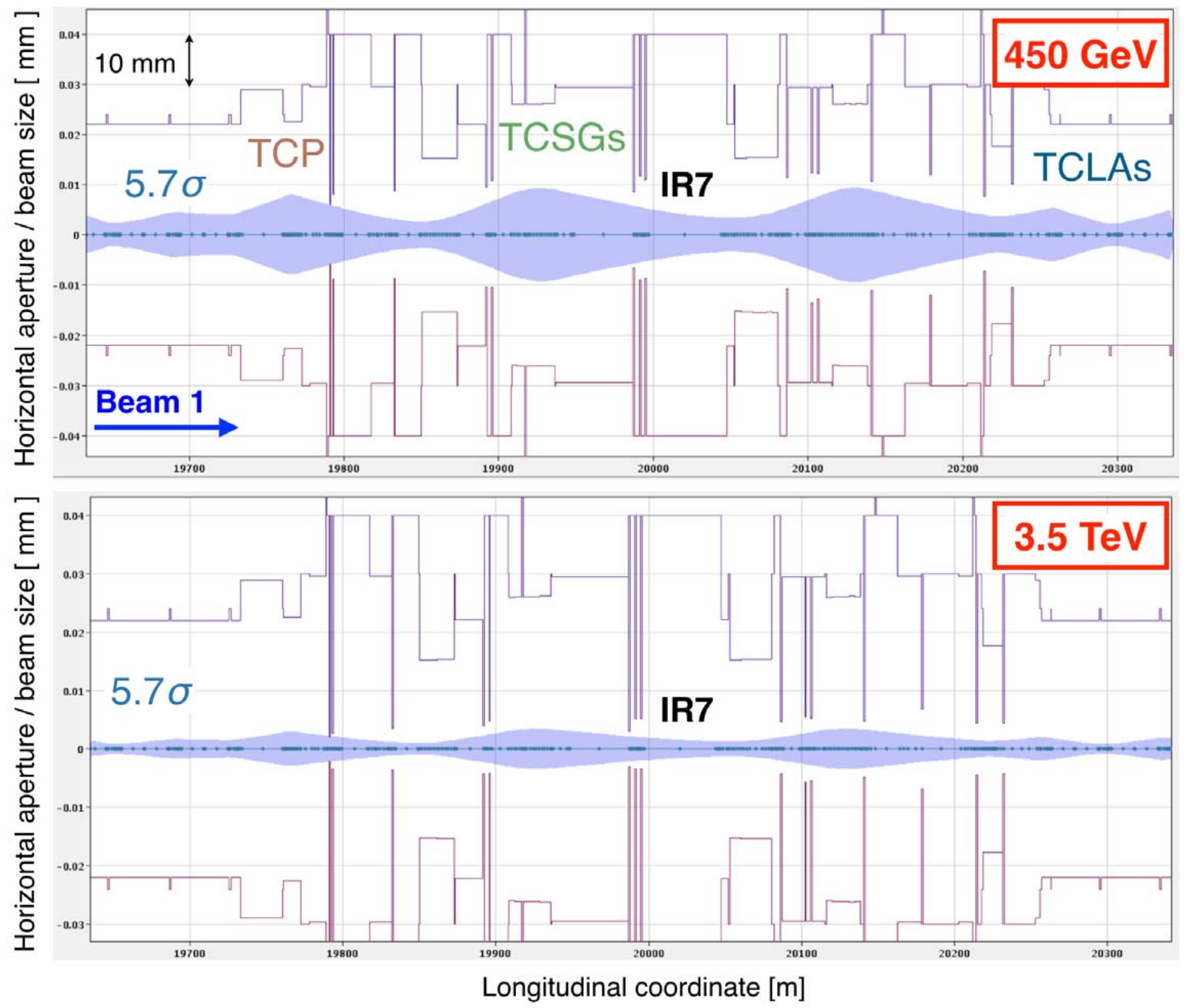}
  \vspace{-0.2cm}
  \caption{Horizontal aperture, collimator jaw positions (vertical bars) and
    $5.7\sigma$ beam envelope at (top) injection and (bottom) $3.5\UTeV$ in
    betatron cleaning (IR7) from the LHC on-line model application
    \cite{hb2010}. IR, insertion region; TCG, target collimator (graphite); TCLA, target collimator (long absorber); TCP, target collimator (primary).}
  \label{figIP7gaps}
\end{figure}

\begin{figure}
  \centering
  \includegraphics[width=65mm]{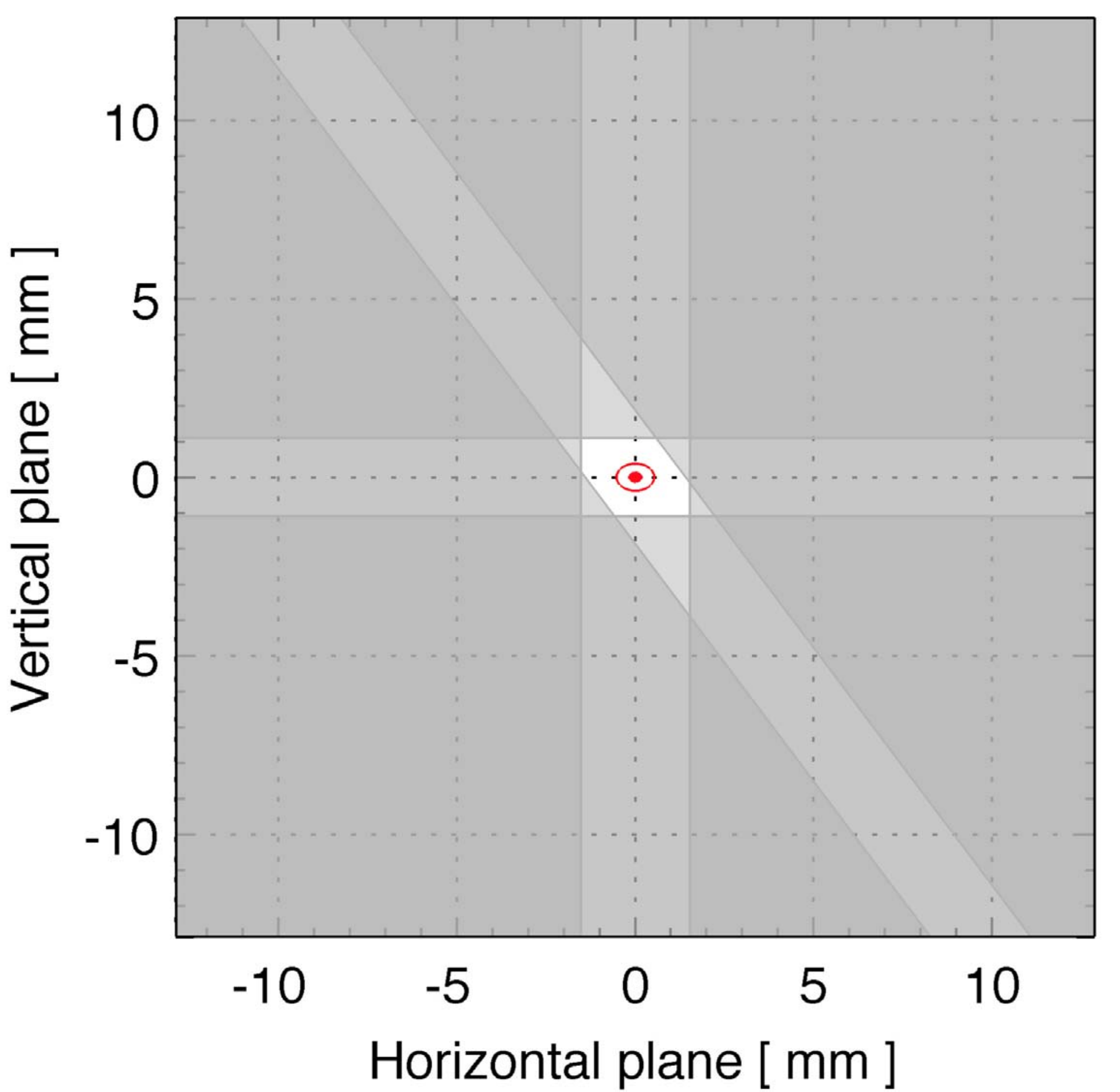}
  \includegraphics[width=80mm]{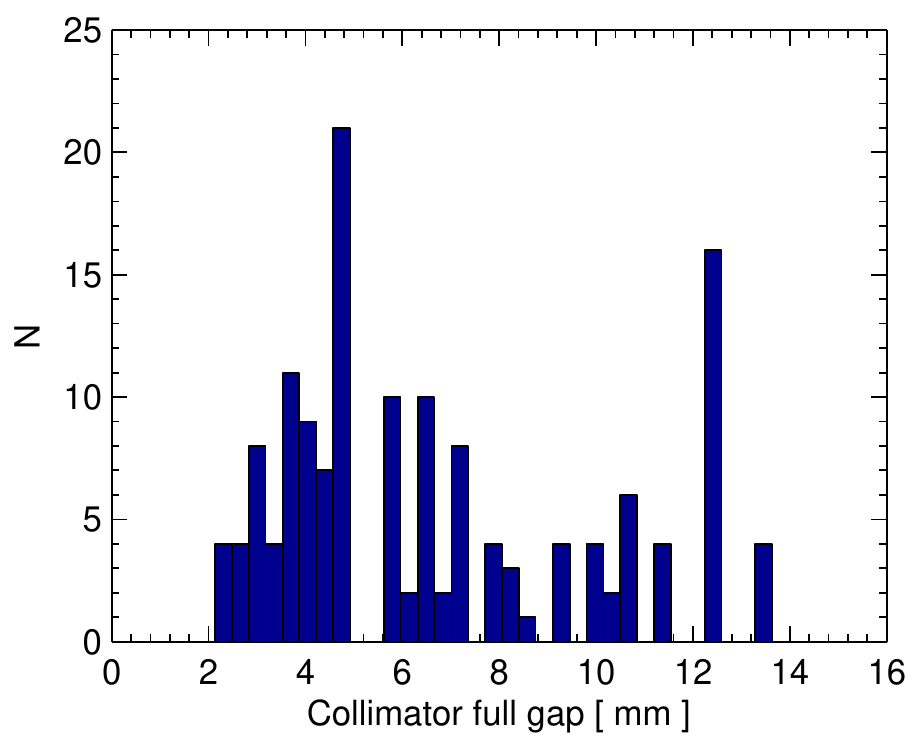}
  \vspace{-0.2cm}
  \caption{Left: Beam clearance for the LHC beams, as defined by the primary
    collimator gaps. Right: Distribution of collimator gaps, as
    adopted for operation at $4\UTeV$ and $\beta^\star=60\Ucm$ in 2012. In 2015,
    the same IR7 settings in millimetres are used for the $6.5\UTeV$ operation.}
  \label{fig_smallgaps}
\end{figure}

\Figure[b]~\ref{figIP7gaps} shows an example of collimator settings at injection
(top) and $3.5\UTeV$ (bottom), taken from the operation configuration of the LHC 2010 run \cite{hb2010}. The horizontal beam envelope at
$5.7\sigma_x$, as defined by the TCP gaps, is shown, together with the values of the collimator half gap projected on the horizontal plane at each
collimator (magenta bars). The TCPs were kept at a
normalized aperture of $5.7\sigma_z$ at all energies. The TCSGs were moved
from $6.7\sigma_z$ to $8.5\sigma_z$, and the TCLAs were moved from
$10.0\sigma_z$ to $17.7\sigma_z$. These relaxed top-energy settings were
conceived to reduce
the operational tolerances in the first year of the run \cite{hb2010} and were then
subsequently tightened to improve the cleaning performance \cite{beta2012},
reaching $4.3\sigma_z$ in 2012.
The collimator gap values in millimetres, as used for the $4\UTeV$ operation at
$\beta^*=60\Ucm$ are shown in \Fref{fig_smallgaps}, where the transverse
clearance left by the IR7 primary collimators and the distribution of gaps
are shown. The smallest gap is $2.1\Umm$.

It is clear from \Fref{figIP7gaps} that a basic requirement for the
LHC collimator design is that the jaws must be movable, as the gaps required
at top energy to ensure optimum performance are not compatible with the
larger beam sizes at injection. The need for small gaps at top energy
also has important effects on the operational strategy of the collimation
system because it necessitates dedicated beam-based alignment procedures, as
collimators cannot be set deterministically to such small gaps without direct
measurements to `find' the local beam position and size.

\subsubsection{Beam-based set-up of LHC collimators}
\begin{figure}
  \centering
  \includegraphics[width=120mm]{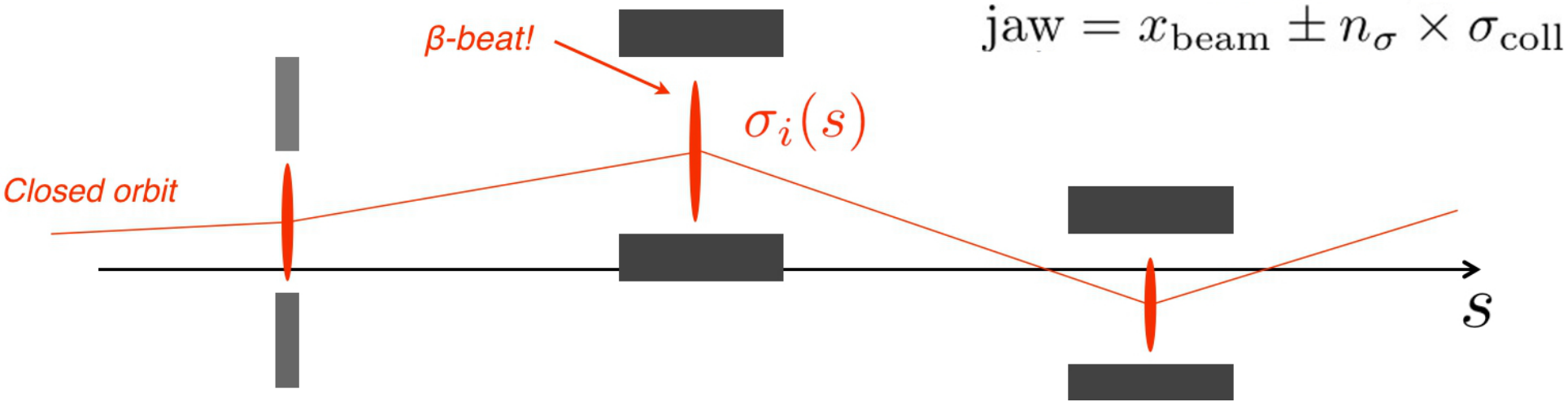}
  \vspace{-0.2cm}
  \caption{Collimator jaw positions at various
    locations in the ring, where the closed orbit and beam size are different.
    Proper collimator set-up requires direct measurements of beam position and size, to ensure that the collimator hierarchy is
    respected.}
  \label{figAlign}
\end{figure}

he LHC collimation system performance relies on respecting the well-defined hierarchy between colli\-mator families. In practice, this involves
knowing the beam orbit and beam size at each collimator, as
shown in \Fref{figAlign}. With beam sizes as small as $200\Uum$ and
orbit offsets of up to 2--3\Umm, and in the presence of collimator alignment errors of up
to a
few hundred micrometres, the determination of optimum jaw positions
can only be achieved through a series of measurements aimed at measuring the
required parameters, which are referred to as \emph{beam-based collimator
  alignments}.

\begin{figure}
  \centering
  \includegraphics[width=130mm]{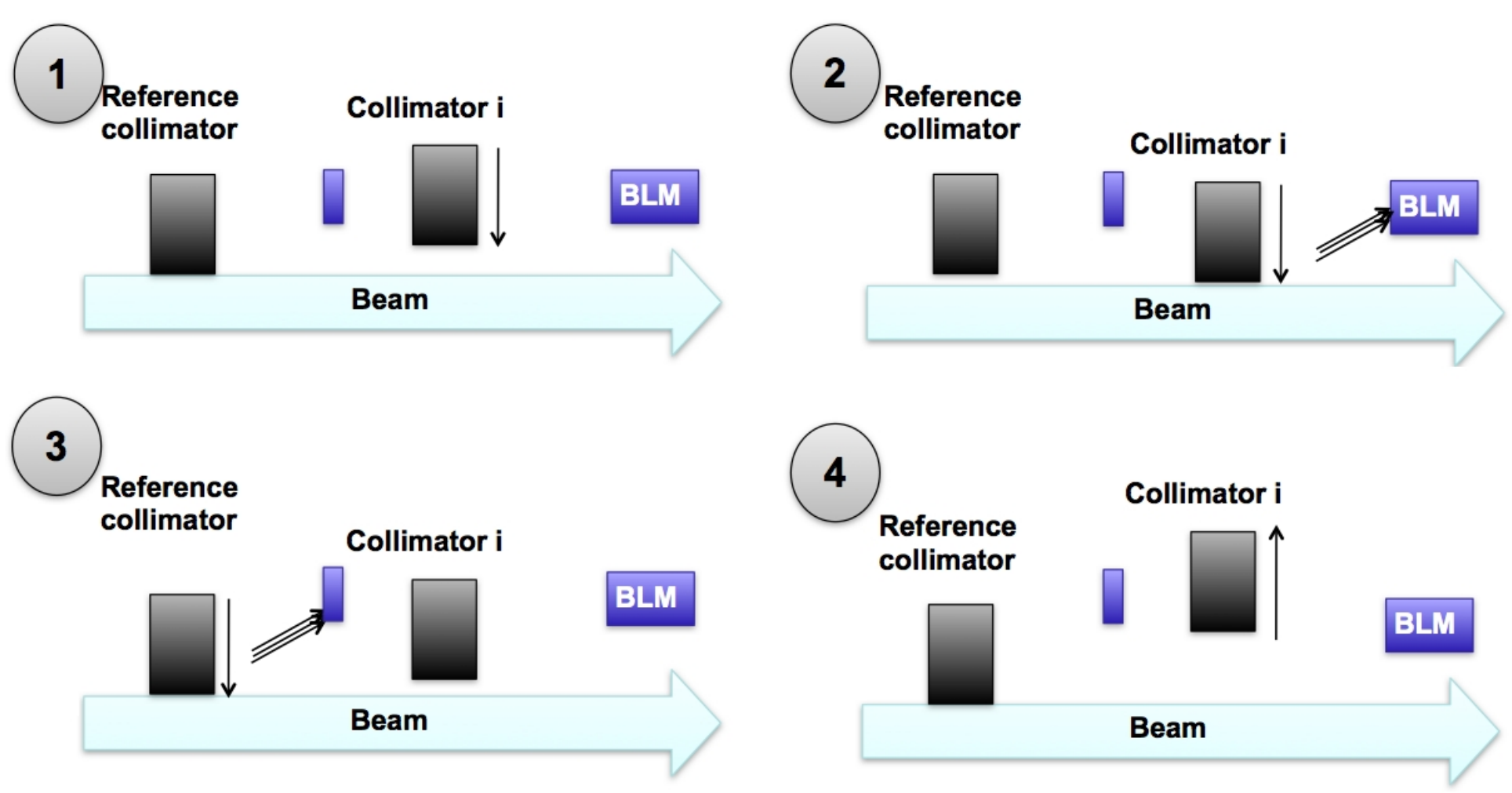}
  \vspace{-0.2cm}
  \caption{The collimator set-up procedure used to determine
    beam orbit and relative beam size to that of a reference collimator,
    for operational settings generation \cite{collIPAC10}. BLM, beam loss monitor.}
  \label{figSetupProc}
  \vspace{-0.2cm}
\end{figure}

\begin{figure}
  \centering
  \includegraphics[width=70mm]{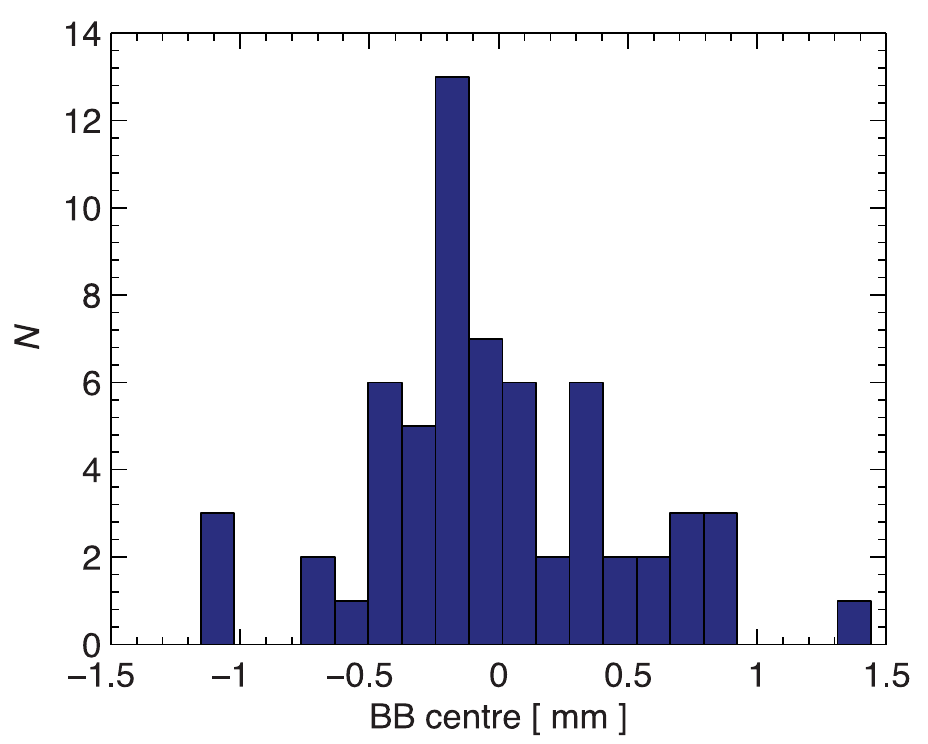}
  \vspace{-0.2cm}
  \caption{Distribution of beam-based centres of LHC collimators as a result
    of the alignment campaign of 2012. Shifts of up to more than $1.5\Umm$ are found
    from the cumulative effects of orbit misalignments, electronics offsets
    of the beam position monitor system, alignment error of the collimators with respect to the
    reference orbit, \etc}
  \label{figBBcentres}
\end{figure}

The procedure for collimation set-up at the LHC
(\Fref{figSetupProc}) was established \cite{hb2010, collIPAC10} based on
ex\-peri\-ence gained with a prototype LHC collimator installed for beam tests
in the Super Proton Synchrotron (SPS) \cite{spsExp}. The beam halo is shaped with a reference collimator
(1), typically a primary collimator, which is closed to a known half gap of $3-5\sigma$. This reference halo is used to cross-align other colli\-mators, by
moving their jaws towards the beam in small steps of 5--20$\Uum$ until the halo is \emph{touched}, with symmetrical beam loss responses from
either jaw (2). This gives the local orbit position. The reference
collimator is then closed further (3) until it touches the halo again: this
enables the gaps of the two collimators to be cross-calibrated. The average of the
initial and final gaps of the reference collimator in units of $n_\sigma$ gives
the normalized gap of the other collimator. Finally, the latter collimator is opened to
its nominal settings (4). This ensures that the relative retraction with respect to the reference collimator is respected, even in the presence of different
beta-beating at the two locations. An example of \emph{beam-based collimator centres} measured in 2012 is shown in \Fref{figBBcentres}. This reinforces the previous assertion that beam-based alignment is mandatory for a proper
collimator set-up at the LHC.

This set-up procedure is precise but time consuming. During the initial
commissioning  in 2010, it was carried out manually for each
collimator. An automated feedback system between collimator movements and the beam loss monitor
signal has been developed, enabling the set-up time to be improved significantly
and dramatically reducing the number of spurious beam aborts from human
error. A detailed treatment of this optimization of beam-based alignment
is beyond the scope of this lecture but can be found in Ref. \cite{thesisGV}.

\subsubsection{Collimator setting generation for operation}

Beam-based alignment must be done for each collimator in the ring, for every
relevant machine con\-figur\-ation (injection, top energy before and after squeeze,
collision). To minimize the risk of damaging the collimators while approaching
them to the beams, the alignment is carried out with the minimum intensity that
allows reliable orbit measurements, \ie with a few bunches of nominal bunch
intensity. Let us now assume that the local orbit, $x_{\text{beam}}$, and beam size,
$\sigma_{\text{coll}}$, are calculated at every collimator in each discrete point
of the operational cycle.

\begin{figure}
  \centering
  \includegraphics[width=80mm]{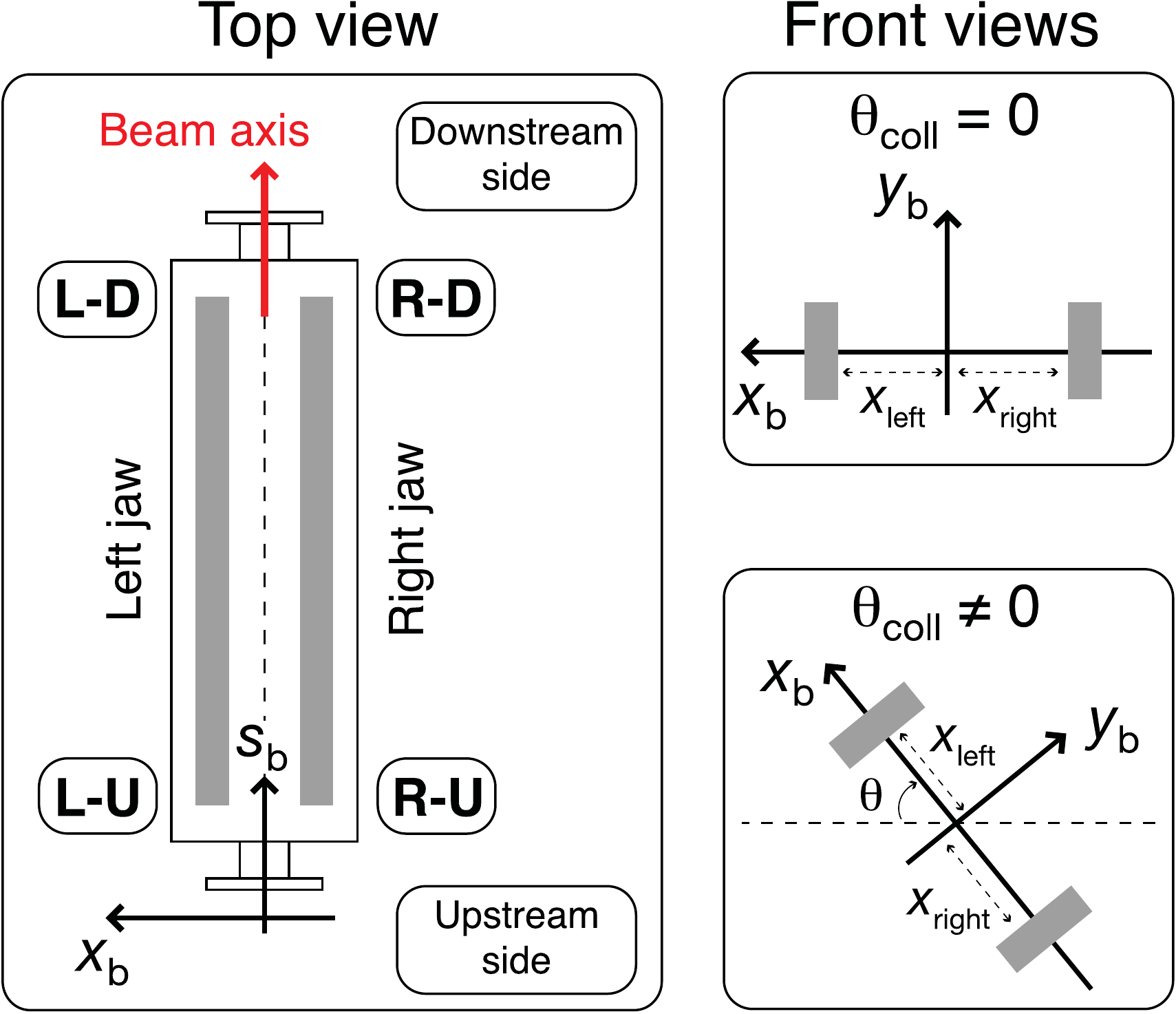}
  \vspace{-0.2cm}
  \caption{Top and front views of a collimator, with labels
    and naming conventions. Each jaw has two motors that move the jaws in the
    collimation plane: horizontal ($\theta_\text{coll}=0$), vertical
    ($\theta_\text{coll}=\pi/2$) or
    skew planes. D,~downstream; L, left; R, right; U, upstream.}
  \label{figNaming}
\end{figure}

While collimators are installed in a variety of azimuthal orientations (see
\Fref{figNaming}), the jaw move\-ment is in one dimension, along the collimator plane. For arbitrary collimator angles $\theta_{\text{coll}}$, the \emph{effective} beam size in the collimation plane, $\sigma_{\text{coll}}$ is computed from the
horizontal and vertical sizes as
\begin{equation}
  \sigma_{\text{coll}}=\sqrt{\sigma_x^2\cos(\theta_{\text{coll}})^2
    +\sigma_y^2\sin(\theta_{\text{coll}})^2}~,
\end{equation}
where $\sigma_z$, $z\equiv(x,y)$, is calculated as in \Eref{sz}. The
collimator half gap is calculated as $h=n_\sigma\times\sigma_{\text{coll}}$ and the
jaw positions around the beam position, $x_{\text{beam}}$, are given by
\begin{equation}
  {\text{jaw}}=x_{\text{beam}}\pm n_\sigma\times\sigma_{\text{coll}}~.
\end{equation}
Note that each jaw has two motors, which allow the tilt angle to be adjusted with respect
to the beam envelope. In the following, the tilt angle is assumed to be zero. Stepping
motors can be driven through arbitrary functions of time. The motion of
collimators around the ring can be synchronized through timing events at the microsecond level \cite{contrIcapeps, collControlsPac09}. This is necessary to
ensure optimum collimator settings during critical machine phases, such as the
energy ramp and the betatron squeeze. To this end, continuous setting functions
must be generated from the beam-based parameters through scaling rules versus
beam energy and optics.

\begin{figure}
  \centering
  \includegraphics[width=100mm]{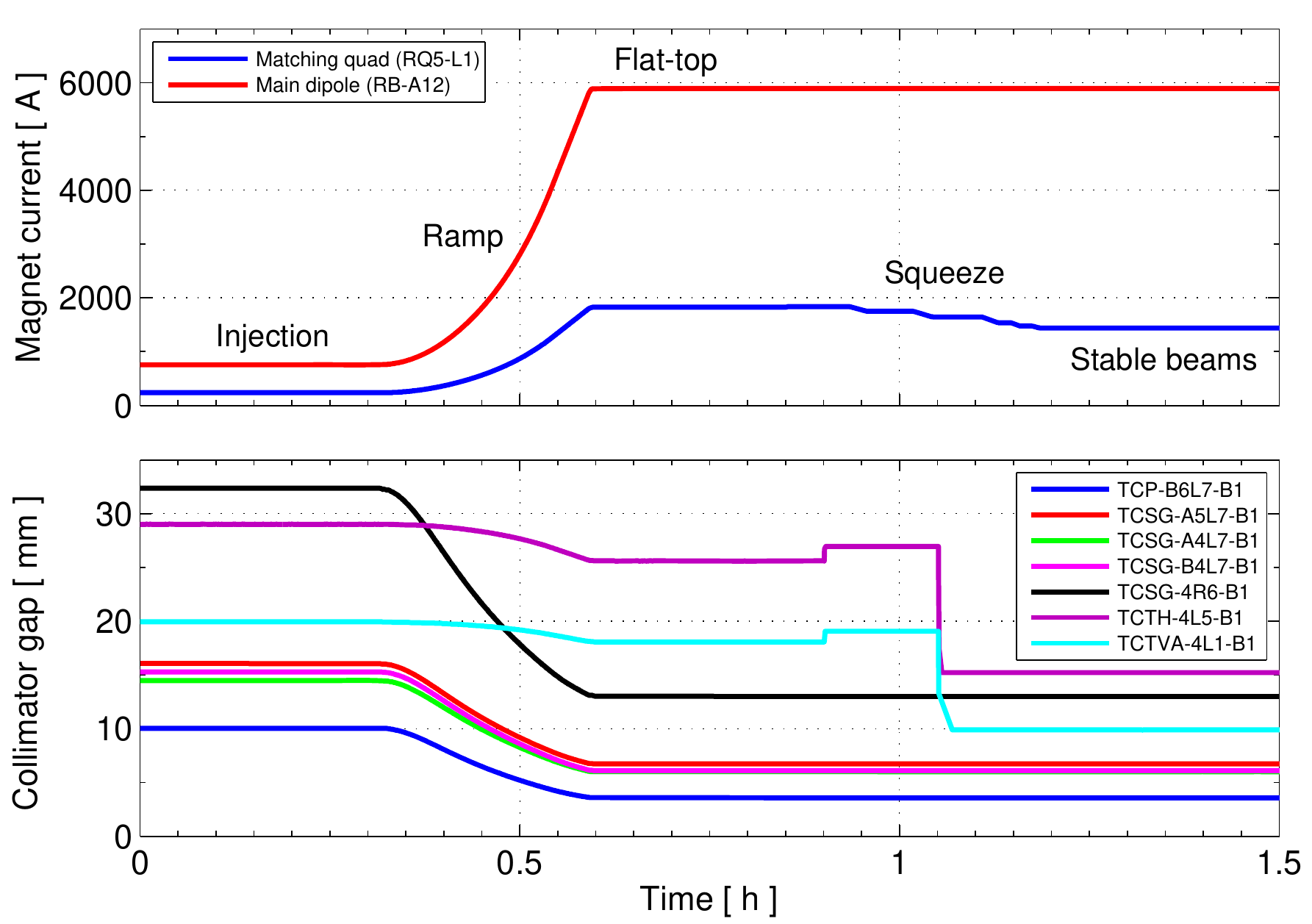}
  \vspace{-0.2cm}
  \caption{Operational cycle for selected collimators for a typical LHC fill.
  Top: Measured magnet currents versus time. Bottom: Collimator gaps versus time.}
  \label{figExFill}
\end{figure}

Let us calculate, for example, the ramp functions, starting from settings
values at injection (`0') and flat-top (`1'). The half gap
during the energy ramp is expressed as a function of the energy:
\begin{equation}
  h(\gamma)=n_\sigma(\gamma)\times\sigma_{\text{coll}}(\gamma)~,
\end{equation}
where $\gamma=\gamma(t)$ is the relativistic gamma function. For the LHC, it
is sufficient to use linear functions in $\gamma$ for $n_\sigma$ and
$\sigma_{\text{coll}}$. A linear interpolation between the beam-based parameters
at injection and flat-top yields:
\begin{equation}
h(\gamma)=
\left[n_{\sigma,0}+
\frac{n_{\sigma,1}-n_{\sigma,0}}{\gamma_1-\gamma_0}(\gamma-\gamma_0)
\right]
\times\frac{1}{\sqrt{\gamma}}
\left[
\frac{
\sqrt{\epsilon_1\beta_1}-\sqrt{\epsilon_0\beta_0}}{\gamma_1-\gamma_0}
(\gamma-\gamma_0)
\right]~.
\end{equation}
%
The beam centre is also expressed as a linear function of $\gamma$ to give
the jaw position as
\begin{equation}
  {\text{jaw}}(\gamma)=\left[x_{\text{beam},0}+\frac{x_{\text{beam},1}-x_{\text{beam}, 0}}
    {\gamma_1-\gamma_0}
    (\gamma-\gamma_0)\right]\pm h(\gamma)~.
\end{equation}
Note that the beam size $\sigma_{\text{coll}}=\sigma_{\text{coll}}(\gamma)$ is also a
function of the optics and therefore might change, typically for the
tertiary collimators in the experimental regions, during the betatron squeeze \cite{srRampSqu}. This notation can be generalized in a straightforward
way by considering functions of $\beta^*$ instead of $\gamma$ for the
parameters involved.
An example of collimator gaps versus time during a full LHC cycle is given in
\Fref{figExFill} (bottom graph), together with the LHC dipole
and matching quadrupole currents, to indicate the times of the ramp and squeeze phases (top graph).

The operation of the collimation system is automated by sequences
that are run at every fill, en\-abling operation crews to run
smoothly through the different sets of the cycle settings. The operation mode can
only work thanks to the excellent stability of the LHC orbit and
optics and of the collimator hardware itself. So far, only one beam-based
alignment per year has been required \cite{collPerf}.




\section{Collimator design for high-power accelerators}
\label{design}
The key parameters for the design of the LHC collimator are
summarized in Table~\ref{tabParam}. The list emphasizes the challenges in terms of quenching of superconducting magnets, damage, heating of components and radiation doses, which must be addressed by a optimized design. It is
important to note that the design must ensure adequate mechanical stability
during jaw position changes and in the presence of important heat loads. Other aspects related to materials choice to ensure robustness and
limited impedance are addressed in a companion paper \cite{ab}. Details of the
final collimator design deployed for the LHC can be found in Refs.
\cite{collDesign, finalColl}. Here, only the main design features are given.

\begin{table}
  \caption{Minimal horizontal and vertical apertures at injection
  ($450\UGeV$) for warm and cold elements}
  \begin{center}
    \begin{tabular}{ll}
      \hline\hline
      \textbf{Parameter} & \textbf{Value} \\ 
      \hline
      High stored beam energy &  $360\UMJ/\text{beam}$\\
      Large transverse energy density& $1\UGJ/\UmmZ^2$\\
      Activation of collimation inserts & 1--15$\UmSv/\UhZ$ \\
      Small spot sizes at high energy & ${\approx}200\Uum$\\
      Collimation close to beam & 6--7$\sigma$\\
      Small collimator gaps & $2.1\Umm$ (at $7\UTeV$)\\
      Big and distributed system & 110 devices, $\approx$500 degrees of freedom\\
      \hline\hline
    \end{tabular}
  \end{center}
  \label{tabParam}
\end{table}


The LHC collimators are high-precision devices that ensure the
correct hierarchy along the $27\Ukm$ long ring with beam sizes as small as $200\Uum$. Each collimator has two jaws, of different lengths and materials,
depending on functionality (Table~\ref{tabList}). Each jaw can be
independently moved by two stepping motors. Key features of the design
are: (1) a
jaw flatness of about $40\Uum$ along the $1\Um$ long active jaw surface;
(2) a surface roughness less than $2\Uum$; (3) a $5\Uum$ positioning resolution;
(4) an overall setting reproducibility below $20\Uum$ \cite{contrIcapeps};
(5) a minimal gap of $0.5\Umm$; (6) evacuated heat loads of up to $7\UkW$ in
a steady-state regime and of up to $30\UkW$ in transient conditions.

Primary and secondary collimators are made of a robust CFC that is designed to withstand beam impacts without significant
permanent damage for the worst failure cases, such as impacts of
a full injection batch of $288\times1.15\times10^{11}$ protons at $450\UGeV$ and
of up to $8\times1.15\times10^{11}$ protons at $7\UTeV$ \cite{collDesign}. Other
collimators made of heavy tungsten alloy or copper, obviously, do not have the
same robustness and are only utilized at larger distances from the circulating beams, where maximum absorption is needed.

\begin{figure}
  \centering
  \includegraphics[width=135mm]{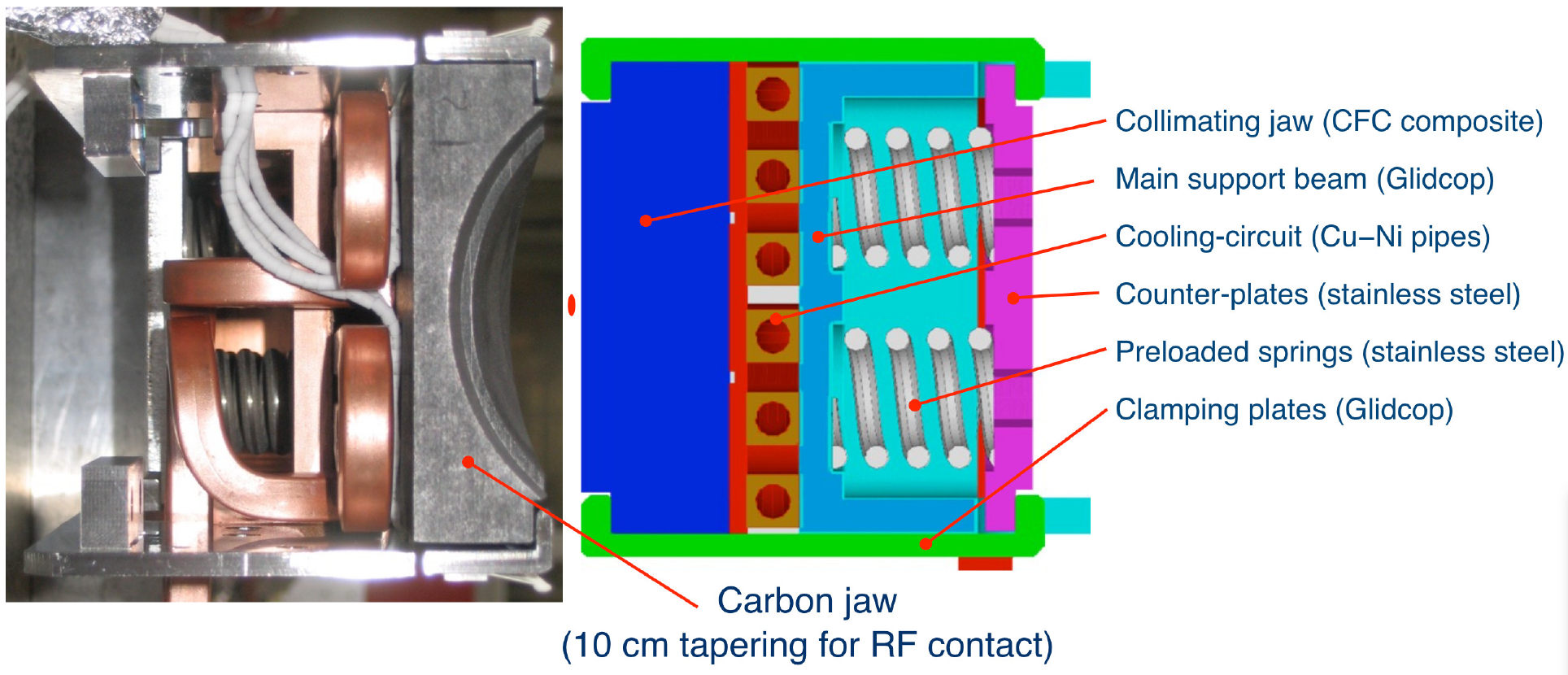}
  \vspace{-0.2cm}
  \caption{Cross-section of the LHC collimator jaws. Left: real prototype. Right: design drawing. The position of the beam is shown by the red
    ellipse, as if the two jaws were those of a horizontal collimator. A
    sandwich structure, with cooling circuits clamped on the
    CFC plate of the active part, is optimized to minimize deformation of the
    structure during steady loss conditions \cite{collDesign}.}
  \label{figXsec}
\end{figure}

\begin{figure}
  \centering
  \includegraphics[width=150mm]{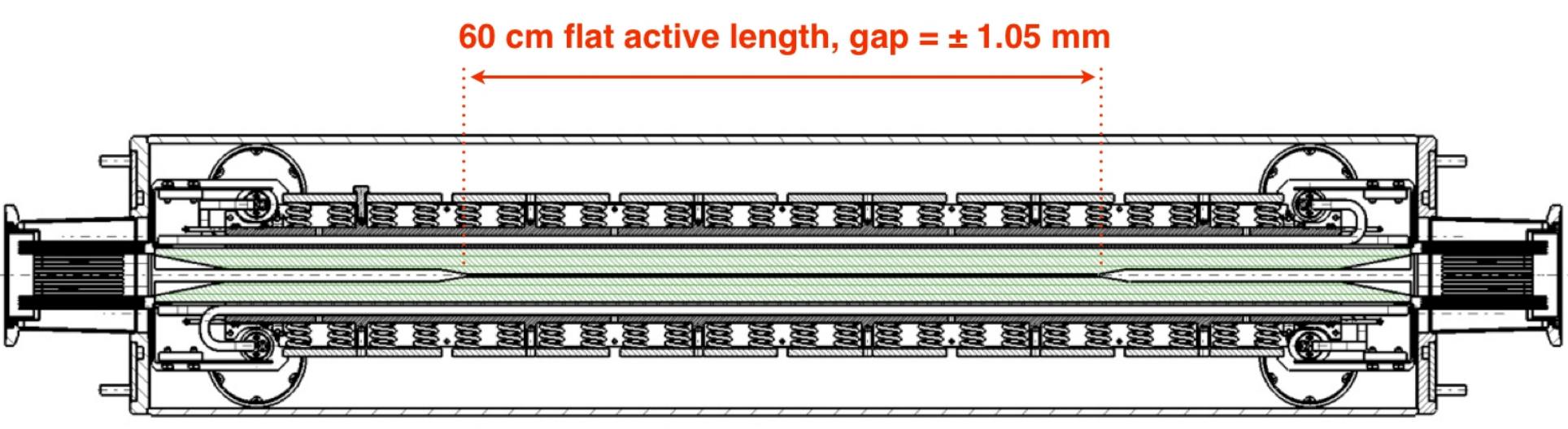}
  \vspace{-0.2cm}
  \caption{Design of the LHC primary collimator. The two jaws can move
  independently, thanks to four stepping motors enabling position and angular
  adjustment with respect to the beam. This design is essentially identical
  to that of the secondary collimator except that the jaws are tapered to an
  effective length of $60\Ucm$ instead of $100\Ucm$.}
  \label{figTCP}
\end{figure}

The cross-section of the primary and secondary collimator jaws, with a $2.5\Ucm$
thick active CFC part and a cooling system underneath, is shown in
\Fref{figXsec}. The design drawing on the right side of the picture is
compared with a real jaw prototype on the left, built  during the initial
production phases to verify the manufacturing quality.
Two parallel jaws are mounted in the vacuum tank, as shown in \Fref{figTCP}
for a primary collimator. 
In \Fref{figTCP}, the jaws are actually shown set to the
operational position for the vertical collimator with the tightest gaps, as in \Fref{fig_smallgaps}.

\begin{figure}
  \centering
  \includegraphics[width=60mm]{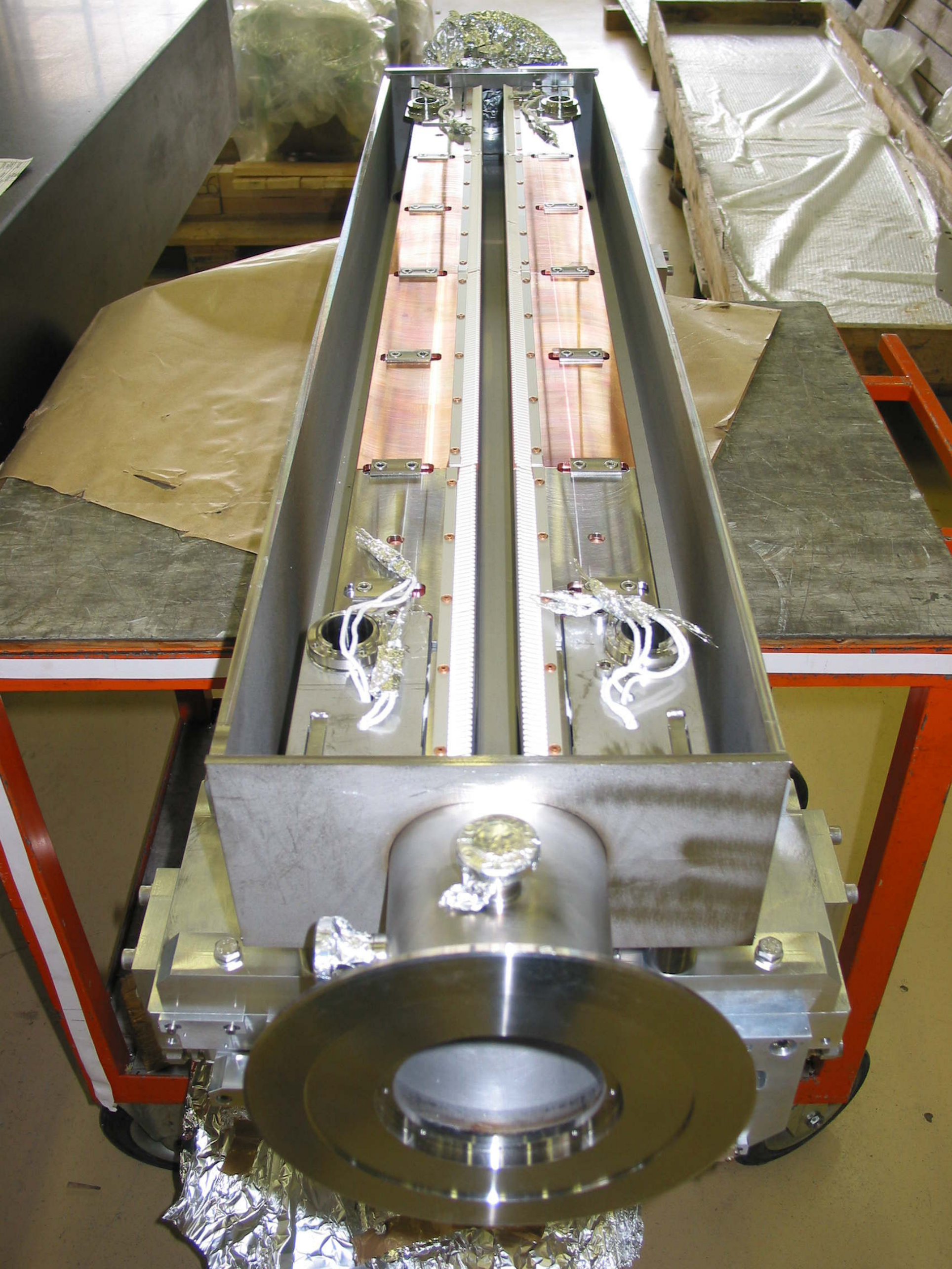}
  \includegraphics[width=80mm]{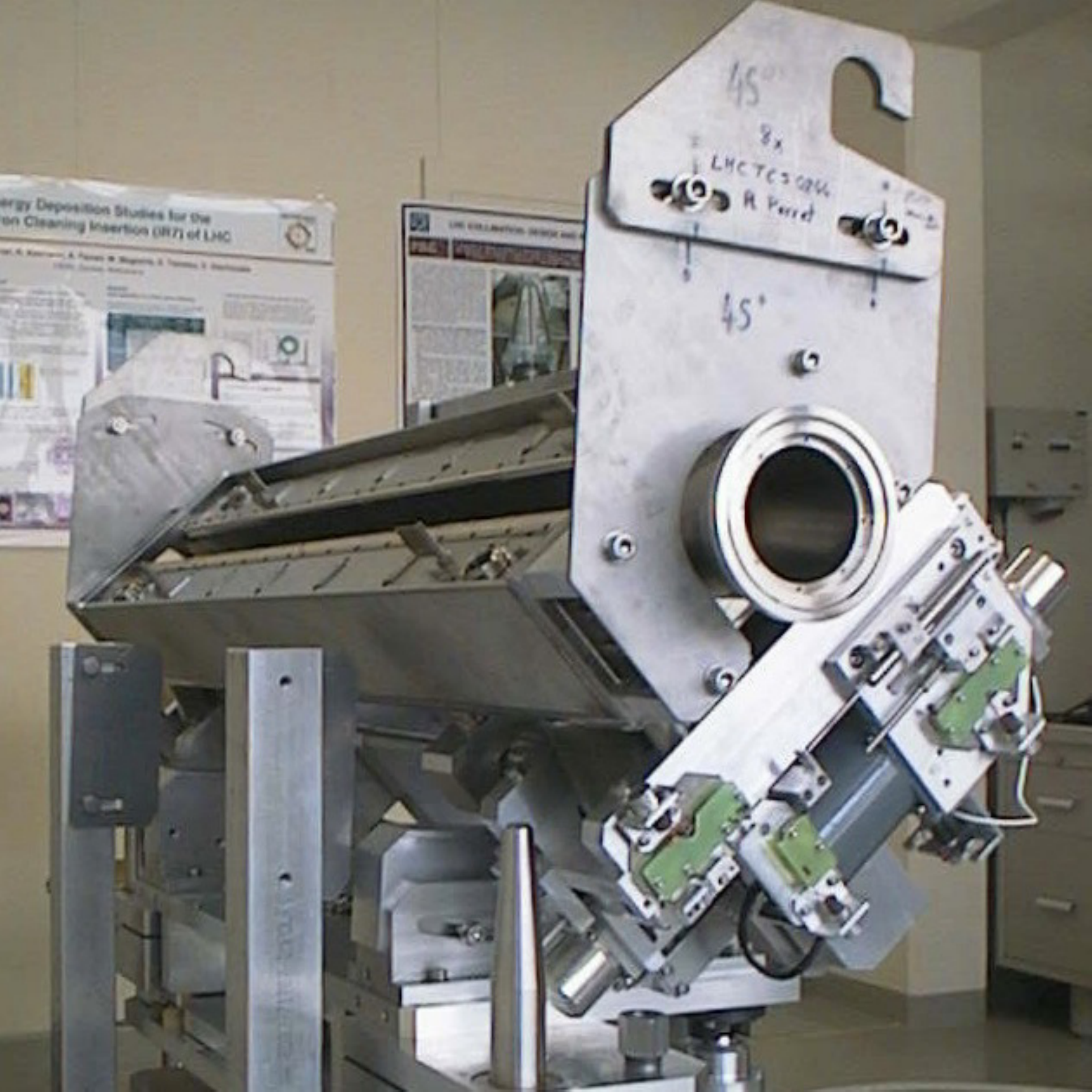}
  \vspace{-0.2cm}
  \caption{Horizontal (left) and skew (right) LHC collimators
    with open tank, showing movable jaws. The support allows assembly
    in the same collimator
    tank of all the required orientations.}
  \label{figCollLab}
\end{figure}

\begin{figure}
  \centering
  \includegraphics[width=80mm]{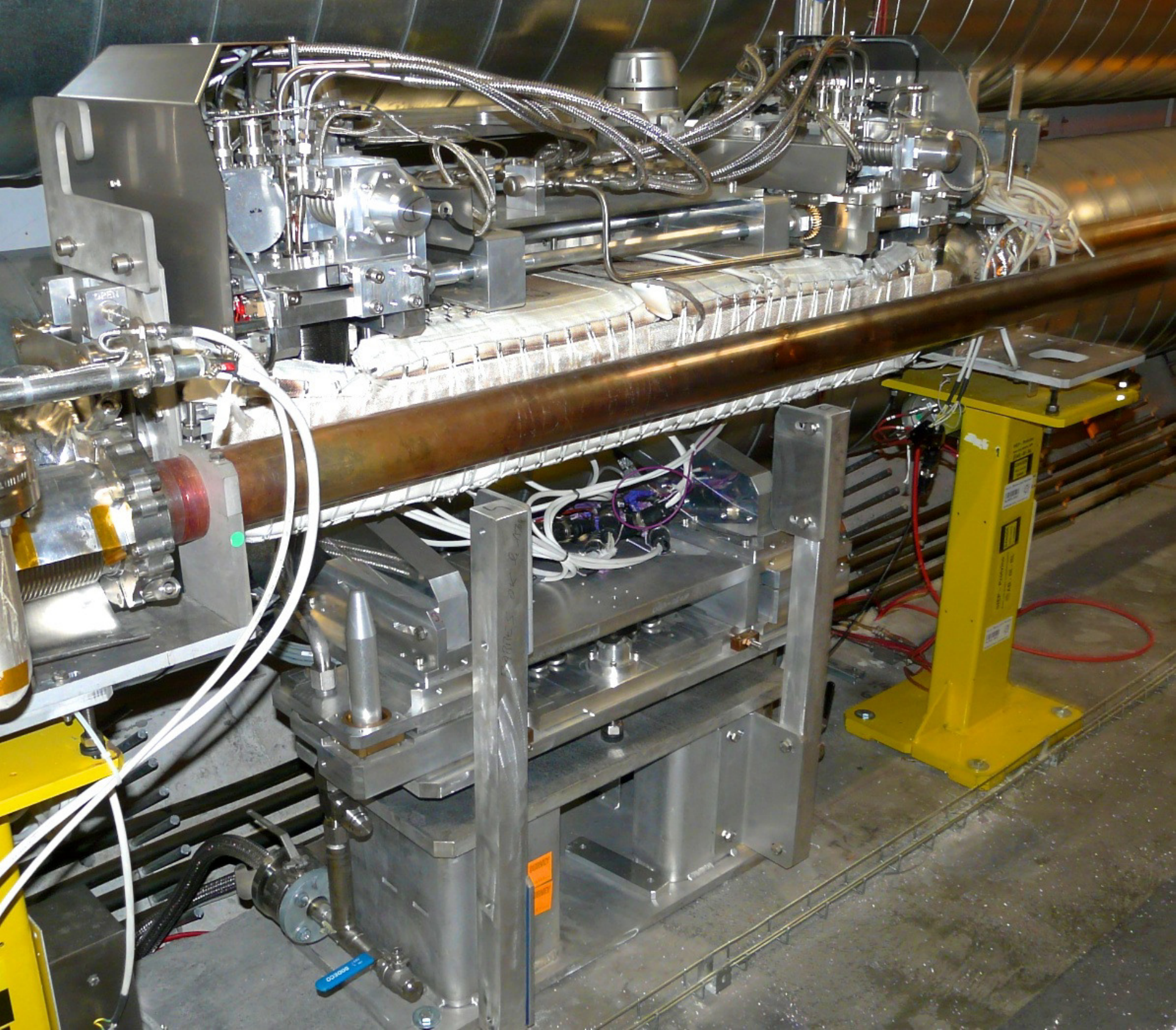}
  \vspace{-0.2cm}
  \caption{Active absorber TCLA.B6R7.B1 as installed in the
    betatron cleaning insert. The stepping motors that control jaw position
    and angle are visible on top of the vacuum tank. The pipe of
    the opposing beam is also shown. }
  \label{figTCLAtunnel}
\end{figure}

\Figure[b]~\ref{figCollLab} shows a horizontal and a $45^{\circ}$ tilted LHC collimator. Their vacuum tank is still open to show the CFC jaws inside. An example of the tunnel installation layout for a IR7 collimator is given in \Fref{figTCLAtunnel}. This is a horizontal TCLA collimator. Notice, next to the collimator, a
yellow support that supports a vacuum pump that is installed next to each collimator. A beam loss monitor, not visible in the photograph, is also
connected to this support, to record losses generated locally when the beam is intercepted by
the collimator jaws.

\begin{figure}
  \centering
  \includegraphics[width=140mm]{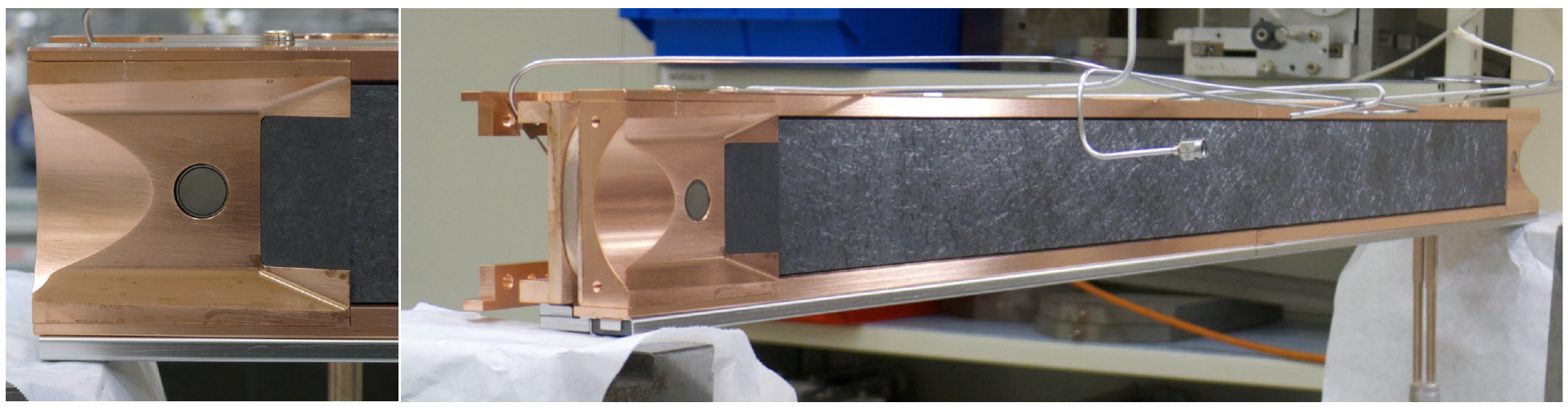}
  \vspace{-0.2cm}
  \caption{New CFC jaw with integrated beam position monitors at each
    extremity for installation in IR6 (see \Fref{figLayout}). A variant
    of this
    design, made with a Glidcop support and tungsten heavy alloy inserts on
    the active jaw part, is used for the new TCTP tertiary collimators in all
    IRs.}
  \label{figTCSP}
\end{figure}

The collimator design has been recently improved by adding two beam position monitors on either extremity of each jaw \cite{carra}.
An example of a CFC jaw prototype with this new design is shown in
\Fref{figTCSP}. This feature allows faster collimator alignment as well
as constant monitoring of the beam
orbit at the collimator, as opposed to the beam-loss-monitor-based alignment that can currently only be performed during dedicated low-intensity commissioning fills. The
beam position monitor buttons will improve collimation performance significantly in terms of
operational efficiency and flexibility, by reducing the machine time spent on
aligning collimators and the $\beta^*$ reach \cite{collDesign}. The beam-position-monitor-embedded
design is
considered the baseline for future upgraded collimator design.

\section{Cleaning performance of the LHC beam collimation}
\label{lhc-perf}
The cleaning performance of the LHC collimation system is measured by
intentionally generating trans\-verse and off-momentum beam losses while
measuring losses around the ring. This is done with low intensities
circulating in the machine. A few bunches are excited by driving the
betatron tune close to resonance or by adding transverse noise with the
transverse damper. The latter method is preferred, as it can act in a
bunch-by-bunch mode so one fill can be used for several loss maps. Large losses of
the momentum cleaning can instead be generated by changing the radio frequency. These so-called \emph{loss maps} are
used to validate, empirically, the response of the collimation system in the presence
of high loss rates. This is an essential part of the validation of the LHC
machine protection functionality, as discussed in Ref.                               \cite{jw}. In particular,
loss maps are used to verify: (1) that the hierarchy is respected, by checking
that the relative loss rates at the different collimators are in agreement with
predictions or within tolerable levels; (2) that the leakage of losses to
the other machine equipment, in particular superconducting magnets, are
as expected; (3) that the system performance remain stable during long
periods when beam-based alignment is not repeated.

\begin{figure}
  \centering
  \includegraphics[width=143mm]{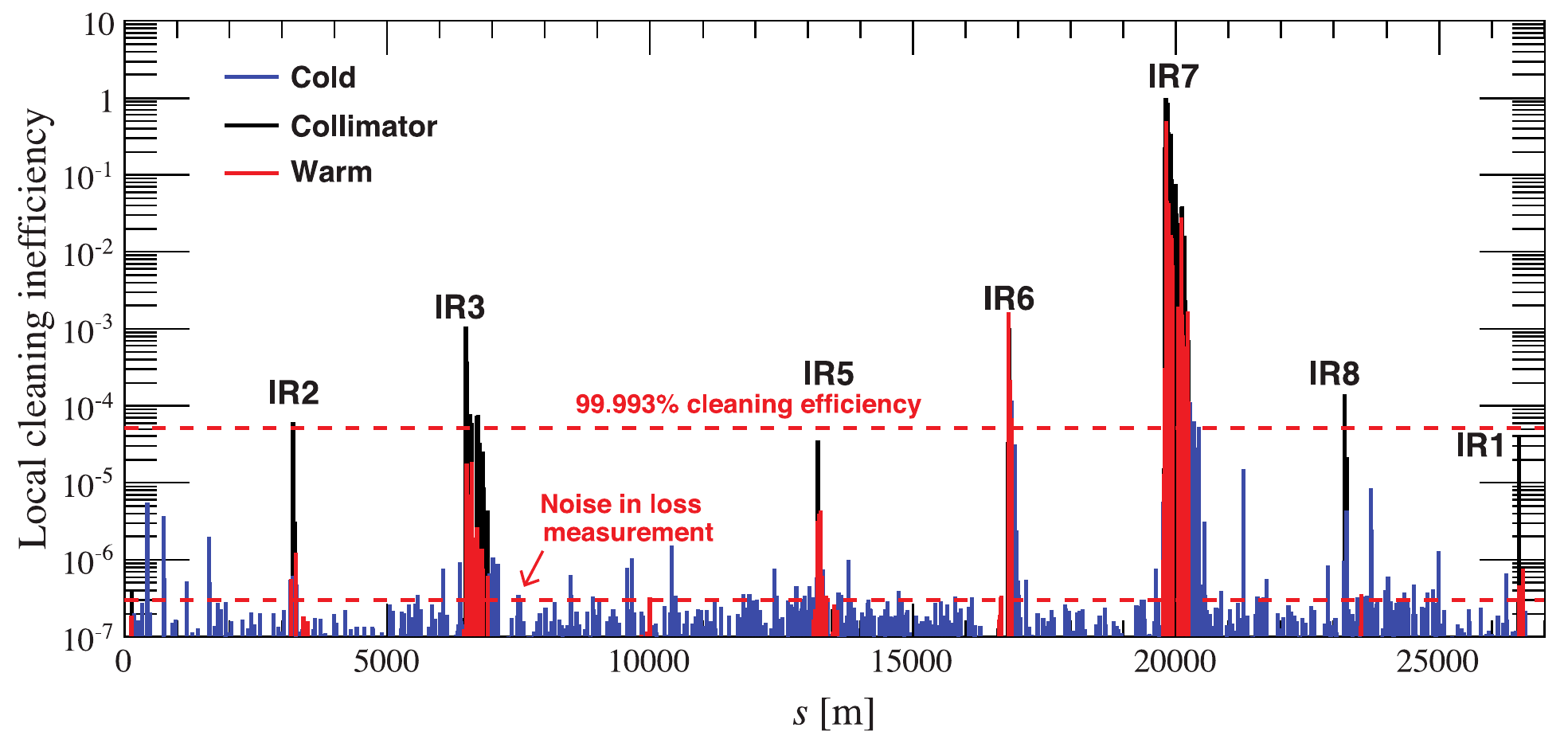}
  \includegraphics[width=143mm]{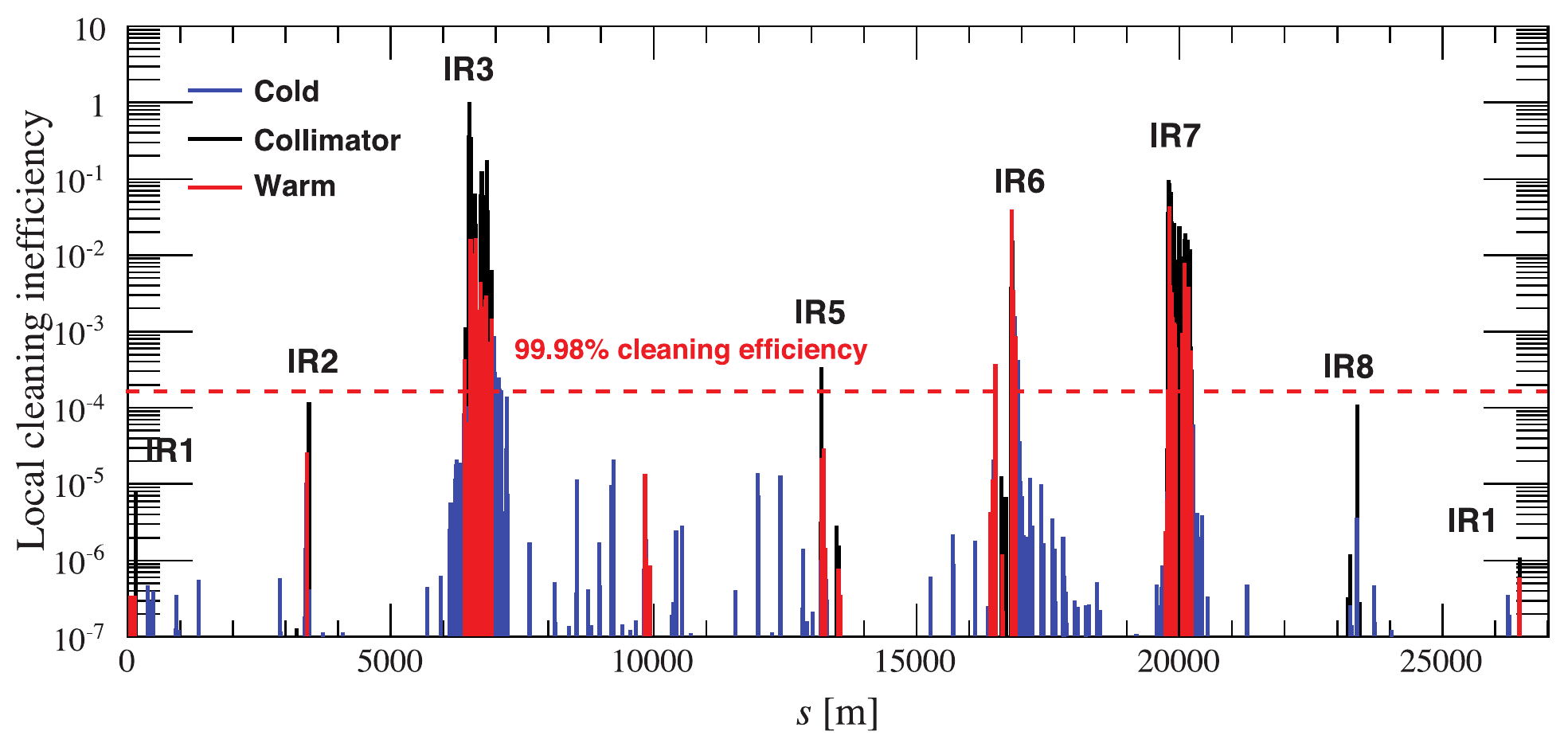}
  \vspace{-0.2cm}
  \caption{Betatron (top) and off-momentum (bottom) loss maps obtained at the LHC at $4\UTeV$ with beams squeezed to $60\Ucm$ in IR1 (ATLAS) and IR5 (CMS), showing the beam losses recorded at about 4000 beam loss monitors around the ring,
    normalized to the highest measured signal.
    Betatron losses are generated in IR7 by adding noise to the kickers of     the transverse damper of clockwise beam 1. IR3 losses are generated by     changing the radio frequency until the full beam is intercepted by the IR3
    TCP. Both beams are excited at the same time as their frequencies are     synchronized. Courtesy of B.~Salvachua \cite{collPerf}.}
  \label{figLM}
\end{figure}

Examples of betatron and off-momentum loss maps are shown in \Fref{figLM}.
These maps were recorded in 2012 at $4\UTeV$, with beams squeezed to $60\Ucm$ in IR1 and IR5. At the LHC, beam losses are recorded by about 3600 beam loss monitors around the ring \cite{bd}. To estimate the cleaning inefficiency, losses at each monitor are normalized to the highest measured signal, \ie next to the primary collimators. This is shown in \Fref{figLM} as a function
of the longitudinal coordinate $s$. It is seen that inefficiencies less than ${\sim}10^{-4}$ were achieved. In all IRs, the largest losses are recorded at the
collimators (black bars), as expected. The cold locations with the highest losses are the dispersion
suppressors downstream of the cleaning insert, as predicted in simulations
(see \Fref{fig_cl}).

\begin{figure}
  \centering
  \includegraphics[width=120mm]{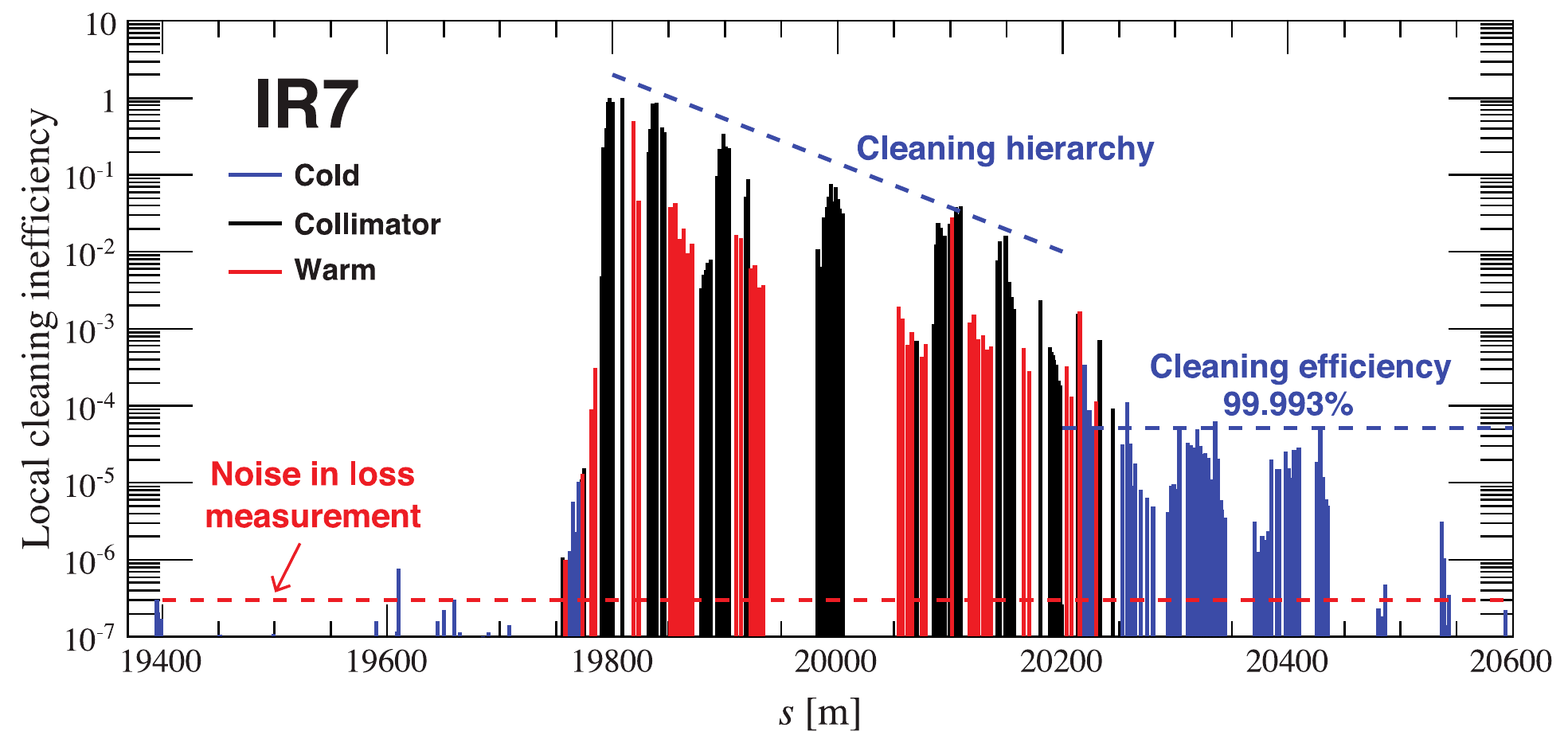}
  \vspace{-0.2cm}
  \caption{Enlargement of the top graph of \Fref{figLM}, showing details of
    losses in IR7. The limiting location for betatron cleaning is given
    by the losses on the cold magnets in the dispersion suppressor immediately
    downstream of IR7.}
\label{figLMz}
\end{figure}

The IR7 losses are given in \Fref{figLMz}. The limiting locations with the worst
cleaning are the disper-sion suppressors on either side of IR7 (the right side for
beam 1). A cleaning efficiency above 99.993\% was achieved. Note that, with the
exception of a few isolated peaks in the dispersion suppressor, the rest of the cold machine
experiences losses that are more than one order of magnitude smaller, \ie close to the noise of the beam loss monitor system.
In simulations, losses are sampled using $10\Ucm$ bins, counting the number of
beam particles hitting the aperture. In measurements, losses are measured
at the discrete locations of beam loss monitors that record the flux of ionizing particles in
the beam loss monitor volume. Clearly, these two quantities cannot be directly compared
without additional simulations of energy deposition, starting from the
multiturn loss pattern. Detailed discussion of this aspect is beyond the scope of this lecture. It suffices to say that the agreement between simulations
and measurements is good \cite{Bruce}.

\begin{figure}
  \centering
  \includegraphics[width=65mm]{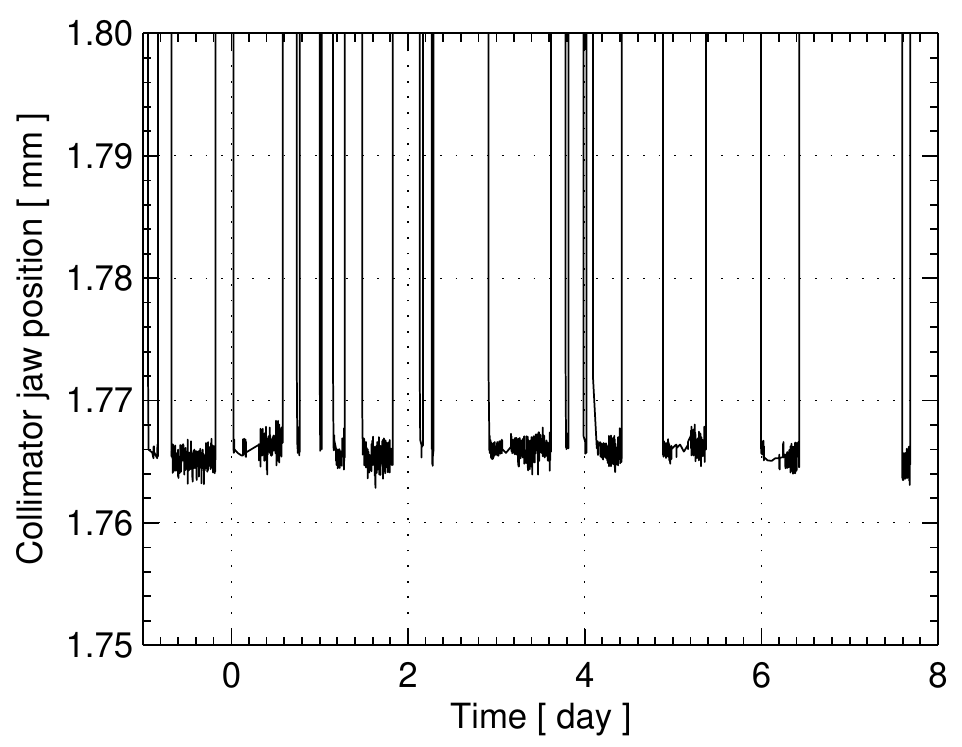}
  \vspace{-0.2cm}
  \caption{End-of-ramp settings for one TCP jaw as a function of
    time over 9 days, showing micrometre repro\-duci\-bility \cite{hb2010}.}
  \vspace{-.5cm}
  \label{figTCPstab}
\end{figure}

\begin{figure}
  \centering
  \includegraphics[width=110mm]{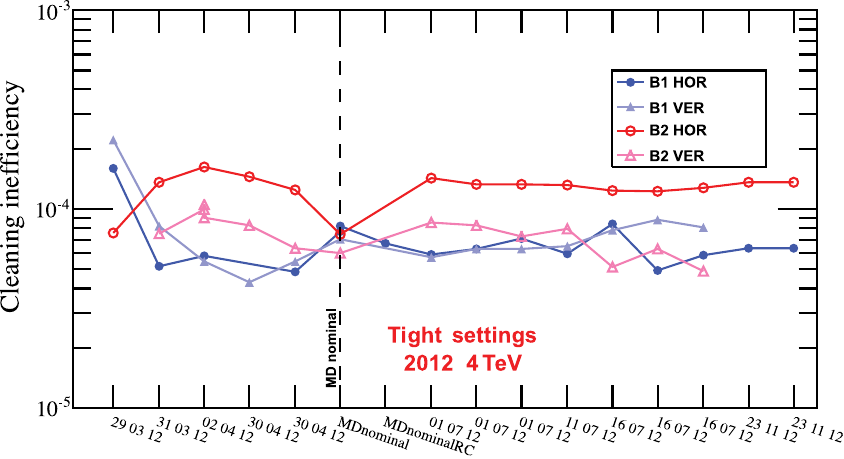}
  \vspace{-0.2cm}
  \caption{Collimation cleaning inefficiency at the worst location in the dispersion suppressors
    at either side of IR7 for both beams and planes, as measured throughout the
    2012 operation with protons ($4\UTeV, \beta^*=60\Ucm$). B,~beam; HOR, horizontal; VER, vertical. Courtesy of
    B.~Salvachua \cite{collPerf}.}
  \label{figCleaning2012}
\end{figure}

The fill-to-fill reproducibility of the collimator positions is of the order of a few
micrometres. A typical example for one jaw of a TCP collimator is given in
\Fref{figTCPstab}. This is a key ingredient for the system performance
because the collimator settings are not realigned. This stability of the hardware, together with the outstanding fill-to-fill reproducibility of optics and orbit at the LHC, makes it possible to maintain excellent collimation performance with one single beam-based alignment per year in IR3/6/7. As an
example, in \Fref{figCleaning2012} the cleaning inefficiencies at the
worst locations in the rings are shown for each beam and loss plane. It can be seen that the stability of the measured cleaning is indeed remarkable.

\section{Advanced collimation concepts for enhanced beam
  collimation}
\label{advanced}
Other advanced collimation concepts have been under study in the last year,
as possible methods of improving the performance of the LHC multistage
system. In this section, the main topics presently under investigation are
introduced. Possible immediate applications of such advanced concepts
are already under consideration for the high-luminosity upgrade of the LHC.

\subsection{Local dispersion suppressor cleaning}
Protons and ions interacting with the collimators in IR7 emerge from the IR
with a changed magnetic rigidity. This represents a source of local heat
deposition in the cold dispersion suppressor magnets downstream of IR7,
where the dispersion starts to increase: these losses
are the limiting locations for collimation cleaning, \ie they are the highest
cold losses around the ring. This may pose a certain risk for inducing magnet
quenches, in particular, in view of the higher intensities expected for HL-LHC.
This problem arises for halo collimation of both proton and heavy-ion beams.


A possible solution to this problem is to add local collimators in the
dispersion suppressors, which is only feasible with a major change of the cold
layout at the locations where the dispersion starts to increase. Indeed, the existing
system's multistage cleaning is not efficient at catching these dispersive
losses. Clearly, the need for local collimation depends on the absolute level
of losses achieved in oper\-ation and the quench limit of superconducting
magnets. In view of the uncertainties in the scaling of the current system
performance for operation at $7\UTeV$, it is important to take appropriate margins,
to minimize the risk of limitation in the future.

\begin{figure}
  \centering
  \includegraphics[width=160mm]{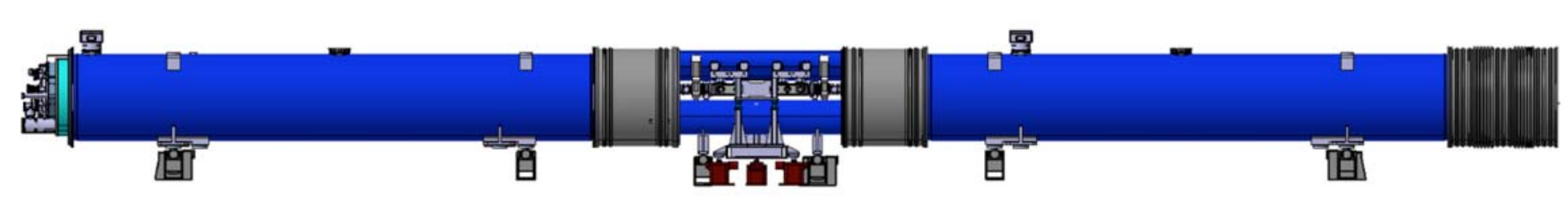}
  \vspace{-0.2cm}
  \caption{Longitudinal integration of a TCLD collimator between two short
    $11\UT$ dipoles. Courtesy of D.~Ramos}
  \label{fig_tcryo1}
  \vspace{-0.4cm}
\end{figure}

\begin{figure}
  \centering
  \includegraphics[width=100mm]{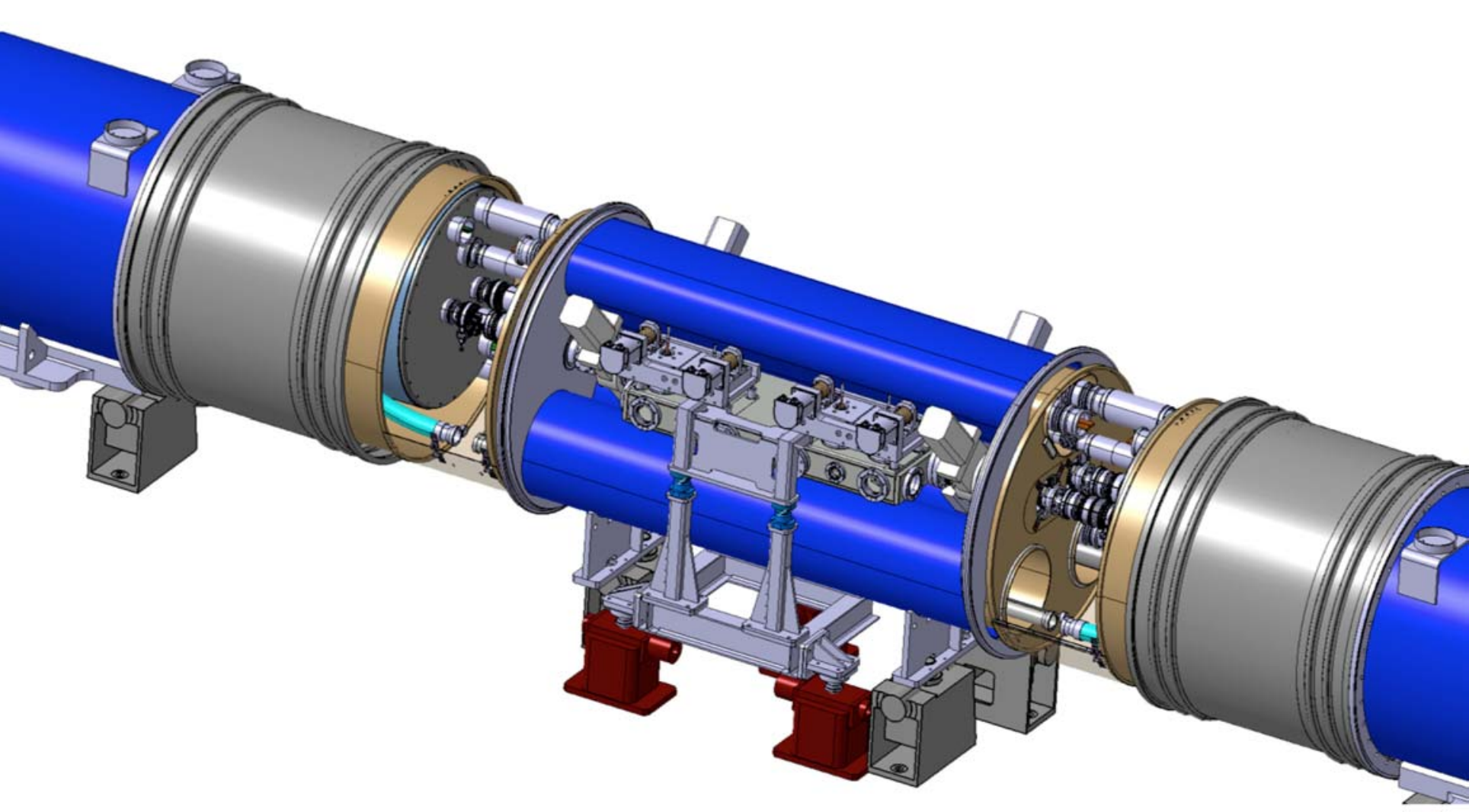}
  \vspace{-0.2cm}
  \caption{Three-dimensional view of the TCLD installation in the cryogenic by-pass region
    between two $11\UT$ dipoles. Courtesy of D.~Ramos and
    L~Gentini.}
  \label{fig_tcryo2}
  \vspace{-0.4cm}
\end{figure}

A solution with minimum impact on the cold section layout is to replace the
existing $15\Um$ long dipoles with two shorter, higher-field magnets, by freeing
enough space to install a warm collimator. This solution is illustrated in
\Fref{fig_tcryo1}. It requires an $11\UT$ dipole field to free sufficient
space for a warm collimator to be installed in a dedicated cryogenics
by-pass system, as shown in \Fref{fig_tcryo2}. Even in this tight space limitation,
an adequate solution can be found. New dipoles and collimators are being
prototyped at CERN, providing a viable solution for IR7 cleaning upgrades
that might already be available for a long LHC stop planned for 2019. Note
that this solution is modular and was designed to be implemented easily
in any existing dipole location. It can therefore also be used to
improve cleaning around collision points, if necessary, as is foreseen
for the ALICE ion experiment \cite{delD55}.

\subsection{Status on research and development on novel collimator materials}

The LHC impedance budget is largely dominated by the contribution of the LHC
collimators. For this reason, the current collimation system was conceived in such a
way that it can be easily upgraded to reduce the impedance \cite{finalColl}: every
secondary collimator slot in IR3 and IR7 features a companion slot for the
future installation of a low-impedance secondary collimator. A total of 22
slots in IR7 and 8 slots in IR3 are already cabled for a quick installation
of new collimators---referred to as TCSPMs---which can either replace or
supplement the existing TCSG collimators. The TCSPMs will include pick-ups for orbit
measurements (`P') and will be based on metal composites (`M').
In addition, limited robustness against beam losses of the present tungsten
collimator has already limited the LHC performance of run~I in terms of
$\beta^*$ reach, because adequate margins had to be taken in the collimation
hierarchy to shield the existing tertiary collimators properly
\cite{roderikBeta}.

A rich programme of research and development was initiated, to find novel material with optimum
response to thermo\-mechanical stress and with reduced impedance, to improve
various limitations of  cur\-rent LHC collimator materials. More details are
given in a companion paper \cite{ab}. Simulations predict that beam stability can be re-established for all
HL-LHC scenarios if the CFC of the existing secondary collimators is replaced, at
least in the betatron cleaning insertion (IR7), with a jaw material having an electrical conductivity a factor of 50 to 100 higher than CFC \cite{delD24}. This improvement could easily be achieved if the jaw material were made of highly conductive metals, such as copper or molybdenum. However, secondary collimators in IR7 also play a crucial role in LHC machine protection and might
be exposed to large beam losses. Therefore, collimator materials and designs must also be robust against beam failure. The driving requirements for the development of new materials are thus: (i) low resistive-wall impedance, to avoid beam instabilities; (ii) high cleaning efficiency; (iii) high geometrical stability, to maintain the  precision of the collimator jaw during
operation despite temperature changes; and (iv) high structural robustness, in
case of accidental events, such as single-turn losses.

\begin{figure}
  \centering
  \includegraphics[width=130mm]{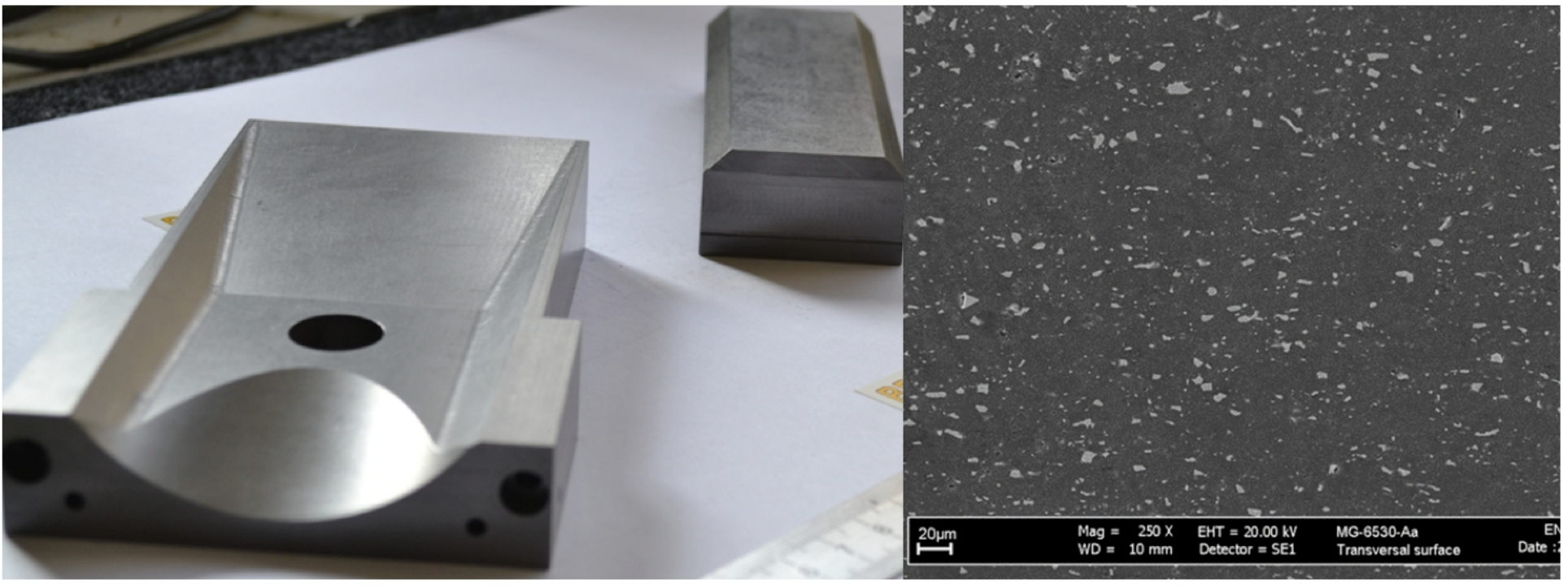}
  \vspace{-0.2cm}
  \caption{Left: MoGr components recently produced by Brevetti Bizz (Italy)
    for a jaw prototype. Jaw extremity (dimensions of $147 \times 88 \times 25\Umm^3$) and
    jaw absorbing block ($125 \times 45 x\times\Umm^3$) are shown. A jaw assembly
    includes two jaw extremities (taperings) and eight blocks. Right: Detail of microstructure; the graphite matrix is visible, together with
    molybdenum carbide grains of about $5\Uum$. Courtesy of A.~Bertarelli.}
  \label{figMoGR}
  \vspace{-0.4cm}
\end{figure}

The current baseline for the upgraded secondary collimators relies on novel carbon-based ma\-terials, such as molybdenum carbide-graphite (MoGr), a ceramic composite jointly developed by CERN and Brevetti Bizz, in which the
presence of carbides and carbon fibres strongly catalyses the graphitic
ordering of carbon during high-temperature processing, enhancing its thermal and electrical properties (\Fref{figMoGR}). To further improve
their surface electrical conductivity, these materials could be coated
with pure molybdenum or other lower-$Z$ refractory coatings. Replacing all existing CFC secondary colli\-mators in both IR7 and IR3 with bulk MoGr or MoGr
coated with $5\Uum$ thick pure molybdenum would reduce the total LHC impedance by 40\% or 60\%, respectively.

The new collimator design, along with novel materials and possible alternative coatings must be validated for operation in the LHC. For these purposes, a rich programme of validation is in progress, in\-volving tests at the CERN HiRadMat facility \cite{hrm}, to address robustness against beam impact,
mechanical
engineering prototyping, beam tests at the LHC, and experimental verification
of the material response under high radiation doses. It is anticipated that this test will be completed, and the production of new,
low-impedance, highly robust collimators in the LHC started, by 2019.

\subsection{Halo diffusion control techniques}
The 2012 operation experience indicates that the LHC collimation would profit from halo control mech\-anisms. The idea is that, by controlling the diffusion speed of halo particles in an aperture range between the core and the TCP opening (${\approx}3$--$5\sigma_z$), one can act on the time profile of the losses. The main goal is to reduce loss rates that would otherwise take place in
a short time, or simply to control the static population of halo particles in a
certain aperture range. For example, it is expected that losses caused by orbit drifts \cite{srHalo} during the squeeze (see Figs. ~\ref{fig_lt_meas} and
\ref{fig_lt_sq}) can be strongly reduced by actively depleting the halo
population.

\begin{figure}
  \centering
  \includegraphics[width=95mm]{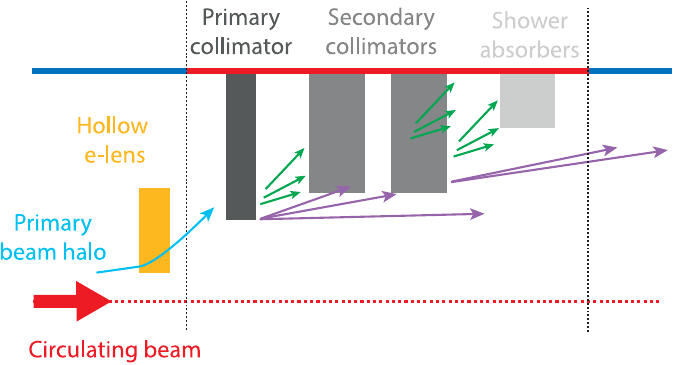}
  \vspace{-0.2cm}
  \caption{Integration of hollow e-lenses as halo diffusers in
    the present collimation system.}
  \label{figHEB}
  \vspace{-0.4cm}
\end{figure}

One of the best candidate techniques for achieving active halo control at the LHC
is to use the hollow e-lens collimation concept \cite{heb, heb2}. A hollow
electron
beam, running coaxially to the proton or ion beam, is used to generate an annular
beam in the transverse $(x,y)$ plane. This hollow beam induces an
electromagnetic field, which affects halo particles above a certain transverse amplitude and can change their transverse speed. The working
principle is illustrated in \Fref{figHEB}. A solid experimental basis
achieved at the Tevatron indicates that this solution is very promising for the LHC. The design for an hollow e-lens for the LHC is ongoing (see
\cite{lhcHEB} and references therein).

Conversely, in the case of loss spike limitations at the LHC during run~II, the hollow e-lens solu\-tion would not be viable because it could only be implemented over a time-scale of a few years \cite{srHEB}. It is, therefore, crucial to work on alternatives that, if necessary, might be implemented on
an appro\-priate time-scale. Two alternatives are currently being considered: tune modulation through noise in the cur\-rent of lattice quadrupoles, as outlined in Ref.~\cite{tuneRipple}, and narrow-band excitation of halo particles, using the transverse
damper system. Though very different from the hardware point of view, both these techniques rely on exciting tail particles through resonances induced in
the tune space. This method works on the assumption that there is a
correlation between halo particles with large amplitudes and
corresponding tune shifts in tune space (de-tuning with amplitude). Clearly, both methods require solid ex\-peri\-mental verification in a very low noise machine, like the LHC, in particular, to demonstrate that these types of excitation do not perturb the beam core emittance. Unlike hollow e-lenses, which act directly in the transverse plane by affecting particles at amplitudes above the inner radius of the hollow beam, resonance excitation
methods require a good knowledge of the beam core and tail particle tunes, even
in dynamic phases of the operational cycle. It is planned to test these
techniques experimentally in LHC run~II.

\subsection{Crystal collimation}
\begin{figure}
  \centering
  \includegraphics[width=95mm]{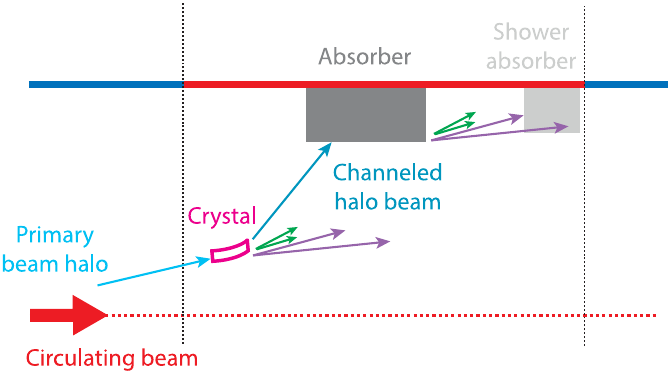}
  \vspace{-0.2cm}
  \caption{Crystal collimation concept foreseen the use of a bent crystal
    to channel halo particle in one single passage to a dedicated absorber,
    reducing significantly the number of secondary collimators.}
  \label{figCRY}
  \vspace{-0.4cm}
\end{figure}

Highly pure bent crystals can be used to steer high-energy particles that become
trapped between the poten\-tial of parallel lattice planes \cite{cry}. Equivalent bending fields of up to hundreds of teslas can be achieved in crystals with a
length of only 3--$4\Umm$, enabling, in principle, halo particles to be steered to a
well-defined point, with obvious potential applications to beam collimation. As
opposed to standard pri\-mary collimators based on amorphous materials, which require several secondary collimators and ab\-sorbers to catch the products developed
through the interaction with matter (\Fref{fig_multi}), one single
absorber per collimation plane is, in theory, sufficient in a crystal-based collimation system. This is shown in \Fref{figCRY}.

In addition to the reduction of secondary collimators, nuclear interactions with well-aligned crys\-tals are much reduced compared with a primary collimator, provided that high channelling efficiencies of halo particles can be achieved (particles impinging on the crystal are to be channelled within a few turns). This
is expected to reduce dispersive beam losses in the dispersion suppressor of the
betatron cleaning insertion significantly, compared with the existing system, which is limited by
the leakage of particles from the primary collimators. Simulations indicate a
possible gain of between 5 and 10 \cite{dmThesis}, even for a layout without an
optimized absorber design. The crystal collimation option is particularly interesting for collimation of heavy-ion beams, thanks to the reduced
probability
of ion dissociation and fragmentation compared with current primary
collimators.

Another potential of crystal collimation is a strong reduction of the machine impedance, since (1) only a small number of collimator absorbers is required and (2) the absorbers can be spaced much farther apart, owing to the large bending angle from the crystal (40--$50\Uurad$ instead of a few
microradians from multiple Coulomb scattering in the primary collimator). Conversely, an appropriate absorber design must be conceived to handle the design peak loss rates, of $1\UMW$ during $10\Us$, expected for the LHC
upgrade \cite{hl}. Other potential issues concern the machine protection aspects of this scheme, which has not yet been studied in detail, and
operational aspects for crystals that require mechanical angular stability in
the
submicroradian range through the operational cycle. Note that
the critical angle beyond which channelling is lost is ${\approx}2\Uurad$ at $7\UTeV$.

Promising results were achieved in dedicated crystal collimation tests at  the
SPS, performed from 2009 within the UA9 experiment
\cite{cry1, cry2, cry3}. However, some outstanding issues on the
feasibility of the crystal collimation concept for the LHC can only be
addressed by beam tests at high energy in the LHC. For this purpose, a study at the LHC has been proposed, and will take place in the LHC run~II
\cite{dmThesis, cryLHC}. Tests at the LHC will address the feasibility of the
crystal collimation concept with LHC beam conditions, in
particular, to demonstrate that such a system can provide better cleaning than
the present high-performance system throughout the operational cycle.




\section*{Acknowledgements}
The material presented here is the result of the work of many people and was
presented on behalf of the LHC collimation project team. Past and present
team members are gratefully acknowledged. In particular, R.~Bruce, D.~Mirarchi,
B.~Salvachua, and G.~Valentino are sincerely thanked, for providing
material for this document and for providing useful comments on the manuscript.



\end{document}